\documentclass[twocolumn,nofootinbib]{aastex6}
\usepackage{graphicx}	
\usepackage{amsmath}
\usepackage{amssymb}
\usepackage{subfigure}
\usepackage{color}
\usepackage{txfonts}
\usepackage{tabto}
\usepackage{braket}
\usepackage{lineno}

\newcommand\nG{$\rm \MakeLowercase{n}G\,$}

\newcommand\rmd{{\rm d}}
\newcommand\aobs{{A_{\rm obs}}}
\newcommand\eobs{{E_{\rm obs}}}

\newcommand\simlt{\lower.5ex\hbox{$\; \buildrel < \over \sim \;$}}
\newcommand\simgt{\lower.5ex\hbox{$\; \buildrel > \over \sim \;$}}

\graphicspath{{figures/}{skymaps/}}
\begin{document}

\title{Treasure Maps for Detections of Extreme Energy Cosmic Rays}
\author{No\'emie Globus\altaffilmark{1,2},  Anatoli Fedynitch\altaffilmark{3,4}, Roger D. Blandford\altaffilmark{5,6}}
\altaffiltext{1}{Department of Astronomy and Astrophysics, University of California, Santa Cruz, CA 95064, USA}
\altaffiltext{2}{Astrophysical Big Bang Laboratory, RIKEN, Wako, Saitama, Japan}
\altaffiltext{3}{Institute of Physics, Academia Sinica, Taipei City, 11529, Taiwan} 
\altaffiltext{4}{Institute for Cosmic Ray Research, the University of Tokyo, 5-1-5 Kashiwa-no-ha, Kashiwa, Chiba 277-8582, Japan}
\altaffiltext{5}{Kavli Institute for Particle Astrophysics and Cosmology (KIPAC), Stanford University, Stanford, CA 94305, USA}
\altaffiltext{6}{SLAC National Accelerator Laboratory, 2575 Sand Hill Road, Menlo Park, CA 94025, USA}

\begin{abstract}
The origin of Ultra High Energy Cosmic Rays is a 60-year old mystery. We show that with more events at the highest energies (above 150~EeV)  it may be possible to limit the character of the sources and learn about the intervening magnetic fields. Individual sources become more prominent, relative to the background, as the horizon diminishes.   
An event-by-event, composition-dependent observatory would allow a ``tomography'' of the sources as different mass and energy  groups probe different GZK horizons. A major goal here is to provide a methodology to distinguish between steady and transient or highly variable sources.
Using recent Galactic magnetic field models, 
 we calculate ``treasure'' sky maps to identify the most promising directions for detecting Extreme Energy Cosmic Rays (EECR) doublets, events that are close in arrival time and direction. On this basis, we predict the incidence of doublets as a function of the nature of the source host galaxy.  Based on the asymmetry in the  distribution of time delays, we show that observation of doublets might distinguish source models. In particular the Telescope Array hotspot could exhibit temporal variability as it is in a ``magnetic window'' of small time delays.
 These considerations could improve the use of data with existing facilities and the planning of future ones such as  Global Cosmic Ray Observatory - GCOS.
\end{abstract}

\section{Introduction}
Understanding the nature and provenance of Extreme Energy Cosmic Rays (EECRs) draws together many scientific communities. To the traditional cosmic ray physicist, it is the culmination of a century's development of experimental technique and stimulates the development of new detectors. To the astrophysicist, it represents the challenge of reverse engineering cosmic accelerators capable of accelerating nuclei to everyday energy, To the particle physicist, it exhibits collisions with center of mass energies $\sim100$ times those attainable at the LHC and the ever-present opportunity to uncover new physics. For the data scientist there is the possibility of deploying new machine learning techniques  to improve the measurement of energy and primary identification from shower data.  To the cosmologist it allows a unique probe of physical conditions within the local and, indirectly, the remote universe. 

In this study, we focus especially on what can be learned from the highest energy cosmic rays with energies in excess of $\sim150\,{\rm EeV}\equiv1.5\times10^{20}\,{\rm eV}\equiv2.4\times10^8\,{\rm erg}$. Extrapolating the spectrum, a very rough estimate of the intensity above 150 EeV is $\sim4\times10^{-22}{\rm cm}^{-2}\,{\rm sr}^{-1}\,{\rm s}^{-1}$. The combined detector exposure per year is $\sim2\times10^{21}\,{\rm cm}^2\,{\rm sr}$ and so the rate of detection is roughly one per year. The associated cosmic ray number and energy densities are $\sim2\times10^{31}\,{\rm cm}^{-3}$ and $\sim{5\times10^{-23}\rm erg\,cm}^{-3}$, respectively. For comparison, the microwave background energy density is $\sim10^{10}$ time larger. (If we were to reduce the energy threshold to $\sim50\,{\rm EeV}$ then these densities increase by roughly ten.) 

However, the lifetime of these particles is short -- $\sim30\,{\rm Myr}$ for protons and iron nuclei -- and so the associated luminosity density can be estimated as $\sim5\times10^{-38}\,{\rm erg\,cm}^{-3}\,{\rm s}^{-1}$, roughly one thousandth the total cosmic ray luminosity density and roughly one millionth the galaxy stellar luminosity density. Clearly, any putative source should have a larger luminosity density and be found within the cosmic ray horizon.

There is an additional requirement on the accelerator. The most efficient accelerators are essentially electromagnetic and there should be electric potential difference $V>150/Z\,{\rm V}$, where $Z$ is the nuclear charge, within the acceleration region. Within a source with magnetic field $B$ and size $L$, the electric potential difference is limited by $V\lesssim300(B/1{\rm G})(L/1\,{\rm cm})\,{\rm V}$. Under electromagnetic conditions this suggests a power associated with the source in excess of $\sim V^2/Z_0\sim2\times10^{45}Z^{-2}\,{\rm erg\,s}^{-1}$, where the impedance $Z_0\sim100\,{\rm Ohm}$ is related to that of free space \citep{2000PhST...85..191B}. This is a serious limitation if the highest energy cosmic rays turn out to be protons. 

Two giant experiments, Telescope Array  (TA) in the North  and Pierre Auger Observatory (PAO) in the South, independently confirmed the existence of a high energy cutoff in the 
spectrum at around 50 EeV \citep{Abbasi:2007sv}, and find compatible results on a mixed mass composition up the highest energies \citep{PierreAuger:2017tlx,Hanlon:2018dvd}. The first signal to pass the $5\sigma$ discovery threshold is a large-scale dipole anisotropy in the distribution of cosmic rays above 8 EeV \citep{2017Sci...357.1266P}. The large-scale anisotropy seems to correlate with the local distribution of matter \citep{2017ApJ...850L..25G,DGF20}. Above 40 EeV, small scale anisotropies start to appear \citep{abbasi2014}, but their statistical significance is much lower and more data is needed to make secure identifications.

Both observatories reported a handful of high energy cosmic rays events above 150 EeV (the most extreme energy EECRs and the one we focus on the paper). At the highest energies, the cosmic rays lose energy and mass by interaction with the extragalactic background light (mostly CMB photons) \citep[see][for a review]{Allard2012}. Therefore, the horizon (defined here as the distance from which 95\% of the cosmic rays of a given energy and composition have been lost) is limited to  our local supercluster ($\sim$40 Mpc) for proton and iron nuclei at 150~EeV. Intermediate mass cosmic rays such as CNO elements are more fragile to photodisintegration on the CMB, and their horizon is smaller. It is likely that the two current observatories will continue to accumulate events for the next decade or more, and EECR events close in arrival time and direction (``multiplets''), might be observed. It is therefore important to understand what such detections can tell us about the nature of the sources.

 \begin{figure*}
\centering
\includegraphics[width=0.49\textwidth]{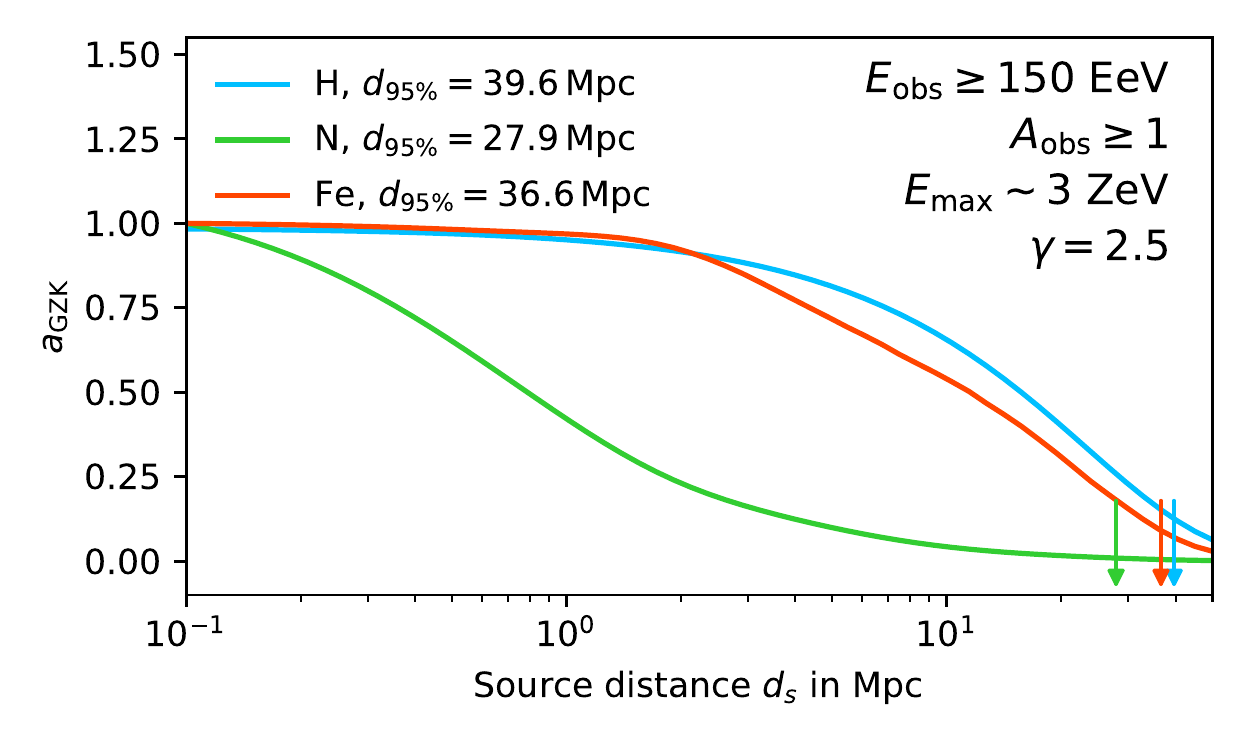}
\includegraphics[width=0.49\textwidth]{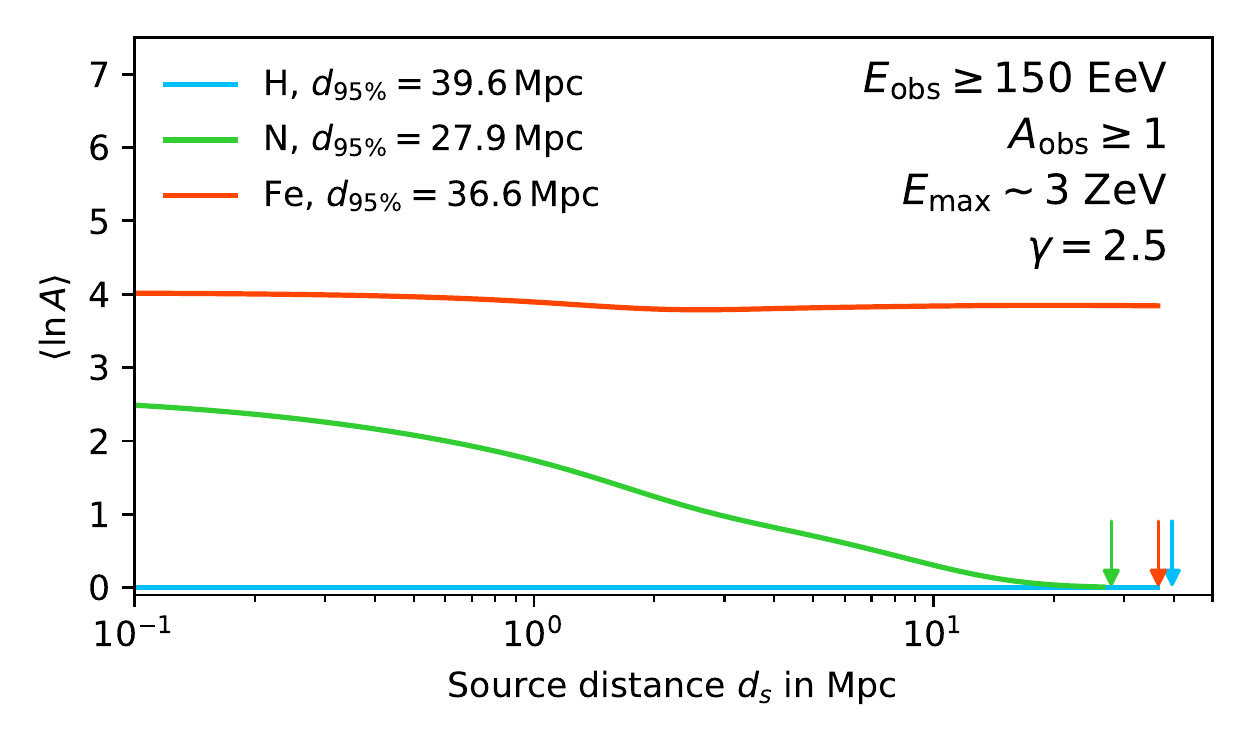}
\includegraphics[width=0.49\textwidth]{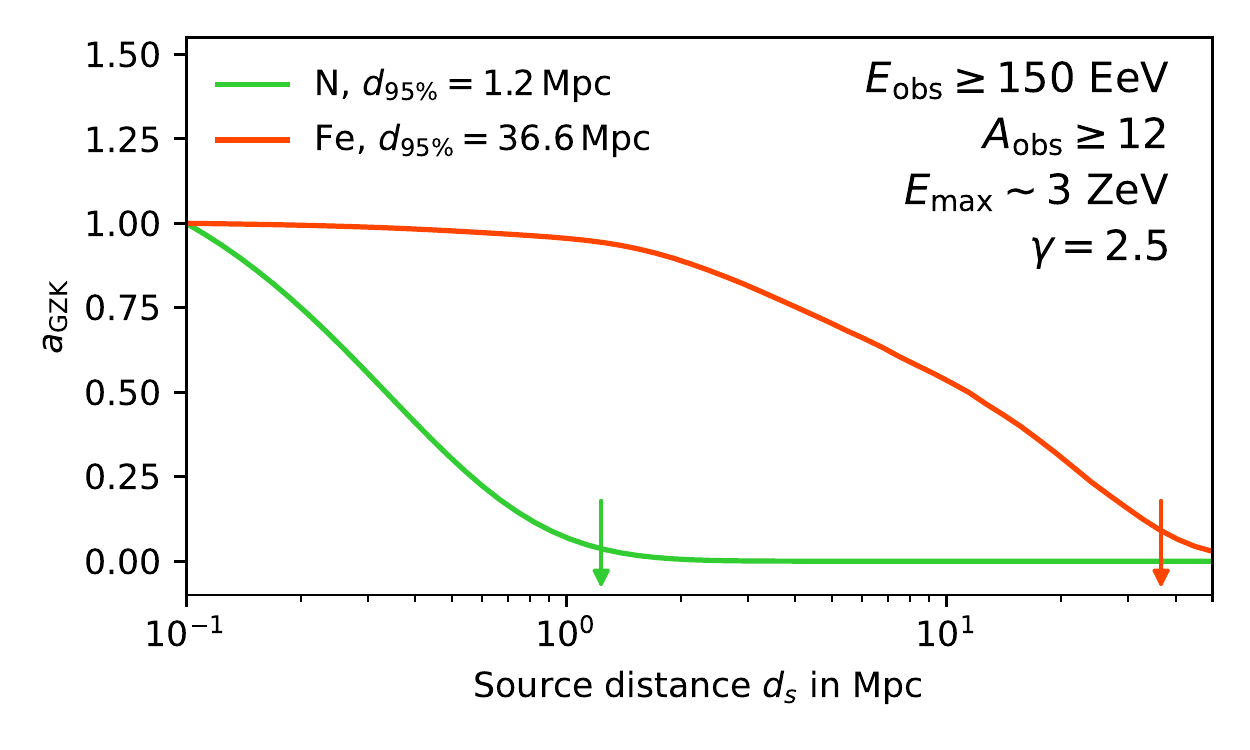}
\includegraphics[width=0.49\textwidth]{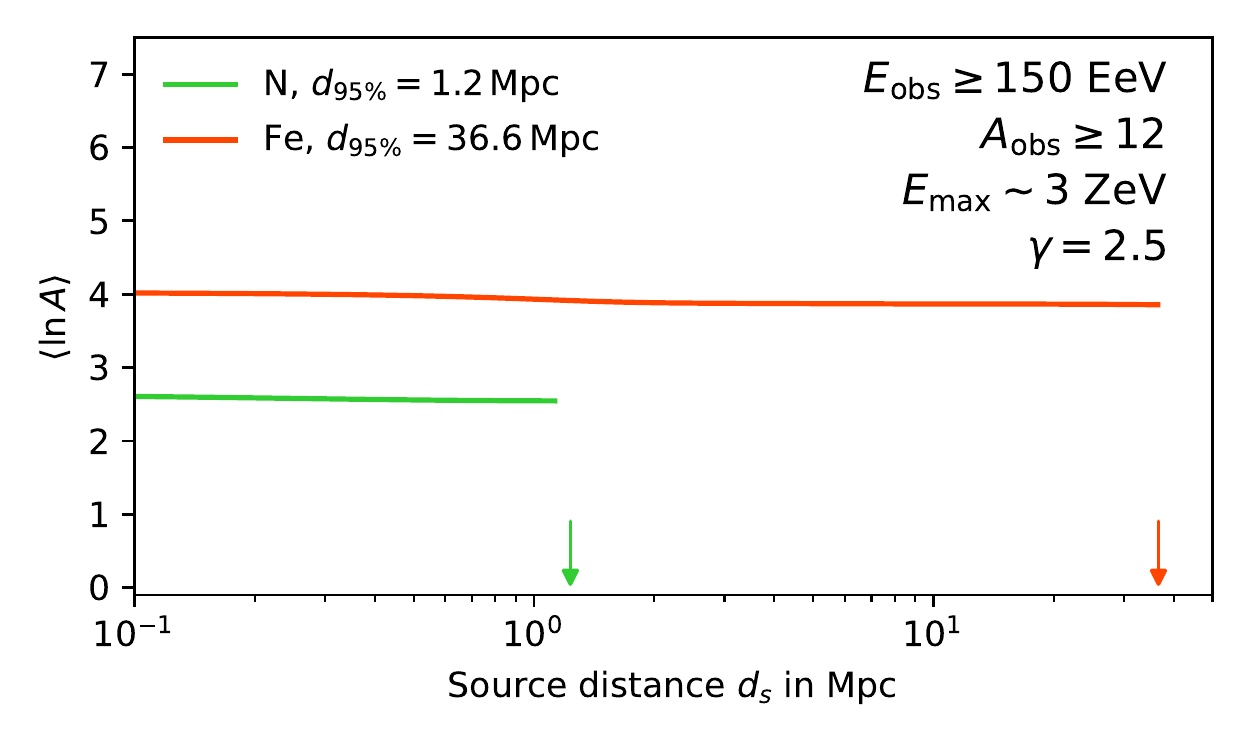}
\caption{Left panel: Loss of number density, $a_\text{GZK}$, for $\eobs\geq150$ EeV. The top  shows the case $\aobs\geq 1$ and the bottom the case $\aobs\geq 12$, to demonstrate the impact of on the observed volume in case of a composition-sensitive detector that can provide a sub-sample of nucleus-like events. On the right, the evolution of the composition, $<\ln A>$, is shown as a function of the source distance, $d_s$.}
\label{fig:containement_figures150}
\end{figure*}

Although many classes of source model have been eliminated, there are several that remain under investigation and which can exhibit very different temporal behavior. For example, there are single, rare, fast, transient events such as Gamma-Ray Bursts, where the delays are due to multi-path propagation to Earth. There are longer duration single sources, such as tidal disruption events, that could be active for years. There are variable point sources, like AGN jets or magnetars, where the cosmic ray luminosity is likely to exhibit flaring. Finally there are steady sources, like intergalactic shock fronts, where no variation is expected and the angular size can be very large. Positive detection of and upper limits to EECR multiplets can be equally prescriptive for further limiting the source options.

The Galactic magnetic field (GMF) is responsible for the point source magnification $M$. There are several effects at work. If there were just an ordered field, then cosmic rays would be mirrored away before they reached Earth, as happens in our magnetosphere. This reduces the flux from a point source. Secondly, the large scale field can both increase and decrease this flux through magnetic focusing/lensing \citep{harari2000}. Here the intensity, the flux per solid angle is the same as it is at the source, however the solid angle changes, thereby changing the flux. Finally, the turbulent magnetic field splinters the particle trajectories into ever-diminishing regions spreading over an increasing area. Intensity will still be conserved along a given trajectory but when averaging over $\sim\delta\theta$, the mean intensity is reduced while the flux is maintained. In addition, it leads to pitch angle scattering and reduces mirroring. It is the second effect that is most important for determining the magnification so long as the deflection is small; when the deflection is large, the cosmic rays from a given sources will be mostly rejected by the Galaxy. However if there are other sources, sufficiently closeby, they will compensate. In the limit that the source is uniform over the sky, all paths will have identical intensity and the cosmic ray number density will be the same as in intergalactic space.

Thanks to the recent progress made in the modeling of the lensing effects \citep{Farrar2015optics} of the GMF, we are  able to determine the extragalactic source direction from the cosmic-ray arrival direction. This allows us to predict anisotropies for a given source distribution presuming the positions of individual sources could be resolved without scattering. The turbulent component of the GMF, responsible for introducing diffusion into cosmic rays transport, is mostly unknown. 
Cosmic rays scattering off magnetic fluctuations will have different diffusion times in the Galaxy, which will introduce a temporal dispersion, $\tau_d$ in arrival times from a given source, depending on its direction.  

Based on the available GMF models, we can provide an estimate of the temporal dispersion as a function of the direction of the source in the sky and the rigidity (momentum per unit charge). Multiplets of EECR events are more likely to be detected in ``magnetic windows'' of the sky where the temporal dispersion, $\tau_d$, is small enough to detect at least one doublet within the typical observation time of an EECR observatory, typically decades. This window has to back-project to   source candidates, which is not guaranteed when the sources are scarce due to the limited horizon. 
 \begin{figure*}
\centering
\includegraphics[width=0.49\textwidth]{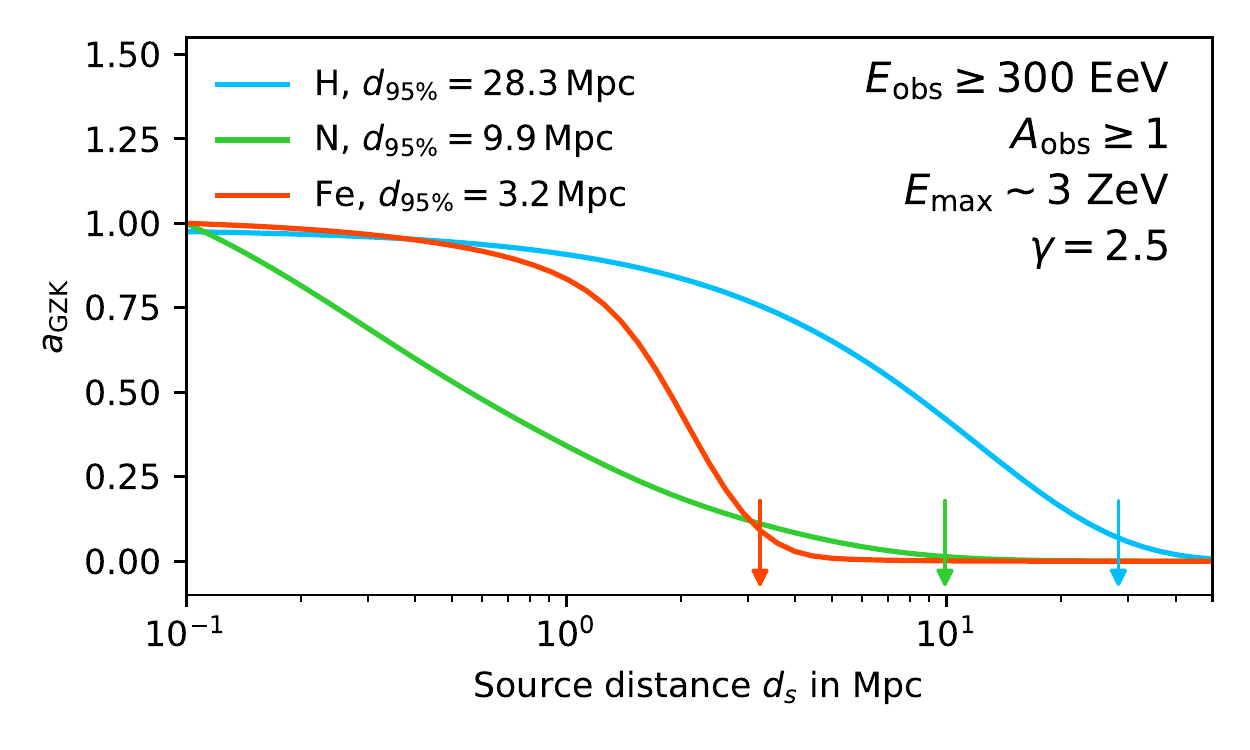}
\includegraphics[width=0.49\textwidth]{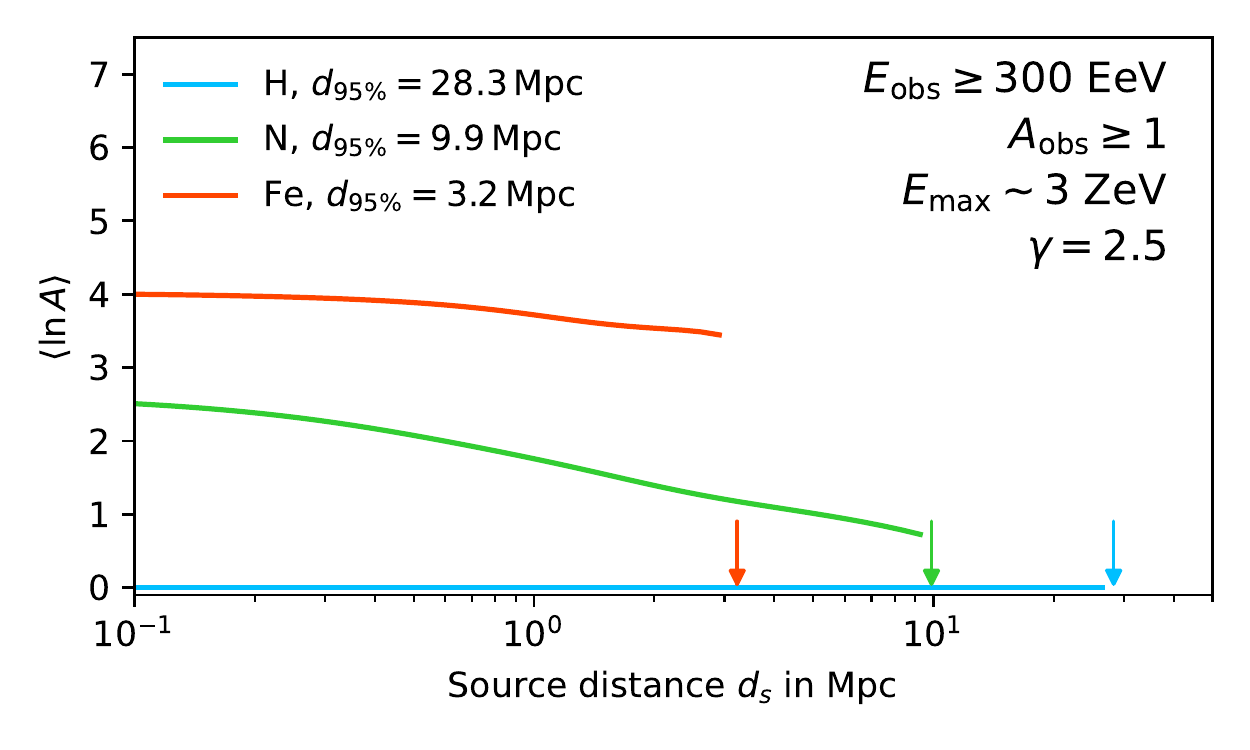}
\includegraphics[width=0.49\textwidth]{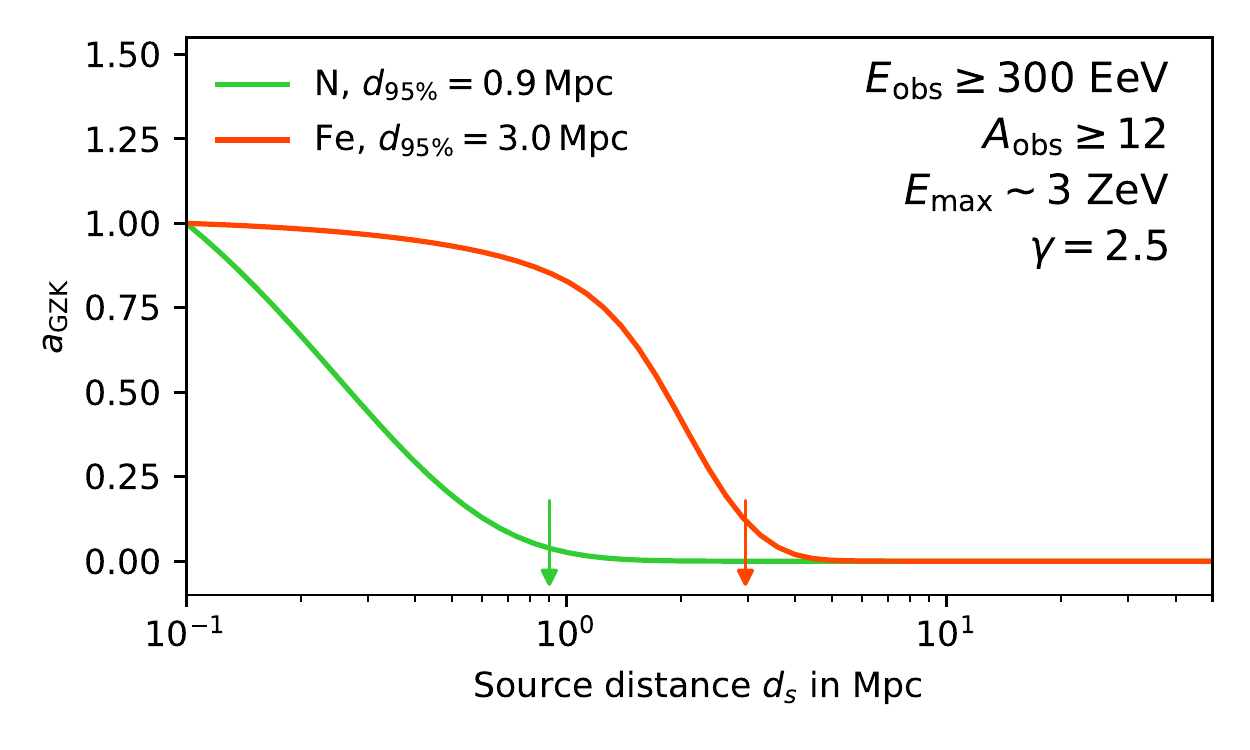}
\includegraphics[width=0.49\textwidth]{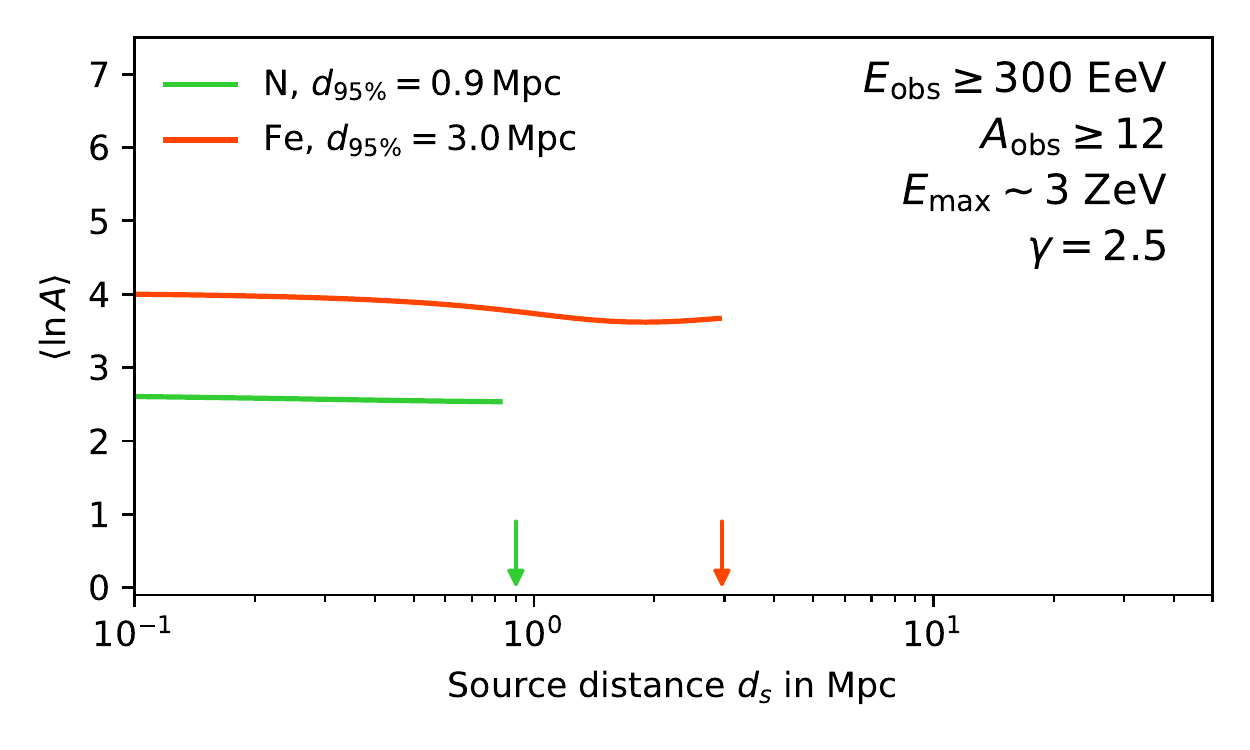}
\caption{Same as Figure~\ref{fig:containement_figures150} for $\eobs\geq300$ EeV.}
\label{fig:containement_figures300}
\end{figure*}

So, the questions arise: ``What directions of the sky are more promising to detect EECR temporal coincidences?'' And, ``what can we learn from future detections of EECR multiplets? ''  We address these questions in this paper, which is organized as follows.   In sec.~\ref{sec:gzk-horizon}, we calculate the GZK horizon for various type of EECR nuclei. 
In sec.~\ref{sec:doublets}, we derive the conditions for detecting more than one EECR from a single source, given the knowledge on the temporal dispersion introduced by the turbulent magnetic fields in the Galaxy and in the extragalactic medium. In sec.~\ref{sec:method}, we present our method to calculate the temporal dispersion of the EECR signal due to the GMF, as a function of the extragalactic source direction, the GMF model and the rigidity of the particle. By combining  sky maps of the temporal dispersion with the true directions  of nearby galaxies, we  derive ``treasure'' maps of the most promising arrival directions for observing EECRs temporal coincidences for PAO and TA sky coverage. We present these sky maps in sec.~\ref{sec:maps}. We  calculate the number of host galaxies candidates for transient able to provide a doublet as a function of the energetic of the transients after taking into account GZK attenuation, field of view of the experiments (TA/Auger) and GMF magnification. In sec.~\ref{sec:time_analysis} we show that a temporal analysis of the events could help us to distinguish between transients and continuous sources. We discuss the implications for the Telescope Array hotspot in sec.~\ref{sec:TAspot}. We summarize our results in sec.~\ref{sec:summary} and discuss the implications for source candidates in sec.~\ref{sec:conclusions}.

\section{The GZK-Horizon and its Implications for a composition-sensitive detector}
\label{sec:gzk-horizon}

Given the tremendous challenges of EECR detection, we must seek compromise whether to optimize for precision in determining the mass  event-by-event, or, if the prospects discoveries are higher by constructing an array of unprecedented size that dramatically increases the exposure beyond the cutoff of energy spectrum.

The ubiquitous Cosmic Microwave Background (CMB) as well as the Extragalactic Background Light (EBL) imposes a distance limit on the source of the highest energy cosmic rays. Protons above $\sim70$ EeV  undergo pion production and swiftly loose energy. This is known as the Greisen-Zatsepin-Kuzmin cutoff as it was predicted by \cite{Greisen:1966jv}, and  \cite{Zatsepin:1966jv}, two years after the discovery of the CMB. In addition, Ultra-high energy nuclei passing through this radiation field will be gradually stripped off their nucleons and lose energy.  Therefore, any observed high energy cosmic ray proton can be the remnant of a heavier nuclei starting its journey at larger distances than the strict horizon for protons. This phenomenon does increase the volume for potential sources; it also suggests that observing heavy nuclei at extreme energies from such sources is less likely. The observational capability for the mass composition above 100 EeV thus carries critical information about the EECR source distance.

The GZK effect was originally defined for protons, and formally, the attenuation of nuclei on the CMB predominantly has a different physical origin in the Giant Dipole Resonance in contrast to the photo-hadronic threshold in the case of protons \citep[e.g.][]{allard2008}. Nonetheless, for the sake of simplicity we call ``the GZK Horizon'' the size of the universe which is transparent to EECRs.

To compute the opacity of the local volume to EECR some assumptions about the sources have to be made. This work mainly focuses on rare transient sources, from which we hope to observe one or more events. (Steady sources will not produce any temporal associations.) Since the nature of these sources is unknown, their spectral characteristics remain in the realm of speculation. Nonetheless, we need to assume either an emission spectrum or an anomalous source that produces a monochromatic pulse of cosmic rays.

For simplicity, we model the spectrum of the cosmic rays emitted by potential sources (denoted by the $s$ index) as a power-law spectrum with spectral index $\gamma$ and a cutoff energy $E_{s, \text{max}}$:
\begin{equation}
\label{eq:source_spectrum}
    \frac{\rmd N_s}{\rmd E_s}(E_s, E_{s, \text{min}}, E_{s, \text{max}}) \propto \left (\frac{E_s}{E_{s, \text{min}}} \right )^{-\gamma} e^{-\frac{E_s}{E_{s, \text{max}}}}, E_s>E_{s,\rm{min}}.
\end{equation}
Due to the small number of observed events, for softer spectra $\gamma \gtrsim 3$ the most likely emission energies would lie close to the threshold ($E_{s, \text{min}}$). For harder spectra $\gamma < 2$ and higher $E_{s, \text{max}}$, the most likely emission energies that would lead to observed events are close to the cutoff, somewhat resembling ``anomalous'' or monochromatic sources.

The extragalactic propagation of EECR between the source location and the Earth is modeled using the code {\sc PriNCe} \cite{Heinze:2019jou}. The source spectrum from Eq.~\ref{eq:source_spectrum} is repeatedly injected at distances $d_s$ for various choices of $E_{s, \text{min}}$, $E_{s, \text{max}}$, $\gamma$, and the several nucleus masses from $A_s=1$ (protons) up to $A_s=56$ (iron). To model these two cases we introduce the parameters $\eobs$ and $\aobs$, which are the lower thresholds for the energy and mass number of cosmic rays observed at Earth. The requirement for imposing a mass threshold $\aobs$ to observational data, implies that the future detector can identify mass event-by-event, or at least can separate proton-like from nucleus-like events on a statistical basis.

We define the spectrum at the distance of the Earth as
\begin{equation}
    \frac{\rmd T_{A_i}}{\rmd E}( d_s, \gamma, E_{s, \text{min}}, E_{s, \text{max}})
\end{equation}
separately for each nuclear mass $A_i \leq A_s$. The lower masses are populated due to the disintegration of nuclei in photo-nuclear and photo-hadronic interactions.

We work towards the definition of the GZK horizon using the \textit{loss of number}, in analogy to an attenuation coefficient:
\begin{equation}
\begin{split}
\label{eq:agzk}
    a_\text{GZK}(&A_s, d_s, E_{s, \text{max}}, \gamma~|~\aobs, \eobs) =\\ &=\frac{\sum_{A_i \geq \aobs}\int_\eobs^\infty \rmd E \frac{\rmd T_{A_i}}{\rmd E}( d_s, \gamma, E_{s, \text{max}})}{\int_\eobs^\infty \rmd E_s \frac{\rmd N_s}{\rmd E_s}(E_s, E_{s, \text{max}})},
\end{split}
\end{equation}
setting $E_{s, \text{min}}$ to $\eobs$. For several choices of $\eobs$, $\aobs$, $\gamma$ and $A_s$, the resulting $a_\text{GZK}$ are shown Fig.~\ref{fig:containement_figures150} for $\eobs\geq150$ EeV and in Fig.~\ref{fig:containement_figures300} for $\eobs\geq300$ EeV. The figures contain two choices for $\aobs\geq 1$ and $\aobs\geq 12$, that demonstrate the impact of composition on the observed volume in case of a composition-sensitive detector that can provide a sub-sample of nucleus-like events.

Although for most of the following arguments we use $a_\text{GZK}$ directly, it is convenient to define a simpler variable $d_{95\%}$ (which can be used as synonym for the GZK horizon) that denotes the distance at which 95\% of the number density is lost during propagation (shown as arrows on the $x$-axes and in the legend of Figs.~\ref{fig:containement_figures150} and \ref{fig:containement_figures300}).

Without imposing a mass threshold at detection, the horizon $d_{95\%}$ for $\eobs = 150$~EeV is of the size of our local supercluster, roughly 30-40 Mpc, irrespective of the composition at the sources. Intermediate mass nuclei suffer from strongest attenuation and mass loss through photodisintegration ($cf.$~upper right panel Figs.~\ref{fig:containement_figures150}). However, for more extreme values of $E_{\rm max}$ and $\gamma$ this distance can substantially grow for CNO, but less so for protons or iron. If the detector is composition-sensitive and the sources accelerate nuclei that are detected as such on Earth, the horizon reduces dramatically to $d_{95\%}\sim1$~Mpc for a higher mass threshold $\aobs=12$. 

At 300~EeV, only protons have a horizon of $d_{95\%}\lesssim30$~Mpc, which includes the Virgo cluster of galaxies at $\sim$~16 Mpc (and the M87 jet or the strong accretion shocks as possible EECR sources). Any intermediate or heavy nuclei would originate from sources at $\lesssim 3$~Mpc. Therefore, if a future composition-sensitive array reports an EECR detection with CNO or heavier masses at $\sim300$~EeV, it will be a strong indication that EECRs sources are transients, since there are no active galactic nuclei or strong accretion shocks within our local group.

From the above definition we find that the GZK horizon can not be rigorously defined and that it almost entirely depends on the choices for $\eobs$ and $\aobs$, and to a lesser extent on the assumptions on the source properties. A composition-dependent observatory would allow a ``tomography'' of the local universe with cosmic-rays, as different mass groups probe different GZK horizons. Hence, the derivation of the horizon size is largely affected by the capabilities of the detector to estimate the mass and energy of the detected events. This means that we can select the horizon in the design stage of a future observatory. This case is significantly different from neutrino astronomy and binary black hole GW sources where there is no practical horizon, and the diffuse background from high-redshift sources is an unavoidable consequence. Thus, EECR astronomy could be a viable path to performing source searches on a dark, highly anisotropic sky if such extreme sources exist within our local volume.

It should be noted that the simulations with {\sc PriNCe} do not include extragalactic magnetic fields, i.e. assuming that EECRs travel on a straight line. Extragalactic magnetic fields can decrease the horizon further \citep[e.g.,][]{2008A&A...479...97G} given sufficient strength. The transition from a ballistic to a diffusive regime occurs when the coherence length of the turbulent field is of the order of the gyroradius. Assuming a turbulent magnetic field of 1~nG and an outer turbulence scale of 1~Mpc, this transition occurs for a rigidity $\sim 1$~EV. At the high rigidities considered here, the propagation is nearly ballistic and the effect of extragalactic magnetic fields on the  size of the horizon is negligible.

\section{Necessary condition for EECR doublets}\label{sec:doublets}

Consider an extragalactic, transient source  at a distance $d_s$, that emits instantaneously  a total energy $U_{\rm CR}(E)$ in cosmic rays of energy $E$.  The presence of non-Gaussian statistics would require detecting more than one cosmic-ray from the same source. This condition can be expressed as \citep{globus2016} 
 \begin{eqnarray}  
 U_{CR}(E) /E \ge  [4\pi\,  d_s^2 (\Delta \Omega/4\pi)/A][\tau_d/\Delta t]\,,
 \label{Eq:Nh}
 \end{eqnarray}
where $\Delta t$  is the period of observation, $\tau_d$ is the dispersion of the flight times  of the cosmic rays propagating from the source to Earth,  $A$ is the time average detector area.  The exposure (area x exposure time x field of view) of the detector is ${\cal E}=A\Delta t$. For PAO, the annual exposure is ${\cal E}_{\rm PAO} \sim 5,500\,\, \rm{km^2\, sr\, yr}$. The solid angle subtended by the beam of CR emerging from the source, $\Delta \Omega$, can be written as: $\Delta\Omega=4\pi U_{CR}/U_{\rm iso}$, $U_{\rm iso}$ being the true source isotropic equivalent energy. Now, we need to account for the flux attenuation due to the GZK effect that depends on the source distance, $a_{\rm GZK}$ calculated in Sec.~\ref{sec:gzk-horizon}, and the magnification (or demagnification) due to the GMF lensing effects, $M(l,b)$, that depends on the source direction and that we will calculate in Sec.~\ref{sec:method} assuming a GMF model. The GMF model will also allow us to calculate the angular separation between the doublets due to the combined effect of lensing and scattering. For the high rigidities considered in this paper, the angular separation is always small, of the order of 10 degrees.   

So, for a transient source at 50 $d_{50}$ Mpc to create a doublet of 2 EECR events at $200 E_{200}$ EeV, the minimum isotropic equivalent energy (denoted as $U_{\rm iso,2}$) would need to be 
\begin{equation}
U_{\rm iso,2} \sim 4.38 \cdot 10^{52}{\rm erg}\, (\tau_d/10^3 {\rm yr})d_{50}^{2}{\,} E_{200}^{\,} {({\cal E}/{\cal E}_{\rm PAO})}^{-1} a_{\rm GZK}^{-1} M^{-1} n_{\rm yr}^{-1} \,,
\label{eq:u_iso}
\end{equation}
where $n_{\rm yr}$ is the number of years of observation. For TA, the annual exposure is ${\cal E}_{\rm TA} \sim 900\,\, \rm{km^2\, sr\, yr}$, so the isotropic equivalent energy  needed would have to be  $\sim$six times higher than for PAO.

The temporal dispersion $\tau_d$ for two cosmic rays with the same rigidity, has two contributions, one  from the GMF, $\tau_{d,\text{GMF}}$, which we calculate in Sec.~\ref{sec:method} assuming a GMF model, and one from  the extragalactic magnetic field (EGMF), $\tau_{d,\text{EGMF}}$. 
Assuming a purely turbulent extragalactic field with a Kolmogorov spectrum with a rms intensity $B_{\rm nG}$ and a coherence length $\lambda_{\rm Mpc}$, the average extragalactic dispersion is \citep{Lemoine1997} 
\begin{equation}
    \tau_{d, {\rm EGMF}}\sim 5\, {\rm yr}\, (E_{200}/Z)^{-2} B_{\rm nG}^2 {d_s}_{\rm Mpc}^2 \lambda_{\rm Mpc}\,.
\end{equation} 

 The total dispersion is calcualted from $\tau_{d} = \sqrt{\tau_{d,\text{GMF}}^2 + \tau_{d,\text{EGMF}}^2}$, since both contributions are uncorrelated statistical processes.
Now EECR doublets will not have exactly the same rigidity and we will address the time evolution in rigidity in sec.~\ref{sec:time_analysis}. However for the sake of simplicity, we consider that the $U_{\rm iso}$ derived in Eq.~\ref{eq:u_iso} remains valid as long as the difference in rigidity between the two events is negligible. 
Note that  if the source is steady at luminosity $L$, then the ratio $\tau_d/\Delta t$ in Eq.~\ref{Eq:Nh} should be replaced with unity and the total energy should be set to $U_{CR} = L\Delta t$.

\section{Models and methods}\label{sec:method}

An important advantage of studying EECRs is the absence of a ``diffuse'' cosmic ray background from distant ($>100$ Mpc) sources that reduces the signal of nearby sources. This background still exists at 100 EeV since at this energy the horizon is roughly 100 Mpc. At present, a significant fraction of Ultra-High Energy Cosmic Rays (UHECRs) can not be associated with individual sources or populations \citep{PierreAuger:2022axr}. The multi-EeV UHECRs seen by current detectors are also a disadvantage for observing transients due the loss of temporal and spatial coherence introduced by the diffusive delays and the magnetic deflections, respectively. Given the apparent advantages of EECRs, we are interested in characterizing how many source candidates from our local volume (or local supercluster) can be expected to be accessible. To answer this question one must consider several arguments:
\begin{enumerate}
    \item[{\small $\bullet$}] Transients sources, such as gamma-ray bursts or tidal disruption events, are, presumably, correlated with the distribution of galaxies. Except for very exotic and equally exciting cases, each galaxy can act as potential host, whereas voids are not expected to contribute.
    \item[{\small $\bullet$}] Due to the deflections within the GMF, parts of the sky are inaccessible to observations since there are no valid trajectories that can connect the Earth to a source. On the other hand, magnetic lensing can magnify certain directions, for which many spatially close trajectories originate from the same direction.
    \item[{\small $\bullet$}] Due to the turbulence of the GMF, cosmic rays arriving from the same direction have slightly different travel times leading to temporal dispersion $\tau_{d,\text{GMF}}$, similar to the arguments outlined in the previous section. It has to be combined with the dispersion due to the EGMF.
    \item[{\small $\bullet$}] The source distances are limited by the attenuation functions $a_\text{GZK}$ due to the GZK effect (see Sec.~\ref{sec:gzk-horizon}). Thus, the number of candidate sources  depends on the choice of $\aobs$, $\eobs$ and $A_s$, and to some extent on the assumed source parameters, like $E_{\rm max}$ or $\gamma$.
    \item[{\small $\bullet$}] One needs to consider the location and zenith angle coverage in case of a ground based detector. In particular, this is important for heavier compositions where the accessible regions of the sky don't coincide with the exposure function at Earth.
\end{enumerate}
The following sections outline the modeling of these arguments.

\subsection{Source catalogs}
\begin{figure}[!h]
\centering
\includegraphics[width=.98\columnwidth]{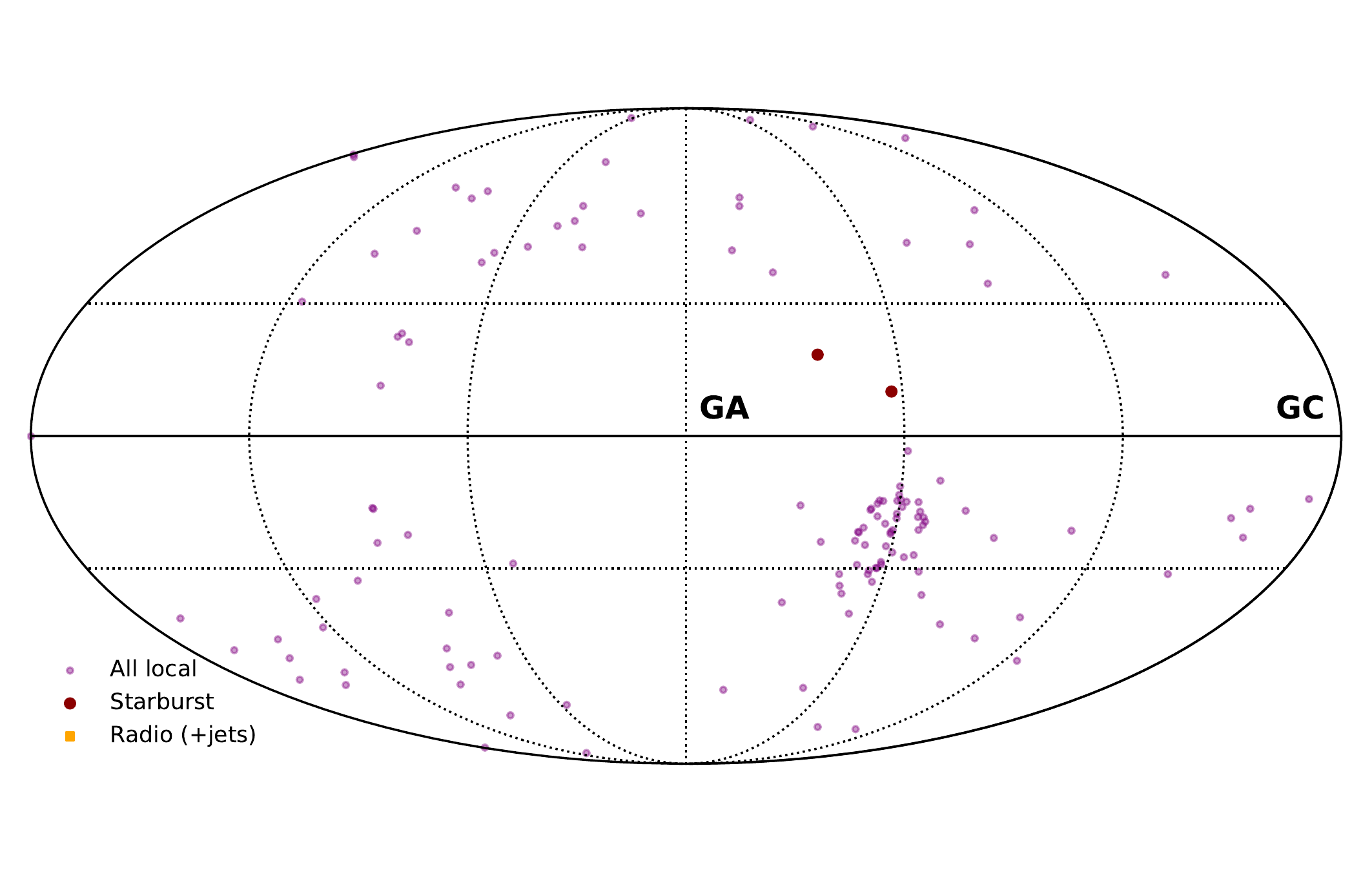}
\includegraphics[width=.98\columnwidth]{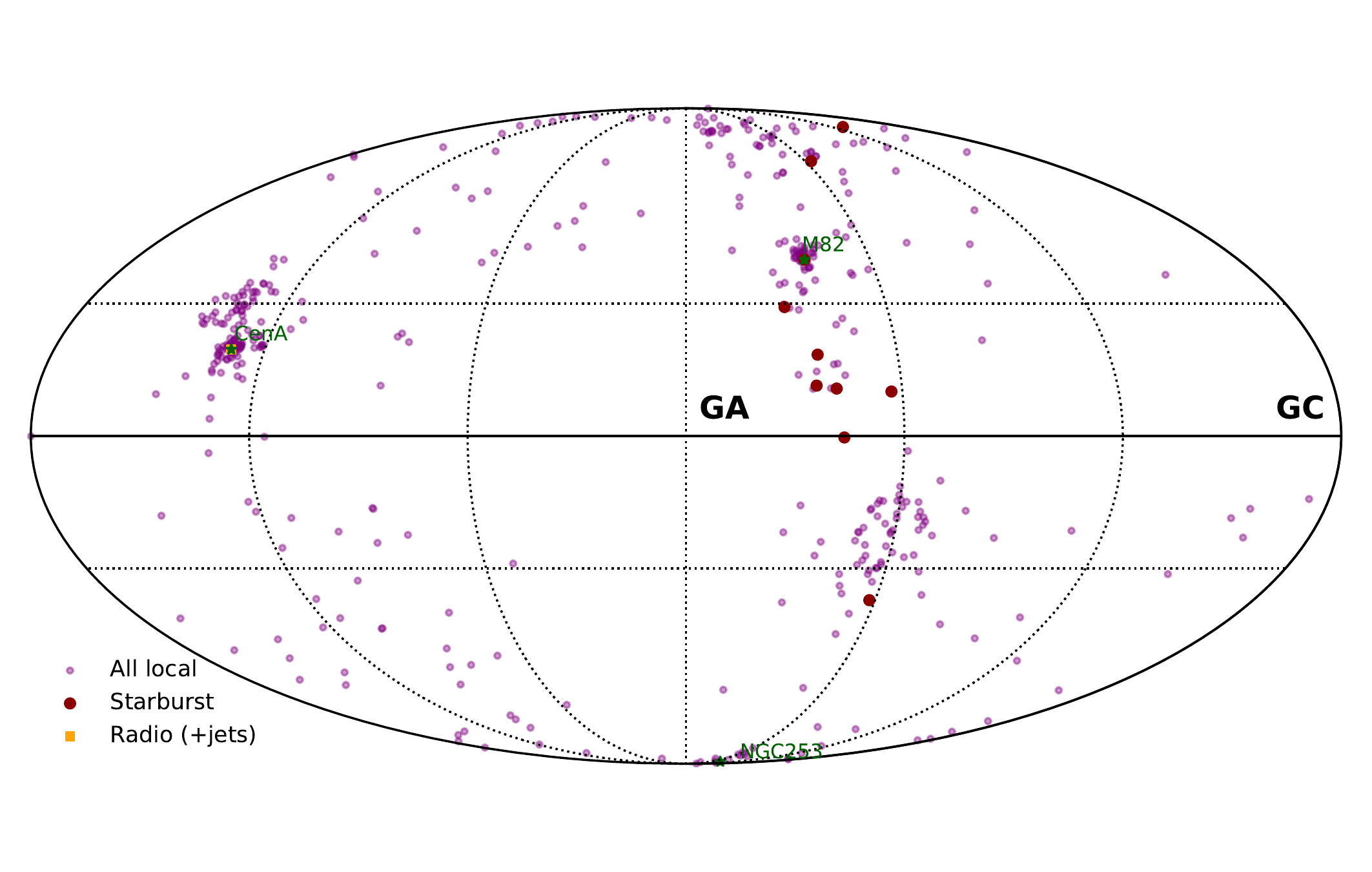}
\includegraphics[width=.98\columnwidth]{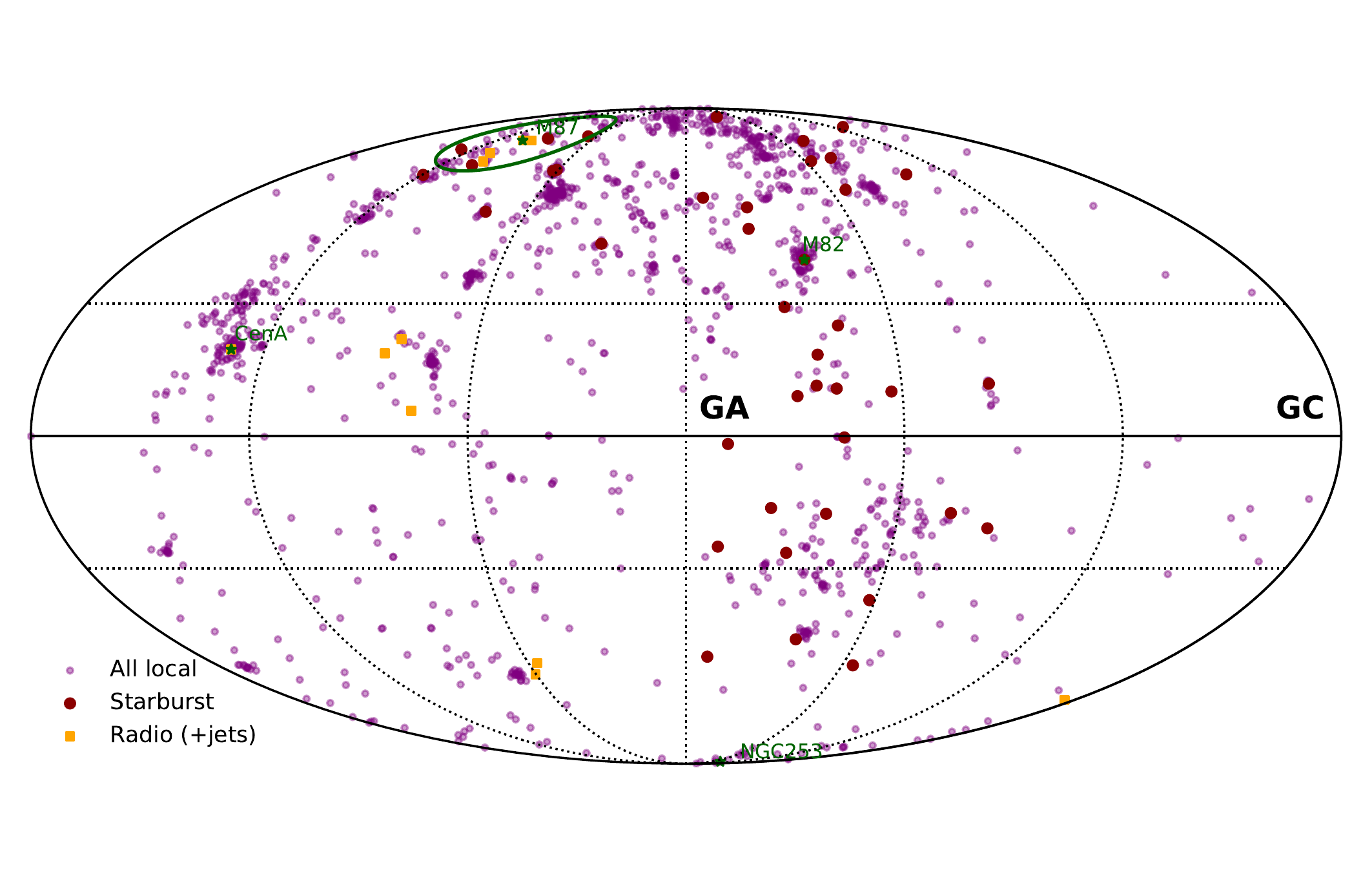}
\caption{Source candidates within 2 Mpc (top), 5 Mpc (middle) and 40 Mpc (bottom). Some popular candidate sources M82, Centaurus A, NGC253, and M87 (including a $10^\circ$ for the expected accretion shock) are shown as green stars to aid with orientation.}
\label{fig:just_sources}
\end{figure}

We use the Updated Nearby Galaxy Catalog maintained by \citet{Karachentsev:2013cva}. As of 2022, the catalog includes 1,420 galaxies with an sufficiently precise distance estimate. With the exception of KK198, which is located at a distance of 49 Mpc, all other galaxies are within a distance of 30 Mpc. We note that 85\% of the entries of this catalog are dwarf galaxies, which follow a similar distance distribution as regular galaxies. For Jetted AGNs we use the a subset of the catalog by \citet{vanVelzen2012} and for Starburst Galaxies (SBG) the catalog compiled by \citet{2019JCAP...10..073L}. The sources catalogs are shown in Fig.~\ref{fig:just_sources}. Source counts within different distances for the different catalogs are given in Table~\ref{counts}.

\begin{table}[h!]
\centering
 \begin{tabular}{||c c c c||} 
 \hline
 $d_{\rm max}$ (Mpc) & $N_s$ (All local) & $N_s$ (SBG) & $N_s$ (radio+jets) \\ [0.5ex] 
 \hline\hline
 2 & 129 & 2 & 0 \\ 
 5 & 440 & 10 & 1 \\
 10 & 1076 & 18 & 1 \\
 20 & 1393 & 34 & 5 \\
 40 & 1419 & 40 & 11 \\ [1ex] 
 \hline
 \end{tabular}
 \caption{Source counts within distance $d_{\rm max}$.}
 \label{counts}
\end{table}

\subsection{Galactic Magnetic Field Models}
We  use the  \citet{JF12} GMF model, with and without the correction introduced with Planck observations \citep{Planck2016} denoted by JF12 and JF12Planck respectively,  and the \citet{TF17} GMF model, denoted by TF17, to calculate the temporal dispersion of the EECR signal due to the GMF  as a function of the source direction and the rigidity of the particle. 

\subsection{Simulation setup}
The propagation of cosmic rays through the GMF is modeled using the  Monte Carlo code CRPropa 3.2 \citep{AlvesBatista:2022vem}. The Earth is at $8.5\, $kpc from the Galactic center. The Galactic boundary for scoring outgoing particle is defined by a sphere with radius $20\,$kpc from the Galactic center. The method of antiparticle tracing \citep{Thielheim1968} is applied.
The sky in Galactic coordinates is subdivided into $N_\text{pix} = $ 49,152 pixels on a HEALPix \citep{Gorski:2004by} grid with \texttt{NSIDE=64}, corresponding roughly to an angular size of $1.7^\circ$. For each combination of magnetic field model, total particle energy $E$ and mass number $A$, we simulate $5\times10^8$ trajectories with directions uniformly sampled from a sphere. At the intersection  of the particle trajectory with the boundary, we record the outgoing angles, the initial angles, the trajectory time, and other potentially interesting variables.

We have checked the choice of the solver for the equations of motion and found that the default rectilinear Cash-Karp method with relative tolerance set to $10^{-4}$ and step sizes between 0.1 -- 100~pc produces sufficiently precise results for our observables.

\subsection{GMF delays}

\begin{figure}
\centering
\includegraphics[width=0.45\textwidth]{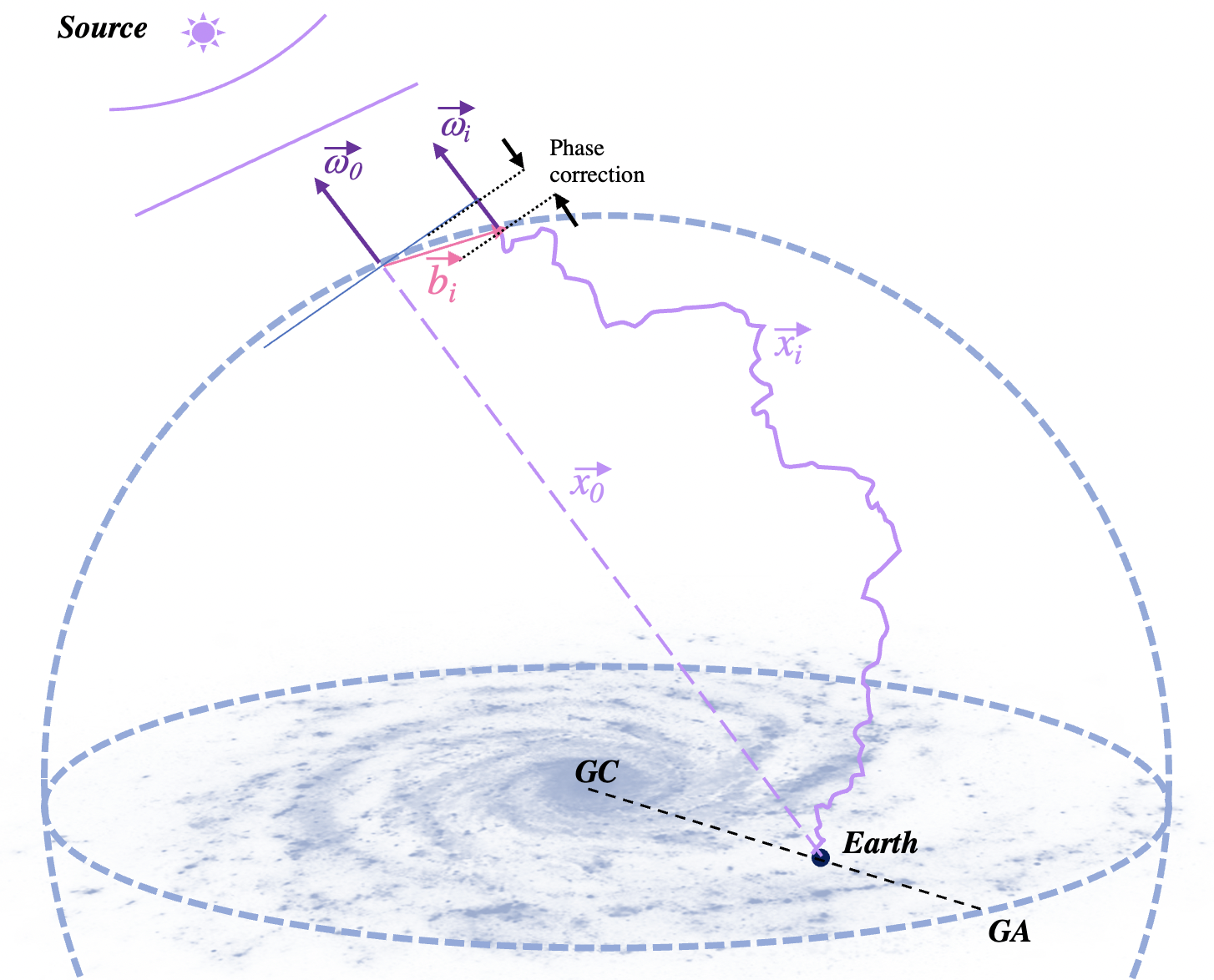}
\caption{Phase-shift correction between trajectories connecting a source to the Earth.}
\label{fig:sketch}
\end{figure}

Since the Galactic boundary is artificially chosen to be a sphere, we need to correct for the phase-shift induced by the curvature and for the fact that cosmic rays arriving at separated pixels on the sphere can point toward the same direction in the sky. The sketch of the geometry in Fig.~\ref{fig:sketch} introduces the geometry and the variables. Since extragalactic sources are far away with respect to the Galactic sphere radius, a pulse of cosmic rays from a transient source can be regarded as a plane wave traveling oriented normal to the directional vector $\vec{\omega}_0$, defined by the first CR event that hits a particular directional pixel on the boundary sphere. The position of this cosmic ray is $\vec{x_0}$ and in general $\vec{x_0}/||\vec{x_0}|| \neq \vec{\omega}_0$, except for the case of rectilinear propagation. With $\vec{b_i} = \vec{x_{0}} - \vec{x_i}$, we can calculate the phase-corrected travel time as
\begin{equation}
    t_i = \vec{\omega_0} \cdot \vec{b_i} / c + t_{{\rm trajectory},i}\,,
\end{equation}
where $c$ is the speed of light and $t_{{\rm trajectory},i}$ the trajectory times for the cosmic rays $i$ arriving in the same directional pixel (angular bin). This correction compensates for the choice of a sphere as intersection surface (instead of \textit{e.g.}~a polyhedron), the curvature induced by the choice of the radius (20kpc), and for the size of the pixels used for counting. For $5\times10^8$ events distributed in $4\pi$, the average count per pixel is $\langle n_\text{pix} \rangle \sim 4\times10^4$ and thus, the statistical error from the Monte Carlo simulation is small enough to not introduce additional bias on the standard deviation
\begin{equation}
    \tau_{d,\text{GMF}} = \sqrt{\frac{1}{N}\sum_i{(t_i^2 - \mu^2)}}\,,
\end{equation}
where $\mu$ is the mean arrival time. The $\tau_{d,\text{GMF}}$'s are calculated for each directional pixel independently.

\subsection{Magnification}
\label{sec:magnification_maps}
To answer the question about the observable directions for EECR events, we can use the simulation setup described above to compute magnification maps. In absence of Galactic magnetic fields, we expect that each pixel will be populated by $\langle n_\text{pix} \rangle$ events that correspond to a magnification factor of $ M = \langle n_\text{pix} \rangle  N_\text{pix} / 10^9 \sim 1$. For 300 EeV protons, with the exception of the Galactic center direction, $M$ is almost everywhere 1. Once significant deflections within the GMF act on the trajectories, $M$ will strongly deviate from 1 and can reach values between 0 and several hundreds. For a given position of the detector, we multiply $M$ by the exposure functions, following \citet{Sommers:2000us}, with maxima re-scaled to one. We choose to use the present UHECR observatories --  PAO and TA --, since it is likely that future observatories can be constructed at the same locations. We want to illustrate the differences for the discovery potential for an observatory located in the northern hemisphere compared to the a southern one. For the maximal zenith angles, we adopt the values $\theta_m = 55^\circ$ for TA \citep{TelescopeArray:2014tsd} and $\theta_m = 80^\circ$ for PAO \citep{PierreAuger:2014yba}, corresponding to the cuts used in anisotropy analyses.

\section{``Treasure Maps'' of the most promising directions for  detecting  EECR doublets}\label{sec:maps}
 \begin{figure*}
\centering
\includegraphics[width=0.45\textwidth]{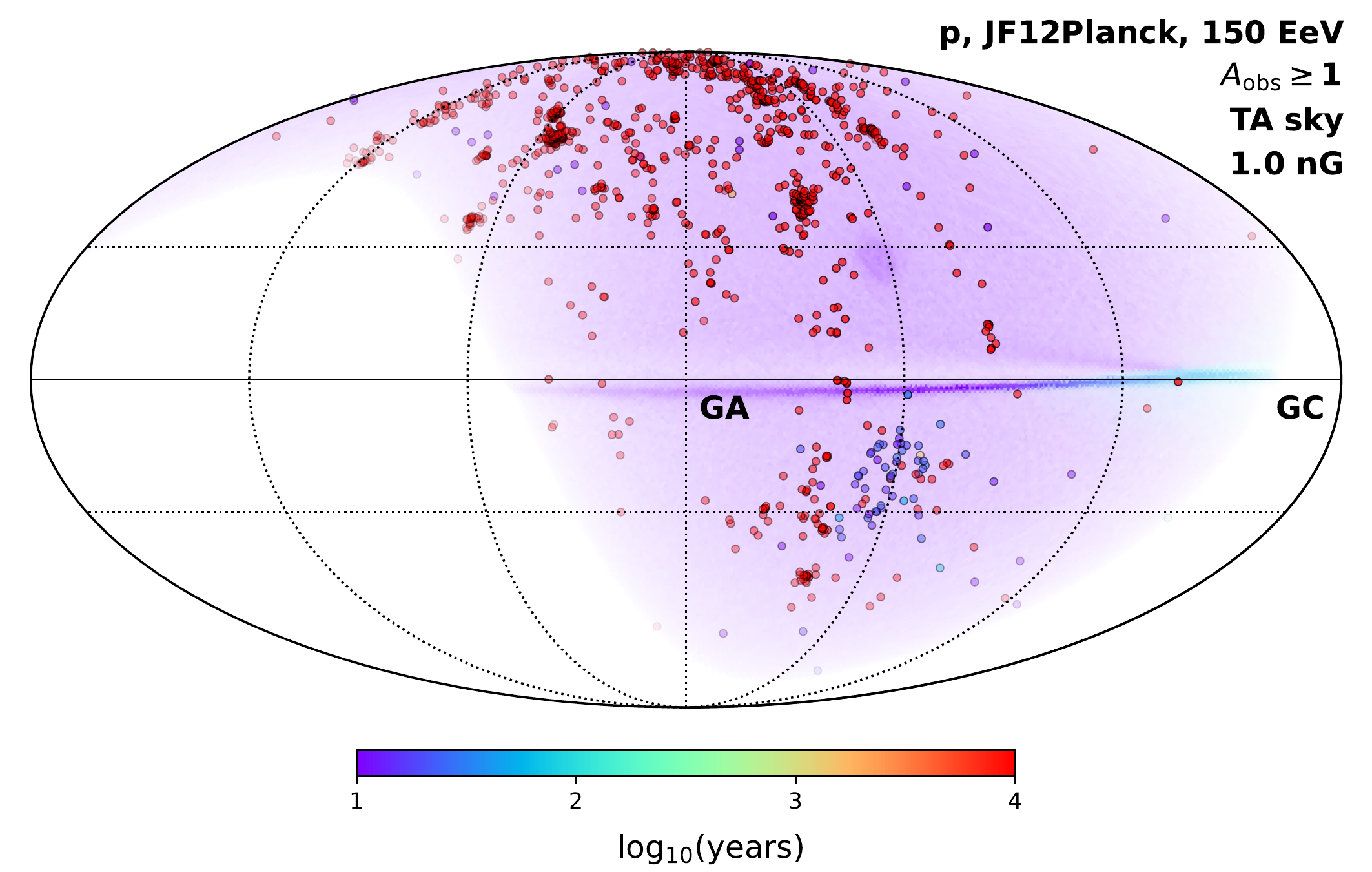}
\includegraphics[width=0.45\textwidth]{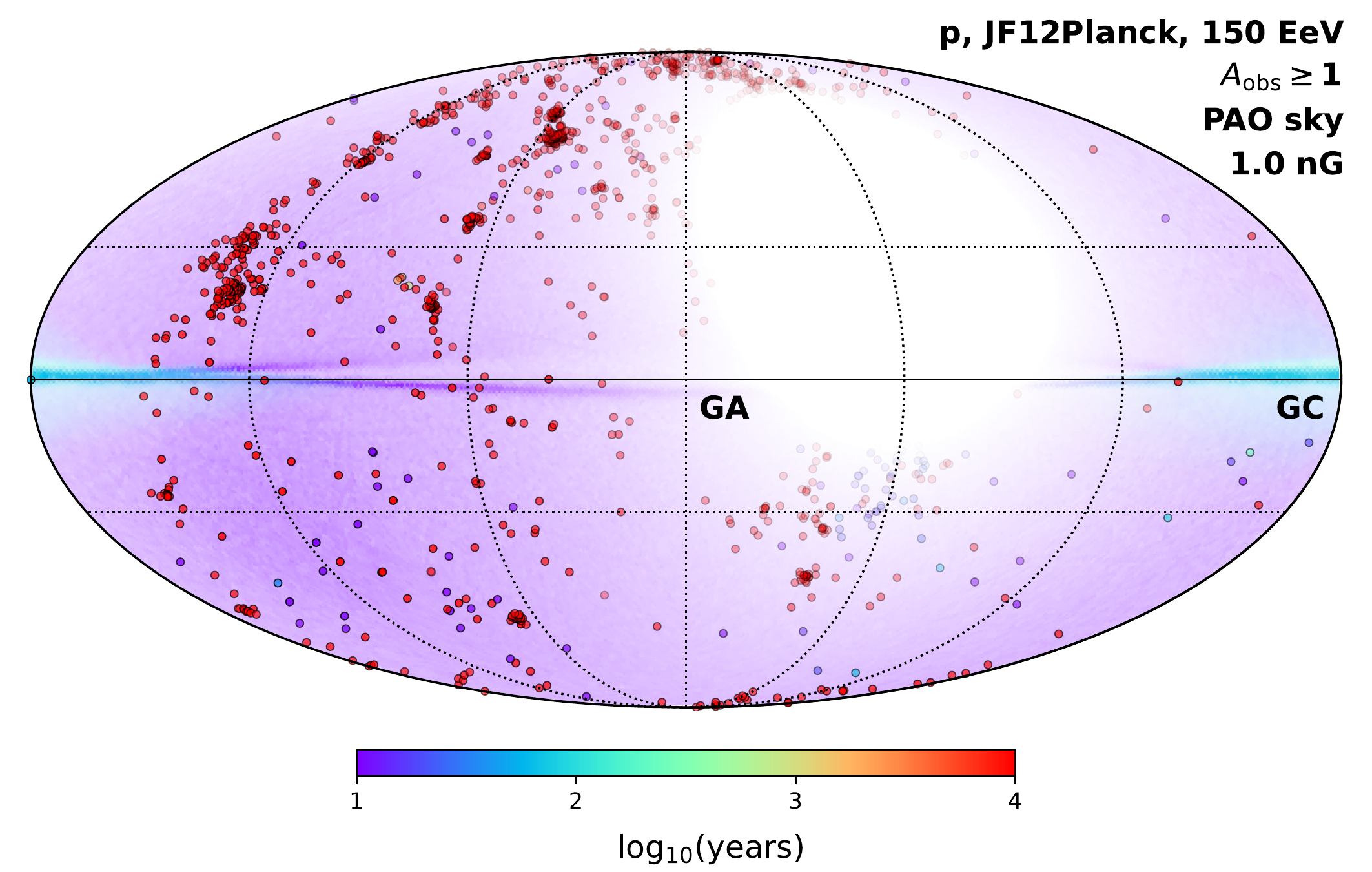}
\includegraphics[width=0.45\textwidth]{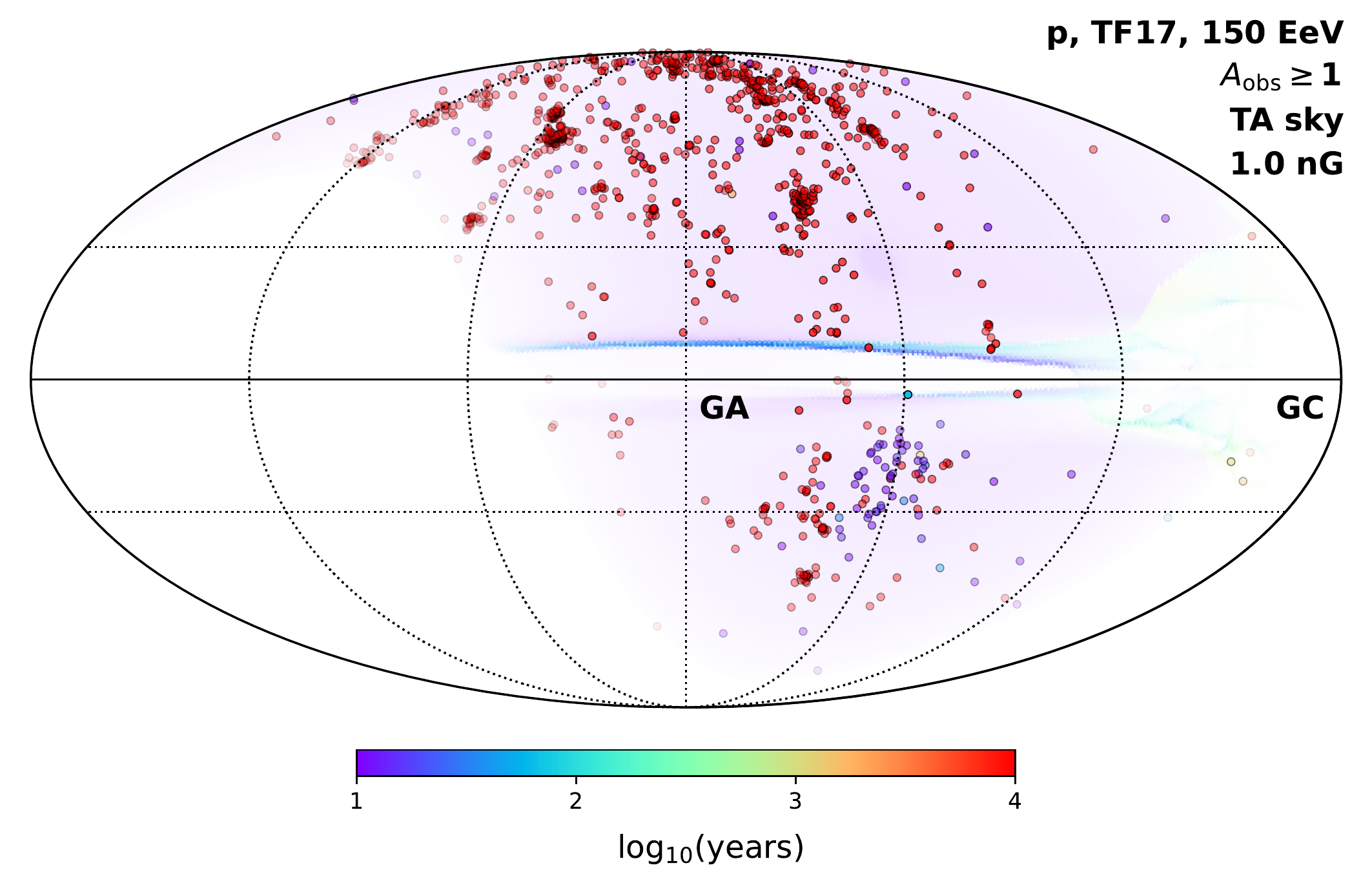}
\includegraphics[width=0.45\textwidth]{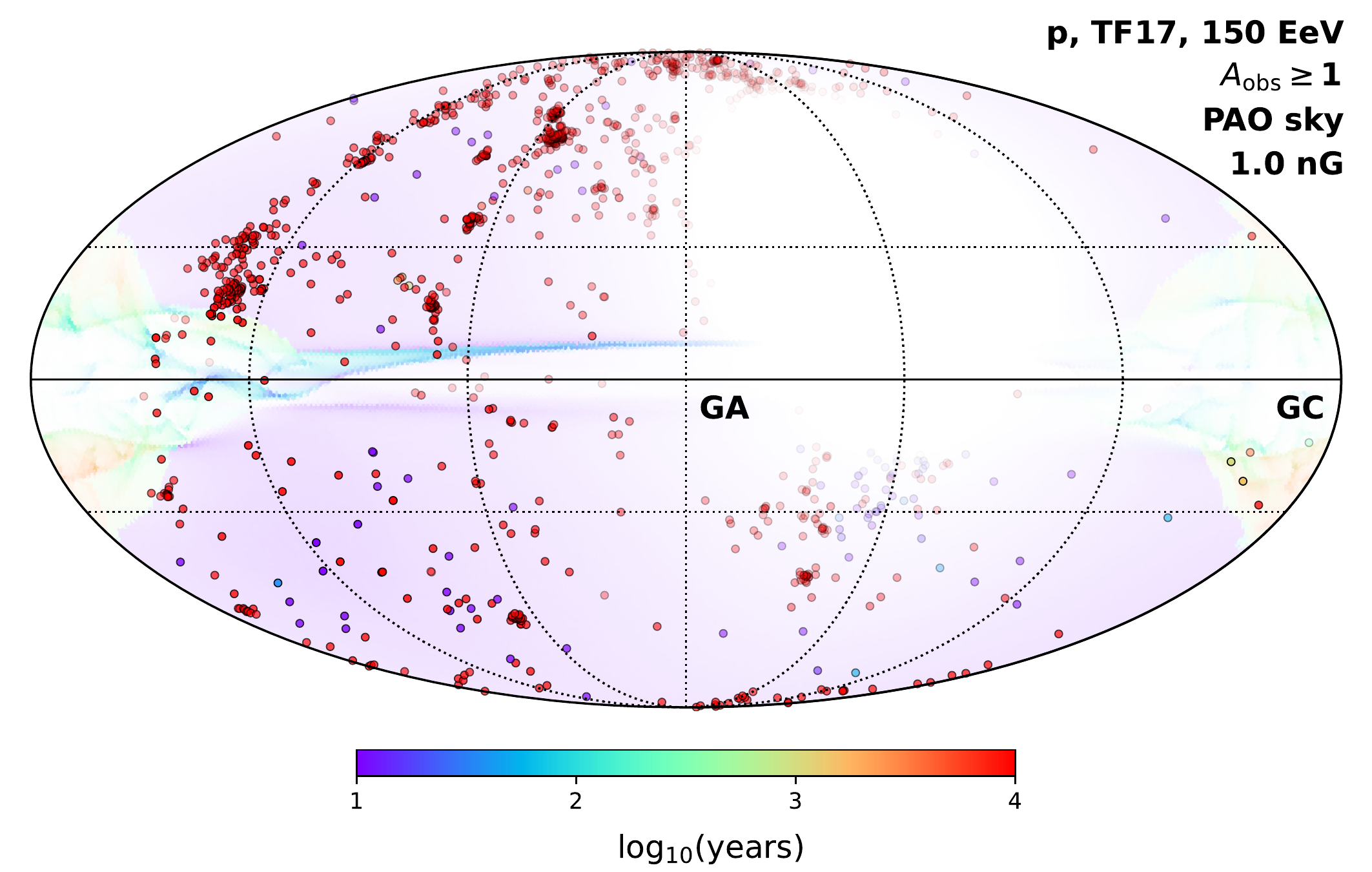}
\caption{Proton, 150 EeV  ``treasure maps'' for TA (left) and PAO (right), and for the JF12Planck (top) and TF17 (bottom) GMF models. The color of the gradients is assigned to $\tau_{d, \text{GMF}}$. The gradient's opacity is controlled by the truncated magnification maps $\min(M, 1)$, which include the detector exposure function. The source colors are assigned to the total $\tau_{d}$ (the EGMF strength is 1 nG as indicated). The source marker opacity is set to $\min(a_\text{GZK}, M)$, i.e.~within the detector's exposure the markers fade mostly due to $a_\text{GZK}$ and due to $M$ outside of it.}
\label{fig:TM_protons_150}
\end{figure*}
 \begin{figure*}
\centering
\includegraphics[width=0.45\textwidth]{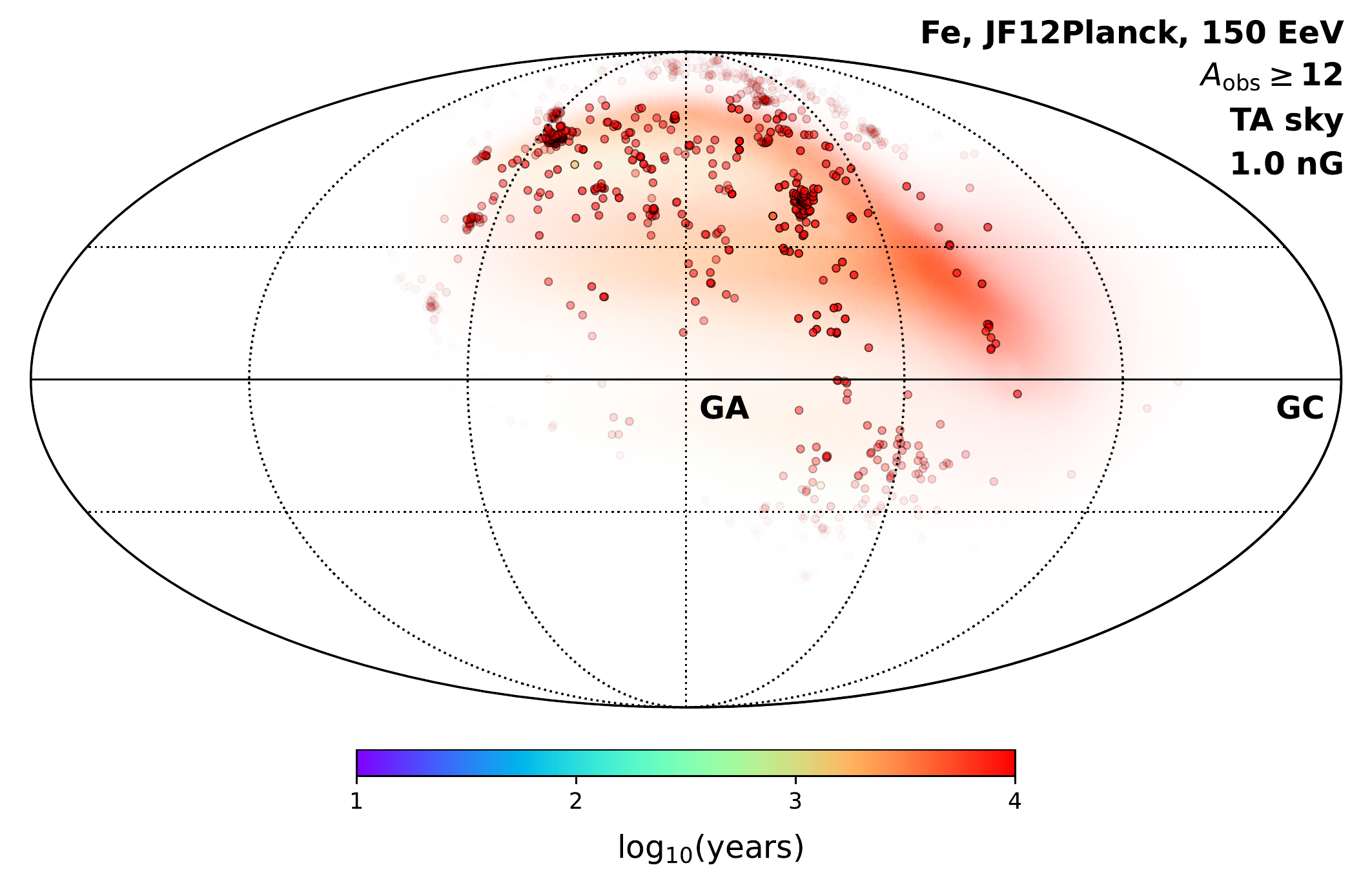}
\includegraphics[width=0.45\textwidth]{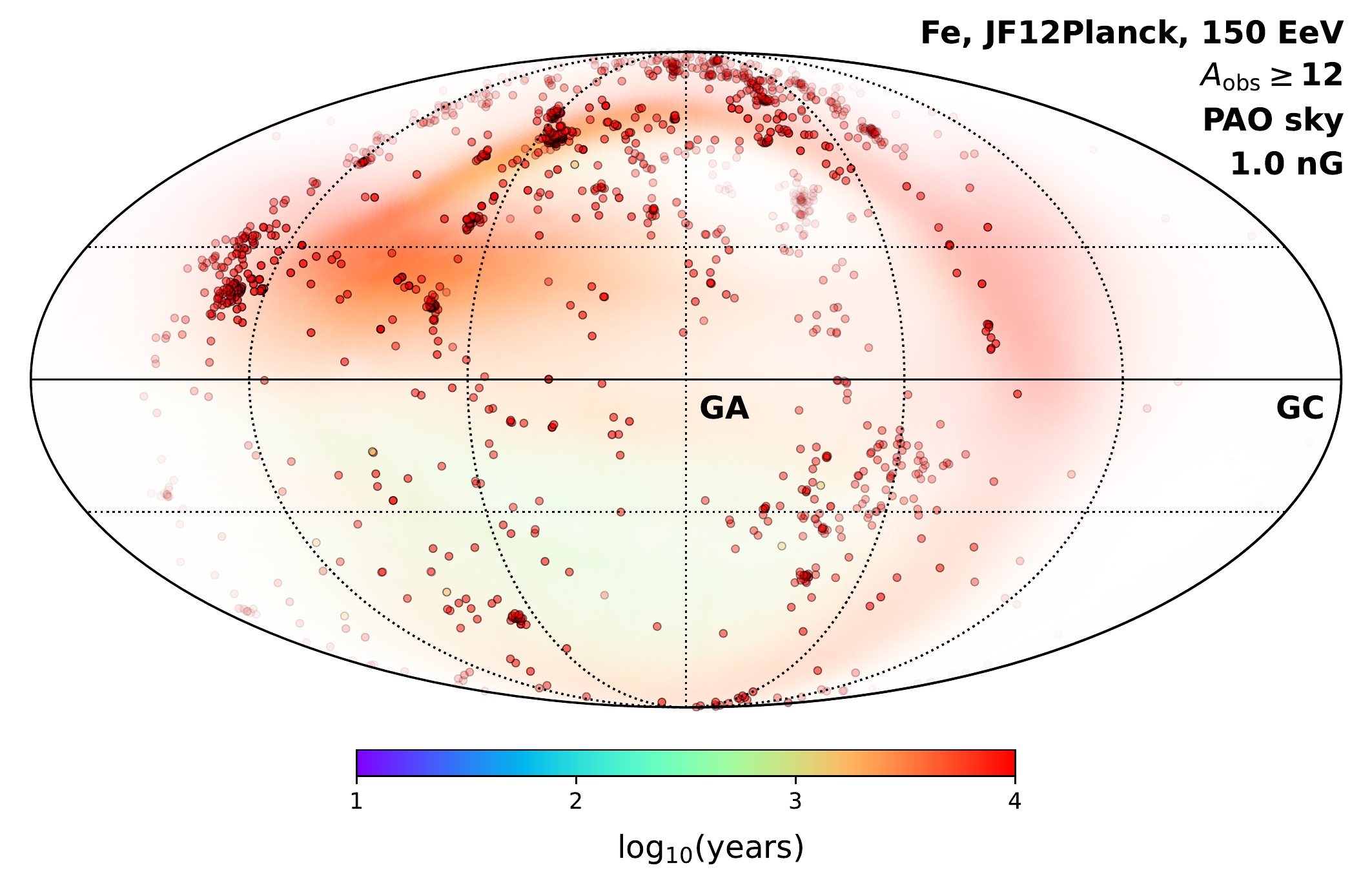}
\includegraphics[width=0.45\textwidth]{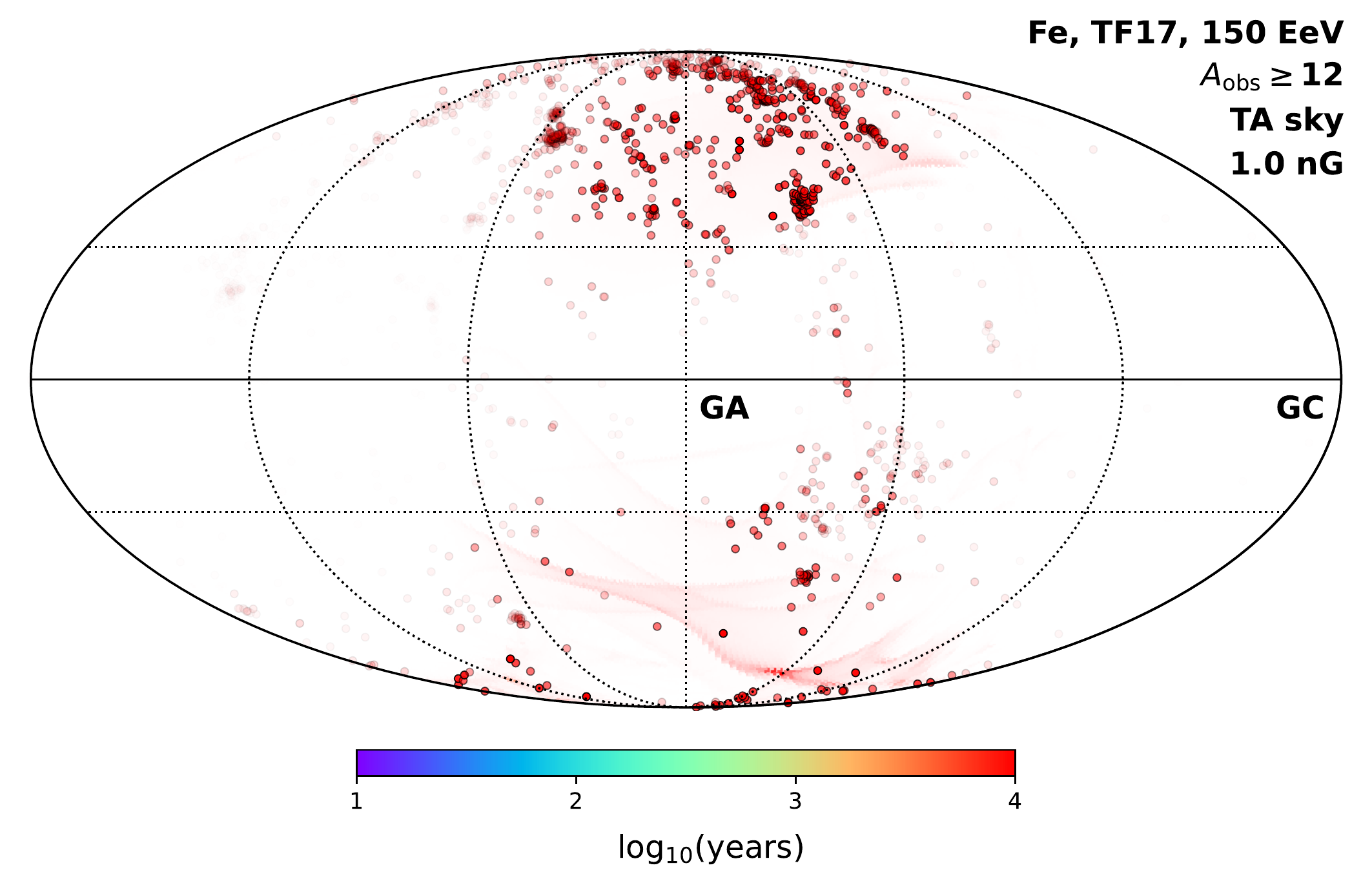}
\includegraphics[width=0.45\textwidth]{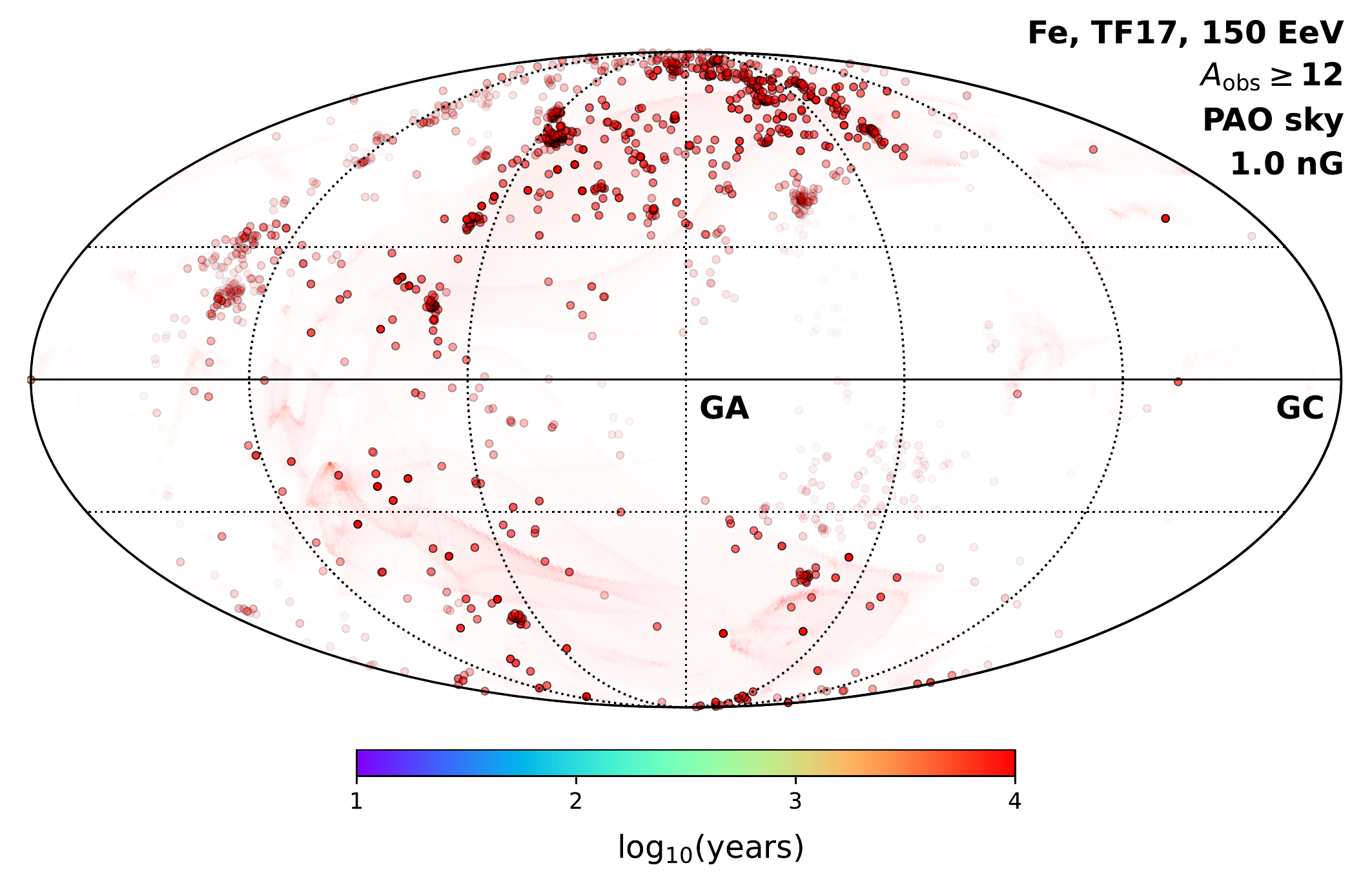}
\caption{Same as Figure \ref{fig:TM_protons_150} for Iron. The threshold mass entering $a_\text{GZK}$ for the source marker opacity is set to $\aobs=12$.}
\label{fig:TM_iron_150}
\end{figure*}

\begin{figure*}
\centering
\includegraphics[width=0.45\textwidth]{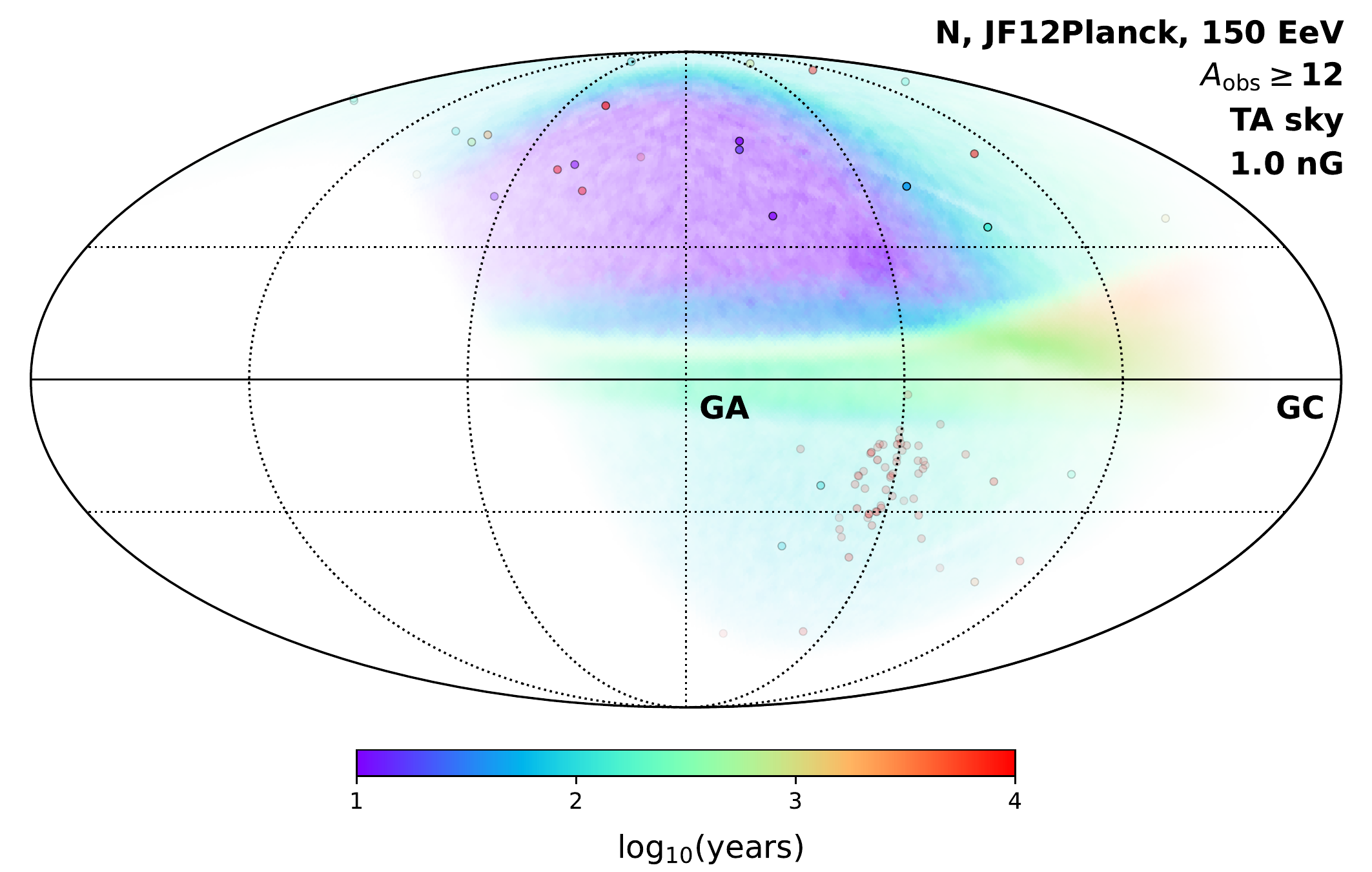}
\includegraphics[width=0.45\textwidth]{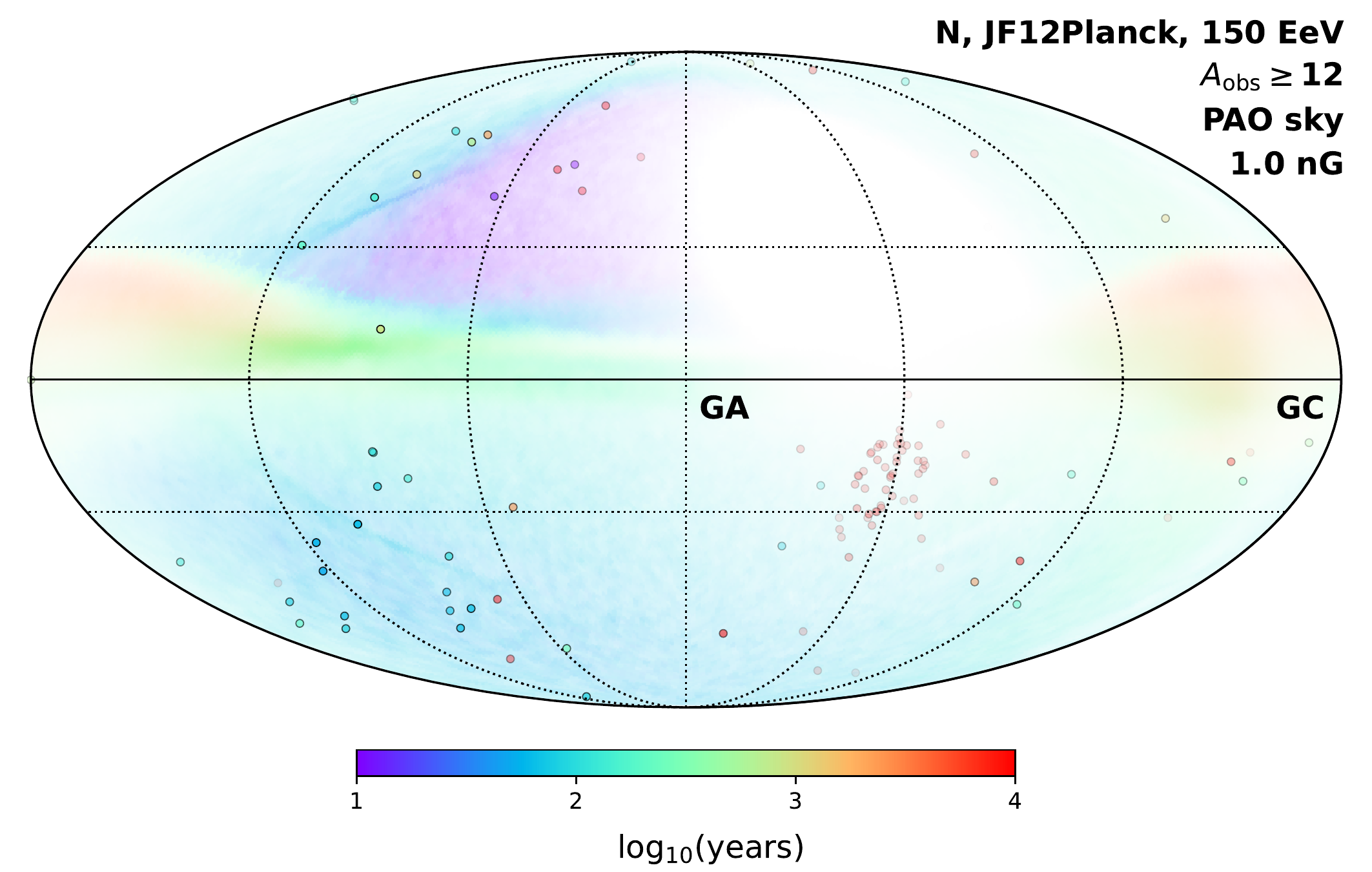}
\includegraphics[width=0.45\textwidth]{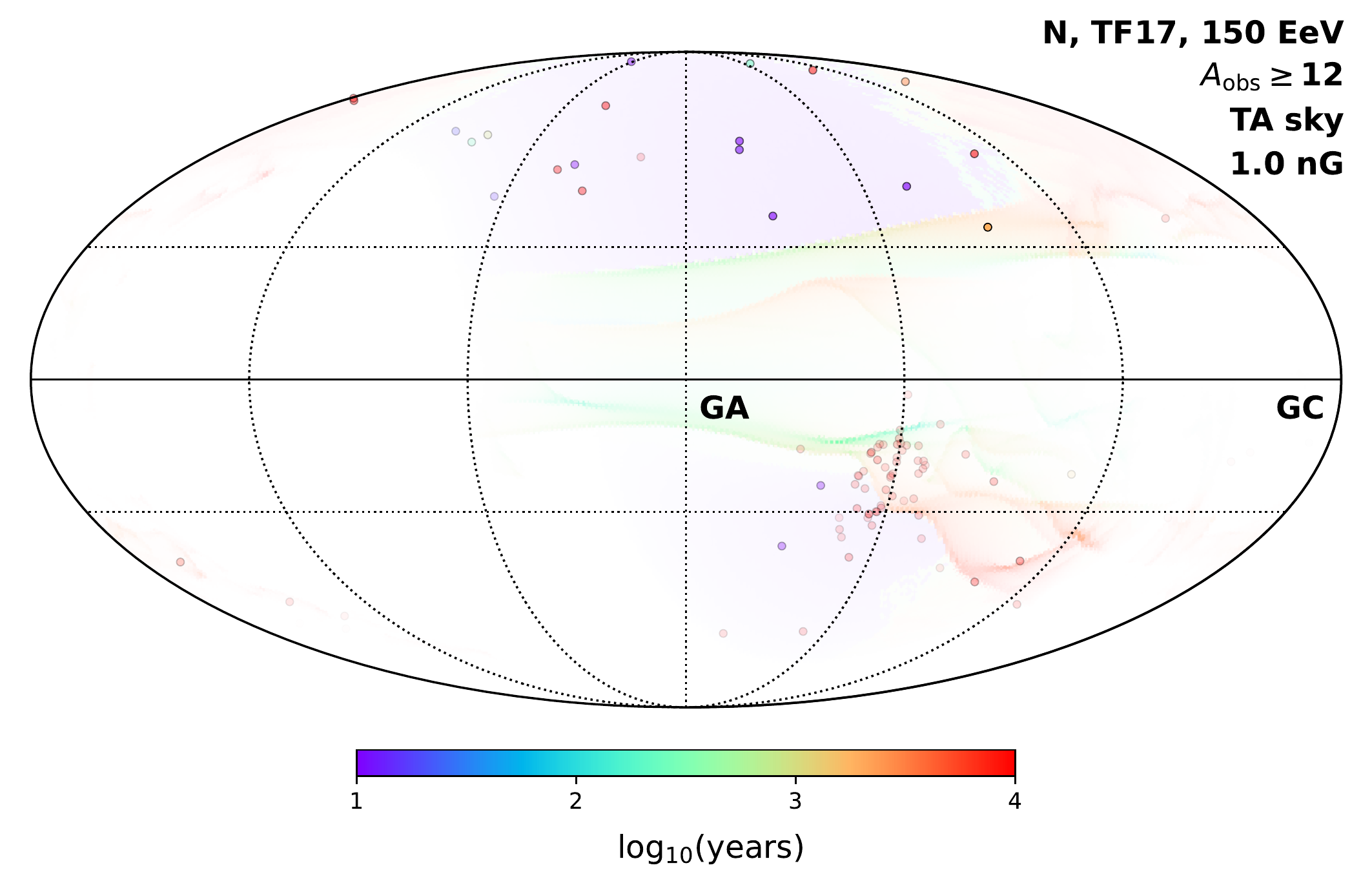}
\includegraphics[width=0.45\textwidth]{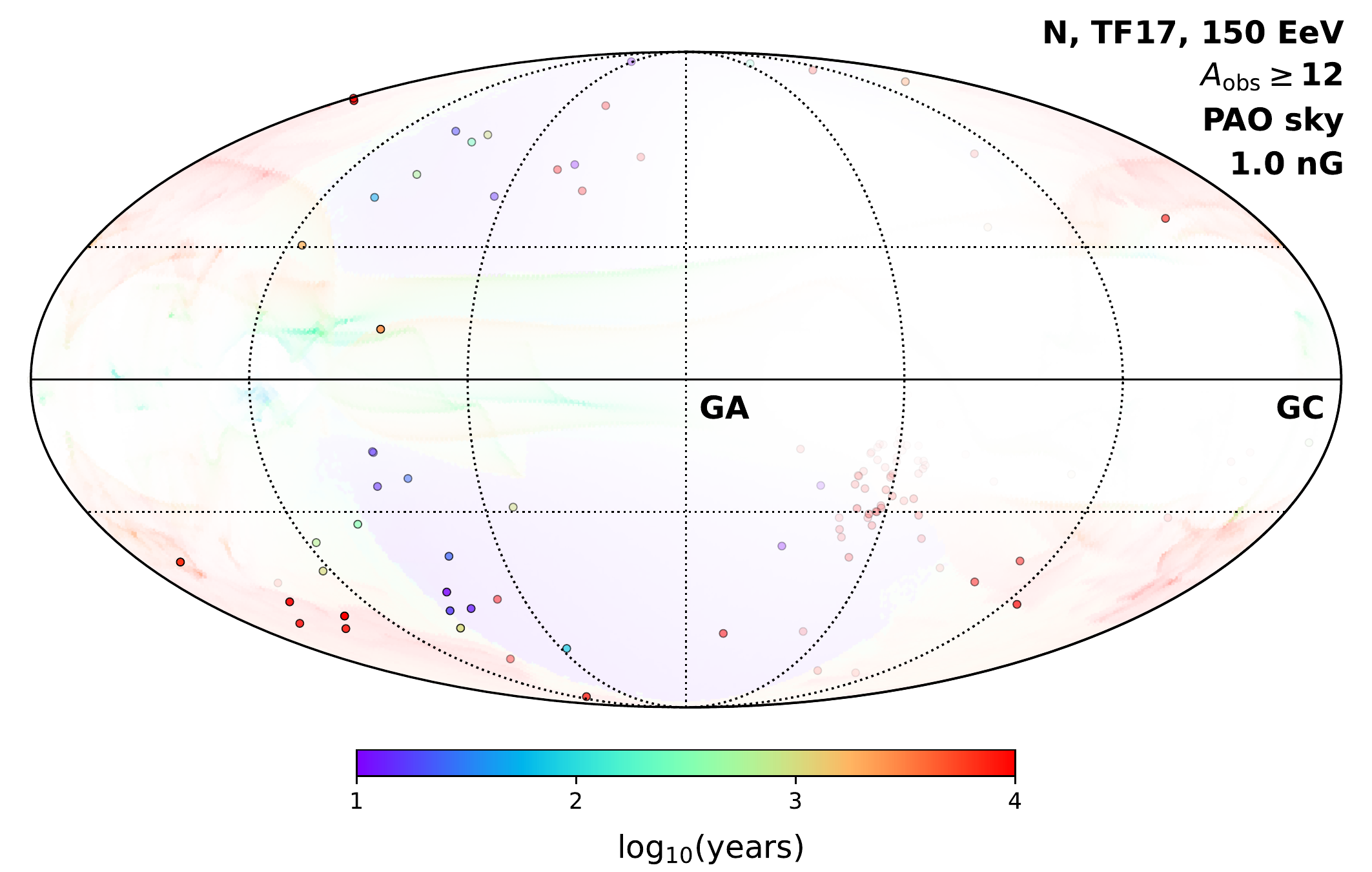}
\caption{Same as Figure \ref{fig:TM_protons_150} for Nitrogen. The threshold mass entering $a_\text{GZK}$ for the source marker opacity is set to $\aobs=12$.}
\label{fig:TM_nitrogen_150}
\end{figure*}

The sky maps, shown in Figs.~\ref{fig:TM_protons_150}, \ref{fig:TM_iron_150}, \ref{fig:TM_nitrogen_150}  aim to visualize and compress the results of the simulations. The maps are in Galactic coordinates and centered on the Galactic anti-center (GA), since for EECRs heavier than protons the directions behind the Galactic center (GC) are unreachable due to the strong deflections, and hence the maps are often blank there. Color coding is used for the time dispersion $\tau_{d, \text{GMF}}$. Transparency is assigned to the magnification map, where $M=0$ corresponding to fully transparent and $M\geq 1$ to solid color. The catalog sources, shown as circles, the color corresponds to the total $\tau_{d}$, and the transparency channel is set to $\min(a_\text{GZK}, M)$. Sources farther than $d_{95\%}$ are not shown for clarity. We refer the reader to Fig.~\ref{fig:just_sources} where some popular candidates sources are displayed for orientation. 

The proton maps are shown in Fig.~\ref{fig:TM_protons_150}. The opacity of the color gradients resemble the exposure function of the two observatories since the GMF deflections are relatively small. The average $\tau_{d, \text{GMF}}$ is of the order of hundreds of years with the exception of the GA that shows values of a few decades. Some magnification (JF12Planck) and de-magnification (TF17) is visible along the galactic disk due to magnetic lensing (edgy contours at the equator of the map). (JF12 and JF12Planck are very similar so we only show JF12Planck here). Although $\tau_{d, \text{GMF}}$ is moderate, most sources show temporal dispersion of $\mathcal{O}(10^4)$ years, which stems from the transport in the EGMF. For an average EGMF of 0.1 nG almost all source colors coincide with that of the background (see Fig.~\ref{fig:EGMF01} for the $B_{\rm nG}=0.1$ case), showing once more that EGMF and GMF both are crucial for modeling $\tau_d$. Despite quite substantial differences between TF17 and JF12, both models reveal a similar set of doublet host candidates for protons. Since the deflections in the GMF are small, the field of view of a TA-like or PAO-like detectors observe relatively distinct regions of the sky. 

The maps for 150 EeV iron are shown in Fig.~\ref{fig:TM_iron_150}.  The (de-)magnification effects are dominant for iron, where the small Galactic longitudes (toward the GC) are not observable at all. The trajectories are lensed toward the GA such that most of the sky accessible from the TA location is also visible from the South. A notable exception is M82 and its neighborhood assuming the JF12Planck GMF. Iron trajectories simulated with TF17 show very strong  lensing visible as sharp contours in the upper panels of Fig.~\ref{fig:TM_iron_150}. We observed similar features in JF12 with the turbulent GMF component switched off.

The maps for 150 EeV nitrogen   are shown in Fig.~ \ref{fig:TM_nitrogen_150}. While the maps for proton and iron showed a similar distribution of sources distance, the GZK attenuation strongly affects the horizon of nitrogen and hence fewer sources appear on the nitrogen maps.

\begin{figure*}
\centering
\includegraphics[width=0.45\textwidth]{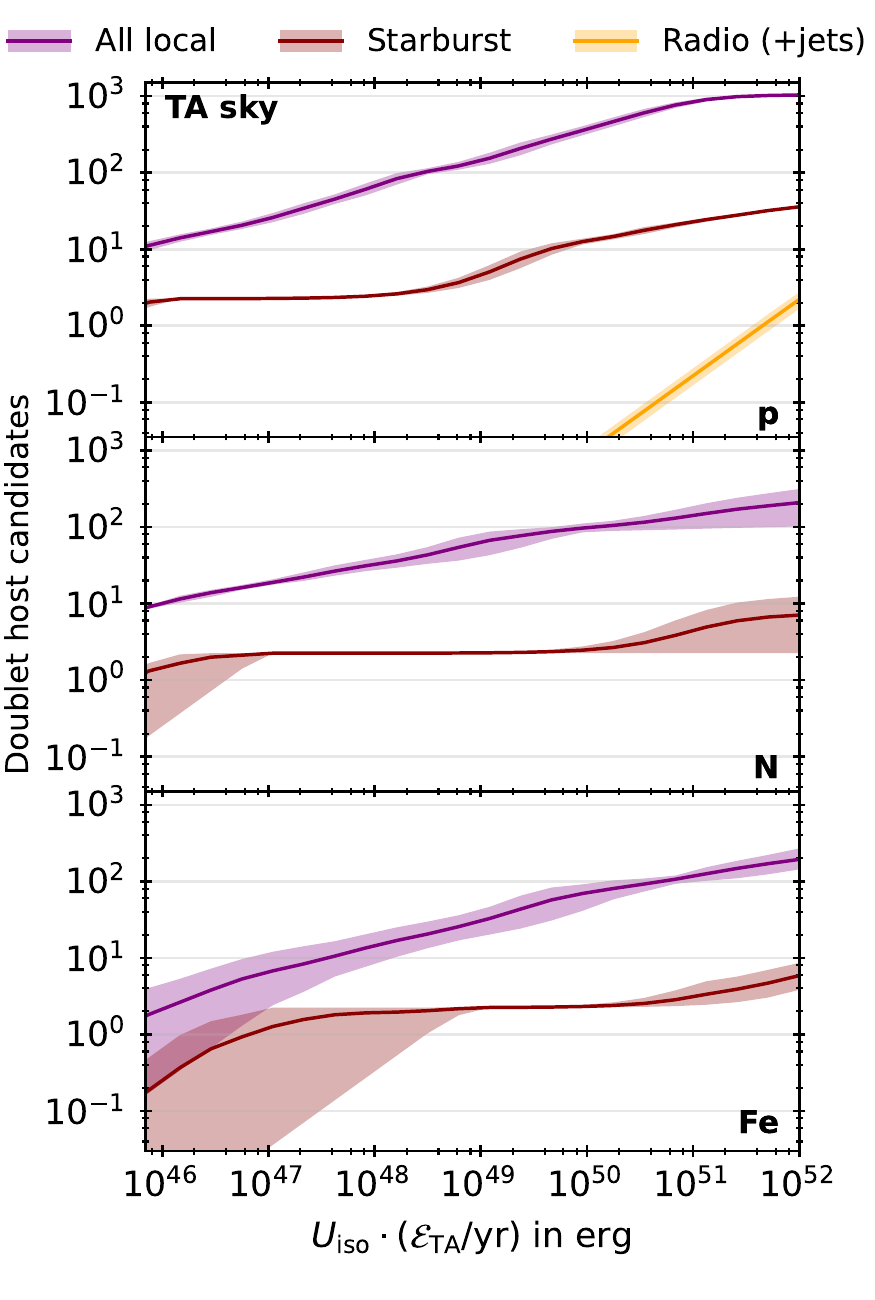}
\includegraphics[width=0.45\textwidth]{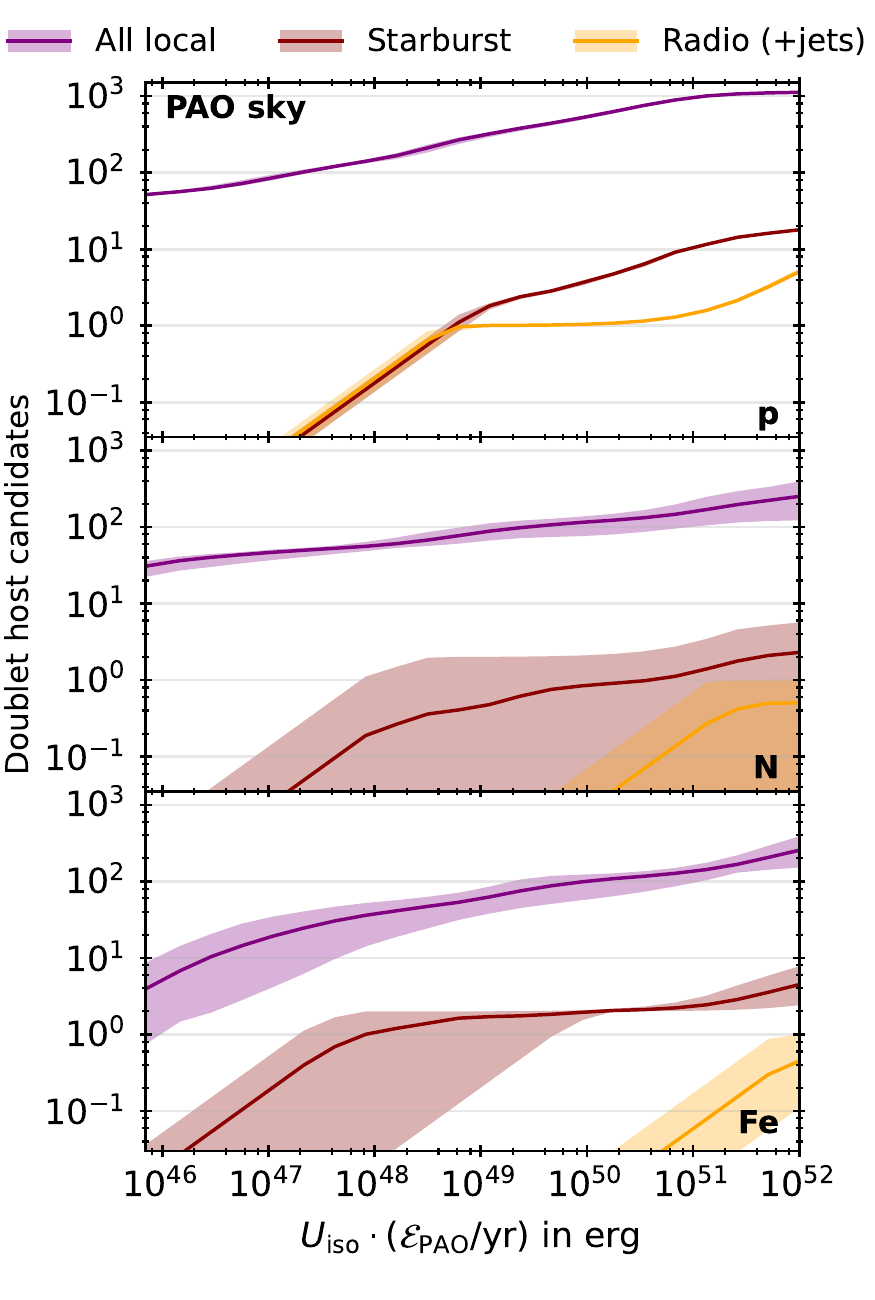}
\caption{Number of galaxies host candidate to detect a EECR doublet as a function of $U_{\rm iso}\cdot {\cal E}$, assuming all local galaxies, Starburst galaxies, or jetted AGNs.  For each of the treasure maps generated with the different choices for $\eobs$, $\aobs$, and GMF models, the host candidate counting is repeated. The maximal,  minimal, and mean are presented as a band and a curve, respectively. The EGMF strength is 1~nG.}
\label{fig:doubletsallsky}
\end{figure*}

These treasure sky maps  demonstrate that to estimate the visibility of specific host galaxies within the field of view of an observatory requires anisotropic, four-dimensional modeling (in direction, distance, and temporal dispersion). The GMF plays a crucial role even at these extreme energies. Thus, neglecting the GMF's impact or approximating deflections by isotropic smearing kernels is inappropriate for source searches under the assumption of light or heavy nuclear composition.

\section{Number of  host candidates for doublets}\label{sec:sources}

A more general result can be derived by counting the number of host galaxies within the field of view that can harbor transient EECR sources able to provide a doublet. The counting is performed using a weighted sum, where for each source in the catalog the weight is defined as
\begin{equation}
  \hat{w}_s = \max{\left(1, \frac{U_{\rm iso}}{U_{\rm iso,2}} \right)}\,,
\label{eq:w_count}
\end{equation}
where $U_{\rm iso,2}$ is given by Eq.~\ref{eq:u_iso} (the energy necessary to provide a EECR doublet after taking GZK and magnification into account) and $U_{\rm iso}$ is the true source energy (which can be larger or smaller). 
We set the $d=d_s$ and $\tau_d$ according to the location of the source, $E=\eobs$, and $B_{\rm EGMF} = 1$\, nG. $M$ are the magnification maps derived in Sec.~\ref{sec:magnification_maps} that can enhance of suppress the number of UHECR due to lensing or de-lensing of the source locations, whereas the GZK opacity $a_\text{GZK}(d_s)$ calculated in Sec.~\ref{sec:gzk-horizon} only suppresses the number. The weights are capped at 1 for the doublet case,  
so once a catalog source reaches this condition it is fully counted, whereas a source doesn't contribute if it is not within the field of view ($M=0$) or beyond the GZK horizon ($d_s >= d_{95\%}$). Transients within host galaxies with $\hat{w}_s$ between 0 and 1 can produce observable doublets with reduced probability.

The total accessible transient host counts are obtained by summing the weights. These are shown in Fig.~\ref{fig:doubletsallsky} as a function of $U_{\rm iso}\cdot \mathcal{E}$. Since in our simple source model, the maximal rigidity or the luminosity are independent of the source energetics, the exposure and $U_{\rm iso}$ are degenerate. For each of the treasure maps generated with the different choices for $\eobs$, $\aobs$, and GMF models (JF12, JF12Planck, TF17), the host candidate counting is repeated. The maximal,  minimal, and mean are presented as a band and a curve in, respectively. (The EGMF is 1~nG. See Fig.~\ref{fig:doublets01} for the $B_{\rm nG}=0.1$ case.)

For less energetic sources a similar number of host galaxies is accessible for the proton and the nitrogen composition assumptions and since very few nearby sources dominate. In the case of very luminous transients, the number of accessible host galaxies is an order of magnitude smaller for the case of nuclei compared to protons, due to the losses imposed by the temporal dispersion and the GZK horizon. 

Despite different detector exposures and field of view, a comparable number of the brightest transients could be seen from the TA site. 

 \begin{figure*}
\centering
\includegraphics[width=0.45\textwidth]{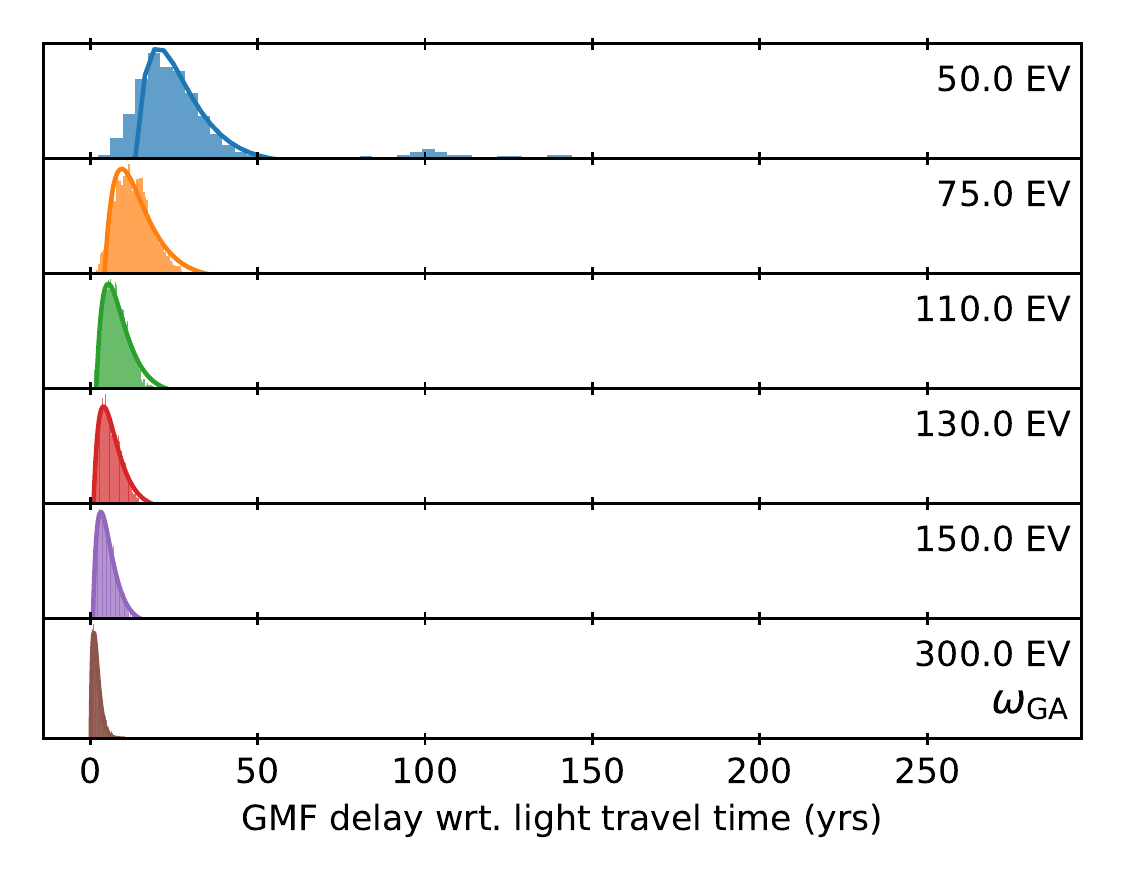}
\includegraphics[width=0.45\textwidth]{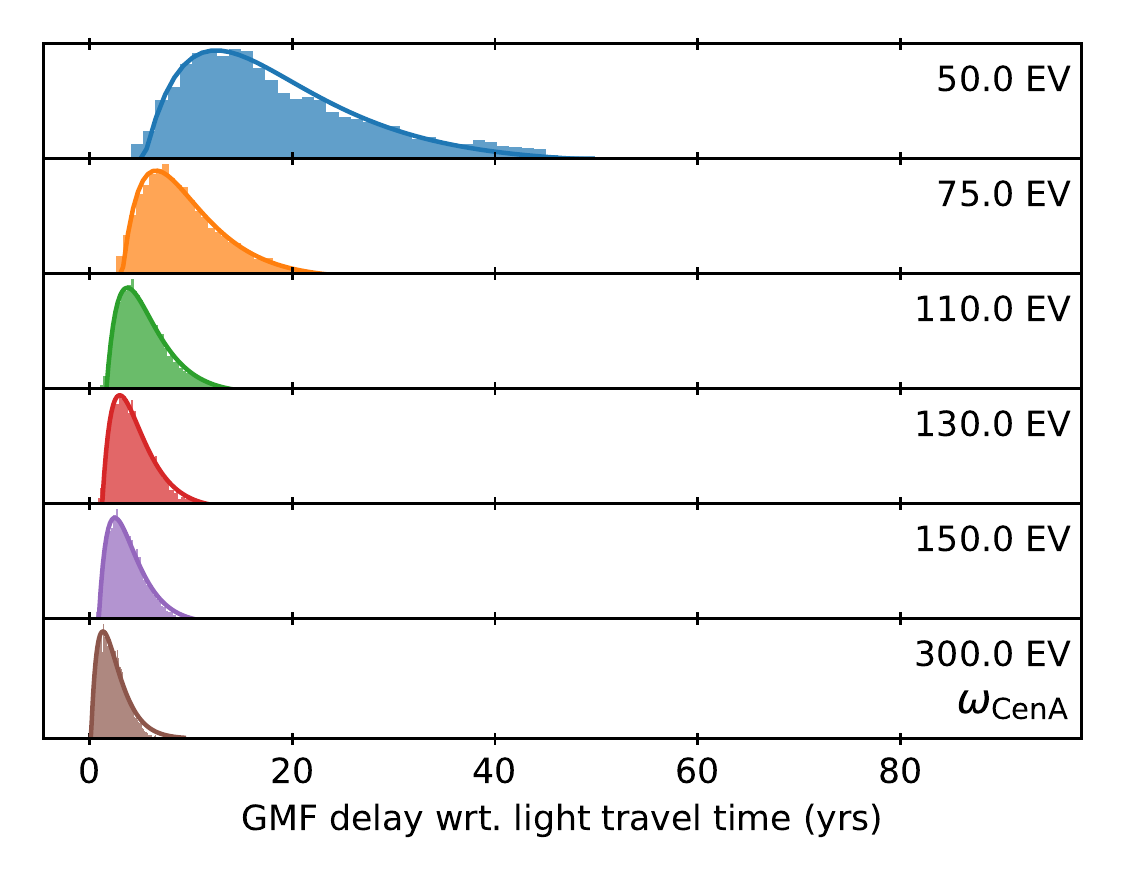}
\includegraphics[width=0.45\textwidth]{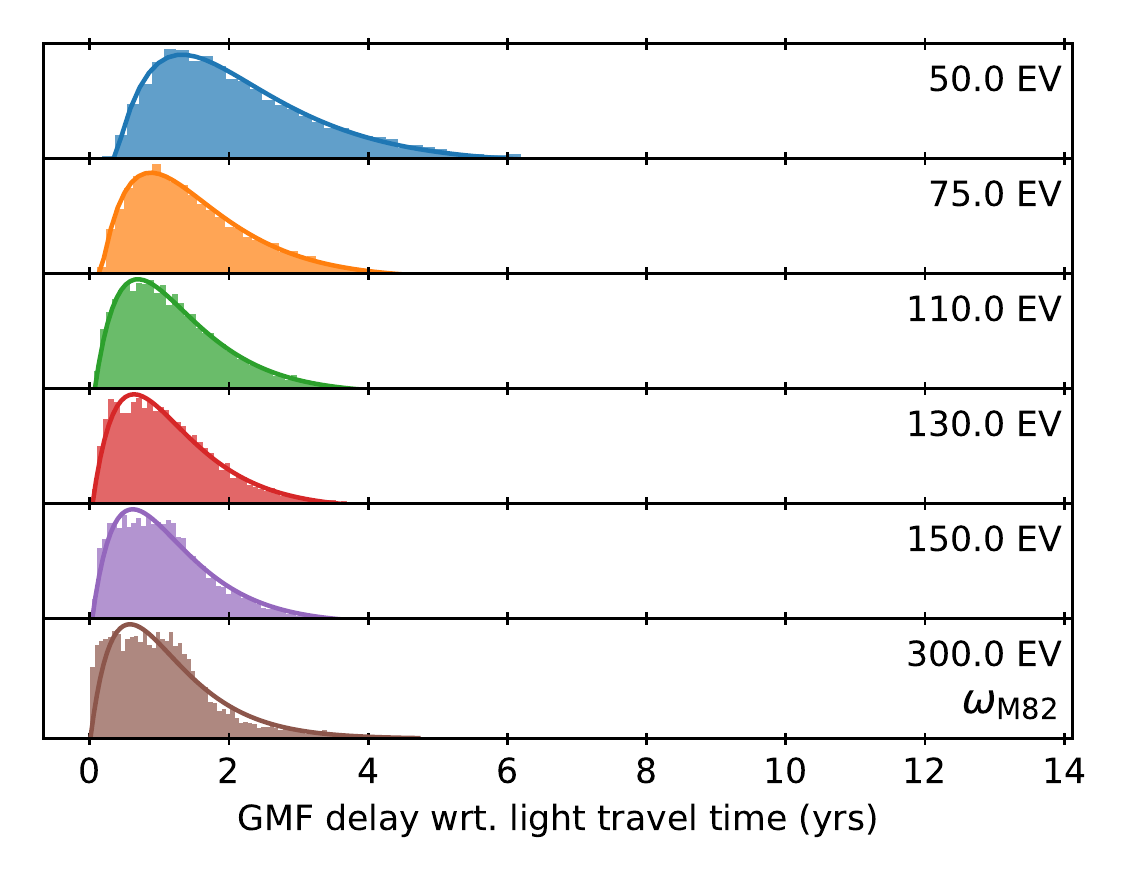}
\includegraphics[width=0.45\textwidth]{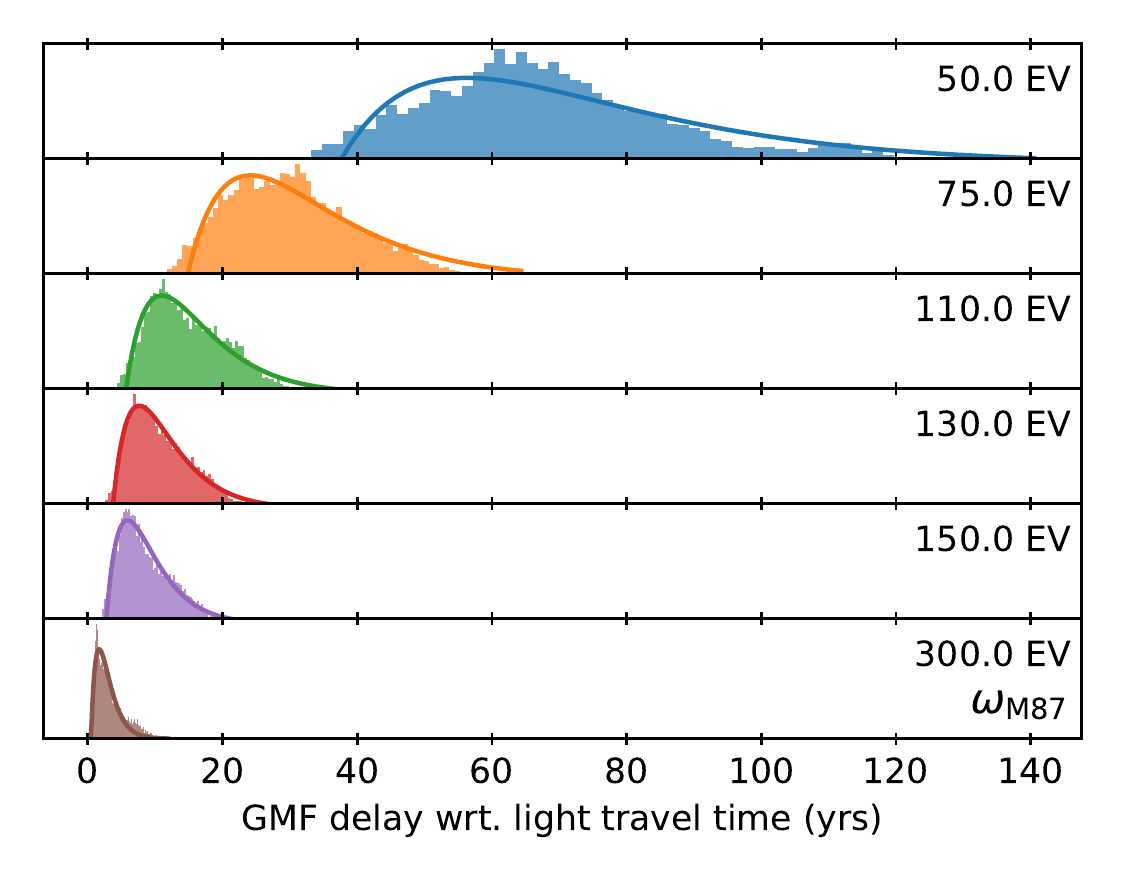}
\includegraphics[width=0.75\textwidth]{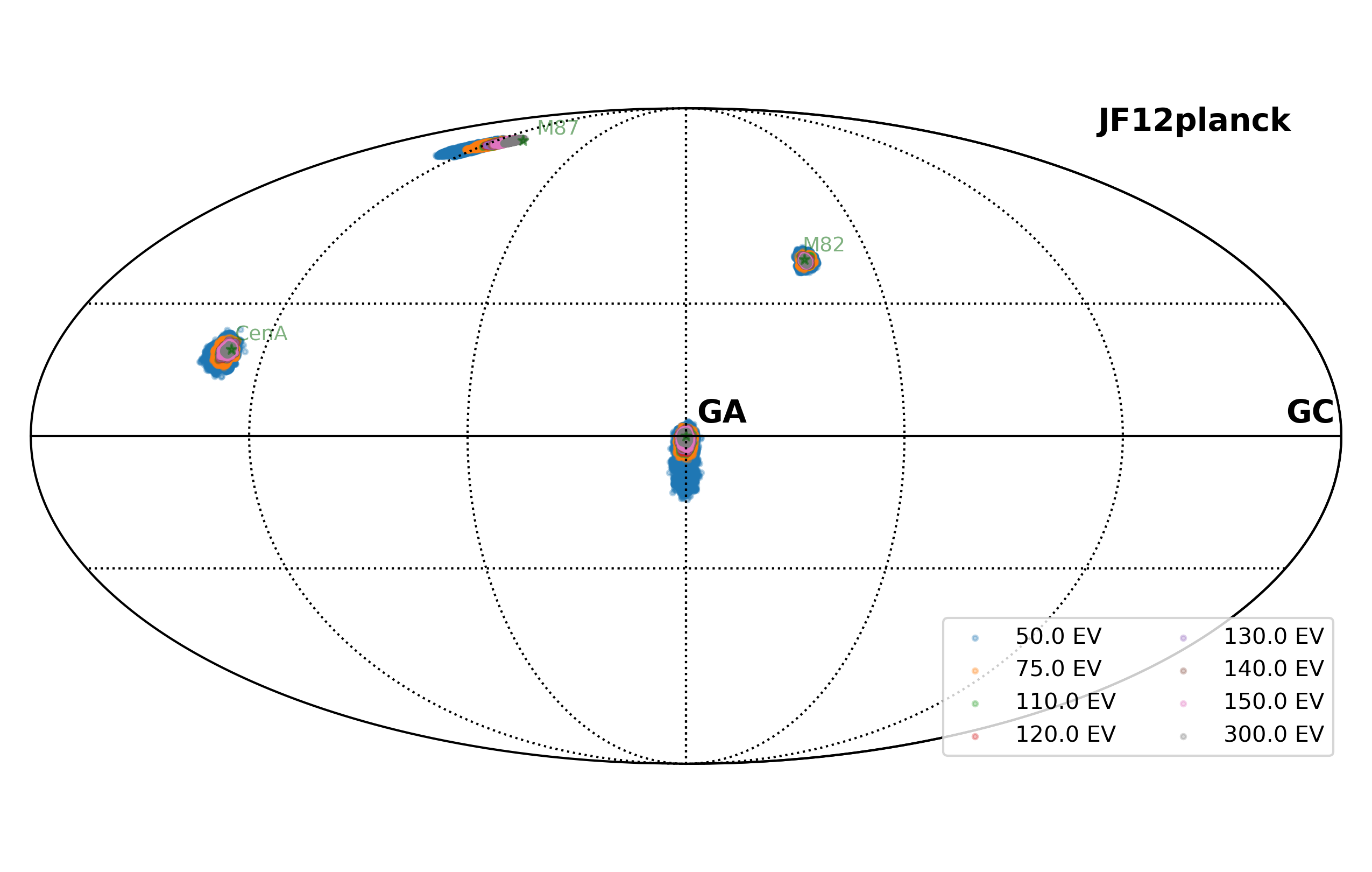}
\caption{Distribution of time delays assuming the JF12Planck Galactic Magnetic Field model and four hypothetical sources plus a distant source behind the Galactic Anticenter. Also shown are scatter plots for arrival directions around these sources at different energies.}
\label{fig:temporal}
\end{figure*}

\section{Temporal analysis of multiplets}\label{sec:time_analysis}
In Fig.~\ref{fig:temporal} we show the positional shifts and distributions of time delays for plausible local sources ``observed'' at different rigidities. To keep matters simple and to maximize the incidence of doublets, we suppose that the EGMF is unimportant. These illustrative calculations bring out several of the features of current and incipient observations and point to opportunities for learning about the sources from even richer datasets.

At present, positions are relatively well-measured (degrees), energies are moderately-measured with fractional accuracies $\sim0.1-0.2$, but nuclide identification can only be performed statistically. As several authors have discussed (Farrar and Sutherland 2017), if we consider the scatter plot of expected positions on the sky (Fig.~\ref{fig:temporal}), the sky positions will exhibit a systematic angular shift $\theta$ that depends upon the GMF model together with a random component that depends upon turbulent field. Both components scale approximately inversely with rigidity, as expected, with the coefficient dependent upon the GMF model and the source location. The rate at which neighboring, back-projected trajectories diverge determines the magnification $M$. Note that the angular spread are different for  Cen A, M82 and NGC253, three sources with similar distance. 

The trajectories can also be used to compute the distribution of delays $t$ (CR travel time - light travel time, cf. Fig.~\ref{fig:temporal}). These generically exhibit a shorter rise time than decay time because there is more phase space associated with larger excursions \citep[c.f.][]{Williamson1974,Alcock1978}. A simple probability distribution function, $P_s=se^{-s}; s>0$ for a scaled observatory time $s$, fits the histogram of delays, $t$, where $s=2(t-t_{\rm GMF})/<t-t_{\rm GMF}>$ where $t_{\rm GMF}$ is the constant delay (i.e. that would be due to the lensing only) due to the GMF. $P_s$ peaks at $s=1$ with mean $<s>=2$ and variance ${\rm var}(s)=2$. For a given source $t_{\rm GMF},<t>\sim x_{\rm GMF}\theta^2/2c$, where $x_{\rm GMF}$ is the effective path length through the Galaxy and $\theta$ is the angular spread. These are found to scale roughly $\propto R^{-2}$ as expected. We assume this distribution in what follows. 

We now turn to the observation of doublets. Consider a transient source of duration short compared with $\tau_d$ and the effective lifetime of a facility, nominally a decade.

Next, suppose that we have two events satisfying the positional requirement and measured rigidities $R_1$, $R_2$, with $R_1>R_2$. If the difference between the rigidities is negligible, then the two events can be treated as independent and the probability distribution of the interval between the two events $\Delta=s_2-s_1>0$ is $P_\Delta=\frac12(1+\Delta)e^{-\Delta}$. The mean interval between the two events is $3/2$ in dimensionless units or $3(<t>-t_{\rm GMF})/4$, which can be estimated from the magnetic field model. 

If $R_1$ is significantly greater than $R_2$, then we expect that the delay will be greater for the lower rigidity particle. The probability that the cosmic ray with rigidity $R_2$ follows the cosmic ray with rigidity $R_1$ can be shown to be $(R_1^2+3R_2^2)R_1^4/(R_1^2+R_2^2)^3$. If $R_2<0.5R_1$, the probability exceeds 0.9. This time-ordering is a quite robust prediction for a doublet from a transient source passing through a turbulent magnetic field. A even stronger statement can be made if we consider these two events as existing as points in a 4D space (the event arrival time, its position on the sky and its rigidity)  and include the scaling with rigidity. The probability density of the point associated with the second event relative to the first is quite peaked and this can be used as a prior in assessing the likelihood that an observed doublet is, indeed associated with a single transient.  

The next stage is to imagine that we have a set of triplets where there are two delays. (Given an expected number of events the distribution of the actual number should be Poissonian.) The asymmetry in $P_s$ implies that the interval between the second and third events should generally exceed that between the first two events. To illustrate, again suppose the three rigidities are very close and introduce $\Delta_1=s_2-s_1>0$ and $\Delta_2=s_3-s_2>0$. The bivariate probability distribution function is $P_{\Delta_1\,\Delta_2}=(4/9+8\Delta_1/9+4\Delta_2/9+2\Delta_1^2/3+2\Delta_1\Delta_2/3)e^{-(2\Delta_1+\Delta_2)}$. From this expression, it can be shown that the probability that $\Delta_2>\Delta_1$ is $50/81$. Again, working in a larger space leads to stronger and more general statements.

More possibilities are opened up if many cosmic rays are detected from a single flare. This could happen using a much larger facility, witnessing a rare, more powerful and closer event or through having over-estimated the strength of the turbulent GMF and EGMF. We can imagine having a set of events at times with delays $t_i$ measured from the first one, over an interval of observation with quantifiable uncertainty in the individual measurements, especially the rigidities. Here the statistical approach is likely to involve the estimation of two, three and four point correlation functions by combining data with simulation using standard techniques.

\bigskip
\bigskip
\section{The case of the TA hotspot}\label{sec:TAspot}
\begin{figure}
\centering
\includegraphics[width=0.45\textwidth]{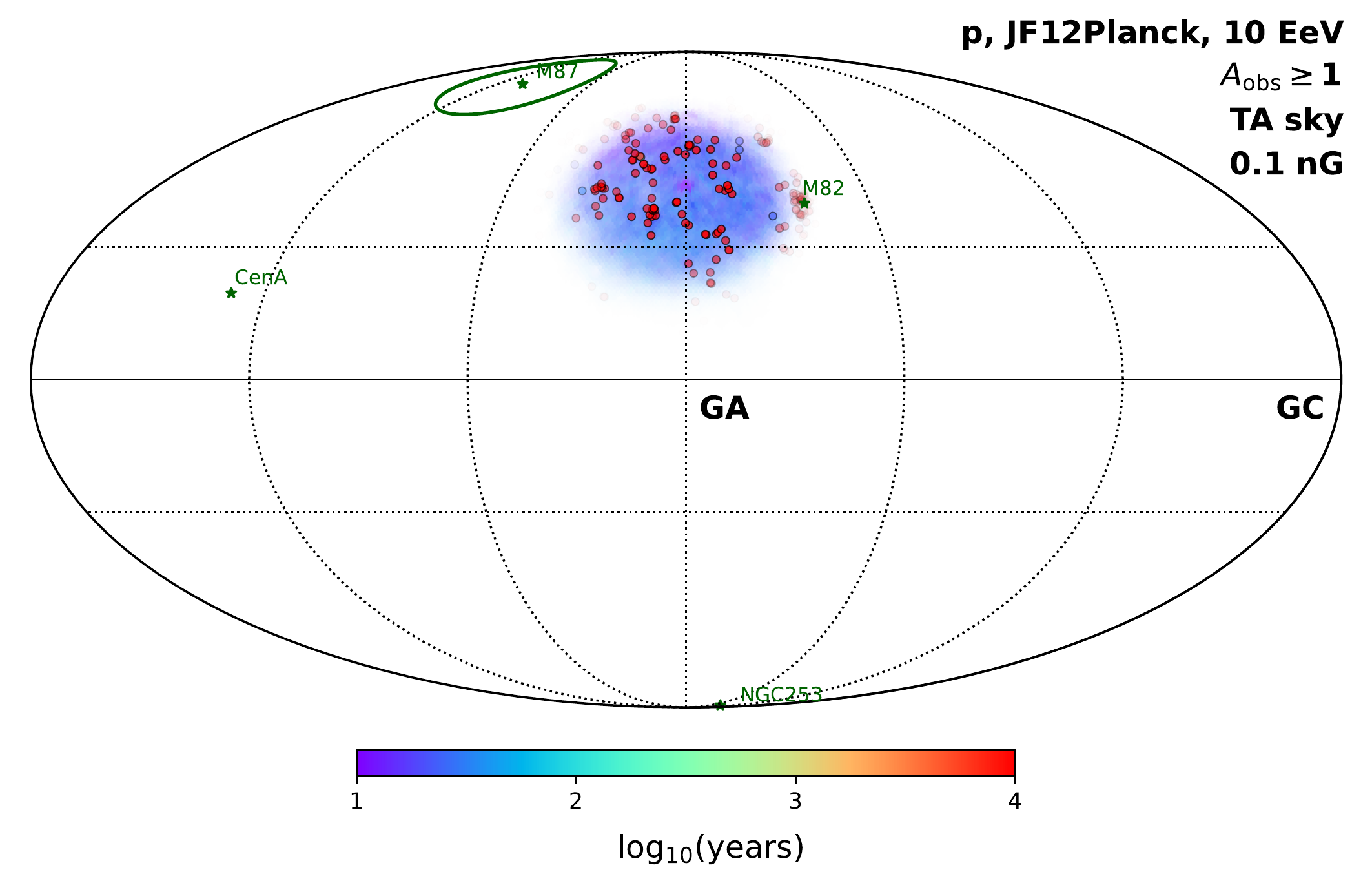}
\includegraphics[width=0.45\textwidth]{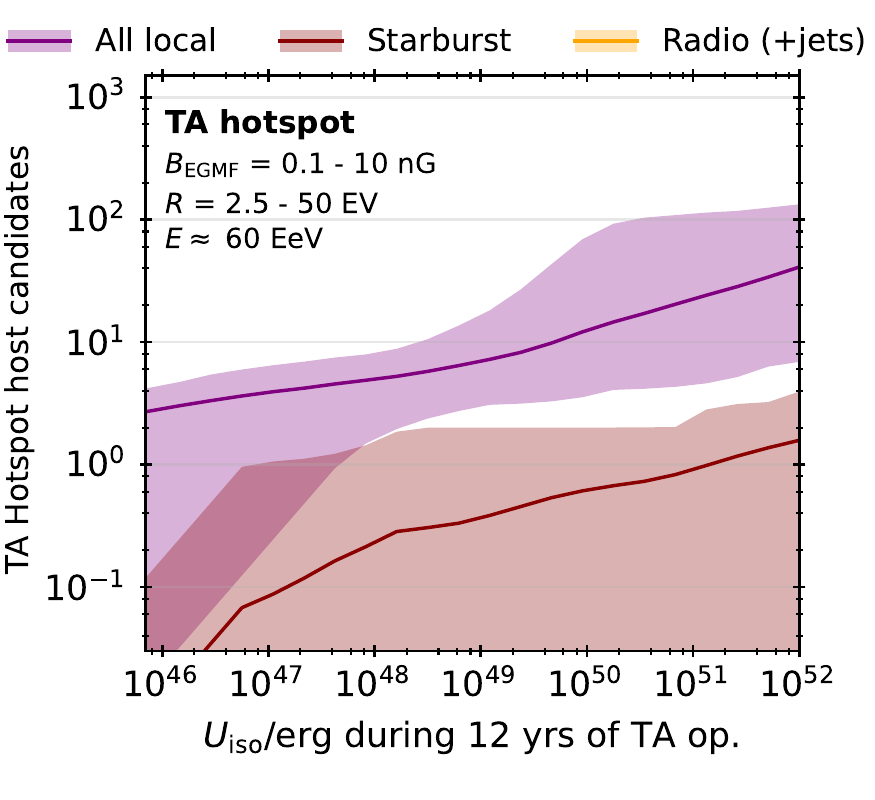}
\caption{Treasure map at 10 EV (top) and host candidate count within the TA hotspot back-projected at the Galactic boundary using the JF12Planck model (more rigidities are shown in Appendix \ref{fig:TA_01}). The local volume galaxies within 50 Mpc are shown color coded in $\tau_{d}$. Even under the assumption of a small EGMF of 0.1~nG, only a few sources with a small temporal dispersion below 100 years are found. These are predominantly Milky Way dwarfs, such as Ursa Major, Segue and Willman 1. The host candidate counts have been computed assuming an approximate energy of 60 EeV for the hotspot excess events and a composition scales the rigidities in the indicated range. The bands enclose the extrema defined by the ranges in $R$ and $B_{\rm EGMF}$, whereas the higher curves always correspond to smaller $B$ values. The assumed number of hotspot excess events is $n=20$.}
\label{fig:hotspot}
\end{figure}
The TA hotspot is an excess of events located within 25 degrees around $b\sim49^{\circ}$, $l\sim177^{\circ}$, indicating a medium-scale anisotropy at 3.2 $\sigma$ post-trial significance \citep{abbasi2014,sagawa2022}. The growth of the significance slowed down within the past 7 years compared to an equivalent period until 2015 and is consistent with the event accumulation rate from an isotropic expectation. One possible interpretation is that the observed clustering was, at least partially, due to a statistical overfluctuation. Alternatively, we might be witnessing the temporal tail from a nearby transient UHECR outburst. The back-projected extragalactic directions of the TA hotspot is shown in Fig.~\ref{fig:hotspot} for 10 EV rigidity given the JF12Planck model and the field of view of TA. Since the composition at the relevant energies $E>$~57~EeV is unknown, we expand our analysis over a range of relevant rigidities from 2.5~EV to 50~EV spanning from protons to iron (at 65~EeV) (see Fig.~\ref{fig:TA_01} in the Appendix). 

As one may observe in the sky map shown in the upper left panel of Fig.~\ref{fig:TM_nitrogen_150}, which corresponds to $R\sim21$ EV, the direction of the hotspot ({\it cf.} Fig.~\ref{fig:hotspot}) is located in a ``magnetic window'', a purple-colored region of the sky with very small $\tau_{d, {\rm GMF}}$ and small delays compared to the light travel time. This magnetic window encloses the shortest paths from the Galactic boundary towards the Earth, where UHECR experience the smallest deflections, time delays and time dispersion compared to other directions on the sky. From there, a cosmic ray pulse is able to travel through our vicinity within a couple of years leaving a potential trace in the data as spatial and temporal clustering. The expected time delays and dispersion are similar to that of $\omega_{\rm M82}$, which is located within this magnetic window ({\it cf.} histograms in lower left panel of Fig.~\ref{fig:temporal}). For a not too heavy composition, $\tau_{d, {\rm GMF}}$ is of the order of 10 years and a decade-scale temporal variability is allowed from GMF-only arguments.

The host candidate counts in the lower panel of Fig.~\ref{fig:hotspot} show that there are sufficient sources within the relevant spatial range. For the lower $U_{\rm iso}$ values, these are mostly local dwarf galaxies like Ursa Major, Segue, Willman 1, and Leo A. Objects that are not within the direct line of sight, such as the Large Magellanic Cloud (although strongly demagnified in the JF12 model so not appearing in Fig.~\ref{fig:TA_01}), can contribute counts due to lensing into the hotspot assuming lower rigidities. The first regular galaxy at lower rigidity is NGC2403 (a starburst galaxy at a distance $\sim3$~Mpc), followed by M82, which disappears from the field of view for $R>10$~EV despite being suggested as the origin of the TA hotspot  \citep{He2016,pfeffer2017,2020arXiv200507312T,2022MNRAS.511..448B}. For $R\gg10$~EV NGC2403 also leaves the field of view and NGC3274 and NGC2903 at $\sim8$ Mpc become the closest candidates. However, to explain a time variability of the hotspot on the scale of decades, the sources can not be distant.  For M82 at a distance of 3.6~Mpc, $\tau_{d, {\rm EGMF}}$ is already 52 years for $B_{\rm EGMF} = 0.1$~nG and 5200 years for 1~nG. The large impact of the EGMF is reflected in the widths of the bands in the lower panel of Fig.~\ref{fig:hotspot}, which enclose values from 0.1 to 10 nG. For starburst galaxies, only NGC2403 can be considered as a potential source under the most optimistic conditions.

\section{summary}\label{sec:summary}

In this paper, we ask what could we learn with future detections of EECR ($\sim$150 EeV) multiplets. The reason to focus on these extreme energies is that  doublets can be used to verify a proposed identification of  a source since the GZK horizon is small enough that the background of most distant sources is negligible.

\subsection{GZK horizon}
Since nuclei can photodissociate, the horizon depends upon the energy and mass threshold at detection, $E_{\rm obs}$ and $A_{\rm obs}$. In sec.~\ref{sec:gzk-horizon}, we determine the horizon by calculating the flux attenuation due to the GZK effect, $a_{\rm GZK}$, for different cases of  $E_{\rm obs}$ and $A_{\rm obs}$ (see Figs.~\ref{fig:containement_figures150} and \ref{fig:containement_figures300}). We find that the horizon is limited to our local supercluster ($d\lesssim 40$~Mpc) for both protons and iron nuclei at 150 EeV. For intermediate mass elements, it is strongly dependent upon the energy and mass threshold at detection since a source of nitrogen can be detected from larger distances if we allows the EECR to be detected as a proton ($A_{\rm obs}>1$). So, at 150 EeV for a nitrogen source, the horizon is 30 Mpc for $A_{\rm obs}=1$ but only $\sim1$~Mpc for $A_{\rm obs}>12$. The horizon also depends on the spectrum and maximum energy at the source. Therefore, knowing whether a EECR is light, medium or heavy on an event-by-event basis at these extreme energies would provide a strong indication of the source distance. For medium mass EECRs (CNO) at 150 EeV,  the sources would have to be located in our local group of galaxies. 

\subsection{Doublets}
In this paper, we provide a new methodology for the detection,  analysis and interpretation of EECR doublet detection. The methodology is as follow:

\begin{enumerate}
   \item[{\small $\bullet$}] We calculate the distribution of time delays in the GMF  for all the  trajectories connecting the observer's sky to an extragalactic source, for all source directions and using three different Galactic magnetic fields models: JF12, JF12Planck and TF17. The temporal dispersion introduced by the GMF, $\tau_{d, {\rm GMF}}$, depends only on the EECR rigidity. It is given by the spread of the distribution of time delays $t$. The distributions are asymmetric, characterized by a fast rise and a slow decay (see Fig.~\ref{fig:temporal}).  
   \item[{\small $\bullet$}] We calculate the  temporal dispersion introduced by the EGMF, $\tau_{d, {\rm EGMF}}$, assuming a purely turbulent magnetic field with a Kolmogorov spectrum, for rms field values of 0.1, 0.5 and 1~nG , keeping its coherence length to 0.2~Mpc (i.e. outer scale of 1~Mpc). 
     \item[{\small $\bullet$}]Taking into account TA and PAO field of view, we show the most promising directions to detect EECR doublets by combining all  four factors ($a_{\rm GZK}$, $\tau_{d, {\rm GMF}}$, $\tau_{d, {\rm EGMF}}$ and field of view), in ``treasure maps''. The treasure maps show the most promising direction to detect multiplets of events given a catalog source candidate and a GMF model.
    \item[{\small $\bullet$}] Assuming different source catalogs, we derived the number of host galaxies able to host a transient with isotropic equivalent energy $U_{\rm iso}$ sufficient to provide a doublet. The source counts statistic shows that exposure function is compensated by the GMF for heavy nuclei.
    \item[{\small $\bullet$}] Estimating the rigidity of both EECRs can allow observers to select the appropriate treasure map, and then select the source candidates within the angular dispersion introduced by the magnetic fields. It would also allow observers to verify that the lower rigidity EECR  follows the higher rigidity EECR.  
\end{enumerate}

\subsection{Multiplets}

While most studies focus on the spatial correlation to find structures in term of multiplets and anisotropies \citep[e.g.][]{2008ApJ...678..606T, 2009ApJ...702..825C, 2014A&A...567A..81R, 2019EPJWC.21001006L} or the temporal correlations between electromagnetic and cosmic ray events \citep[e.g.,][in the case of Gamma-Ray Bursts]{1995ApJ...449L..37M}, we ask what can be learn by studying the temporal and rigidity structure of the multiplet of events. The asymmetry in the distribution of time delays implies a specific temporal evolution in the EECR arrival times as a function of their rigidity.

 We show that an analysis of the multiplet arrival times would allow to distinguish between transient and continuous sources. Essentially, time arrival of events with a similar rigidity can be drawn independently from the distribution of time delays (Fig.~\ref{fig:temporal}) and compared with the data, {\it provided that we know the rigidity of the events}, i.e. provided that we are able to have a better characterization of the mass of the high energy cosmic-ray events. 

Since a ``hotspot'' (a multiplet of events) has been reported in the TA sky \citep{abbasi2014}, we also provide treasure maps for the back-projected hotspot window at  rigidities 2.5 to 25~EV. We can see that for all these rigidities the temporal dispersion introduced by the GMF is small in this direction, a ``magnetic window''. Now, taking into account the EGMF dispersion,  only a few sources are near enough to have a decade-scale temporal variability. Such sources are promising host candidates for a bursting source explanation for the origin of the hotspot. It would also explain the its disappearance after few years if we are already getting events from the tail of the temporal distribution. Such magnetic windows might explain the presence of a multiplet of events arriving within the lifetime of an array.

\section{Implications}\label{sec:conclusions}

The minimum isotropic energy $U_{\rm iso}$ in EECRs above 150 EeV to get a doublet from a single transient source at a distance $d_s$, in the direction $(l,b)$, depends on the total temporal dispersion $\tau_d$ (in the EGMF and the GMF), on the GZK attenuation $a_{\rm GZK}(d_s)$, on the magnification $M(l,b)$, and the field of view of the experiment, all displayed in the treasure maps. We then count the number of sources satisfying this condition. For the most optimistic case where the EGMF is negligible, we get 100 galaxies satisfying this condition for $U_{\rm iso}=10^{47}$~erg, and 1000 for $U_{\rm iso}>10^{49}$~erg. So, even if we have 1000 candidate galaxies, we have to take into account the rate of the transients to estimate the probability to observe a doublet during 10 year observation period, typical for an UHECR observatory. The minimum rate is then $10^{-4}$yr$^{-1}$galaxy$^{-1}$. 

We can compare this rate to the rates of known transient sources. For example high power long and short Gamma-Ray Bursts have $U_{\rm iso}\gtrsim10^{49}\,{\rm erg}$ and an observed burst rate ${\cal R}\sim10^{-7}yr^{-1}\,{\rm galaxy}^{-1}$ \citep{2010MNRAS.406.1944W}. These are too rare to expect a single doublet in a decade of observing. \citep[These are also too rare (by a factor $\sim 80$) to account for the luminosity density needed to account for the  highest energy cosmic rays, see][]{stv893}. Low luminosity GRB (LLGRB) are estimated to occur a hundred times more frequently with a hundred times smaller $U_{\rm iso}$ \citep{Nakar_2015}. These could produce observable doublets. Tidal Disruptions Events occur at a similar rate to low luminosity GRBs \citep{2014ApJ...792...53V} but it is unclear how they could accelerate $\sim150\,{\rm EeV}$ cosmic rays.

Another type of source undergoes repeated flares. Examples include powerful AGN \citep{Farrar2009} and magnetars \citep{Arons2003,Fang2013}, both of which can, in principle accelerate $\sim150\,{\rm EeV}$ cosmic rays. However, suitable jets are too distant to be invoked to account for the highest energy cosmic rays. This shortcoming is exacerbated if, as is likely, the cosmic rays are beamed.  Magnetars are created at a rate $\sim 10^{-3}\,{\rm yr}^{-1}\,{\rm galaxy}^{-1}$, but may only last for a hundred years before their magnetic energy, at least $10^{46}\,{\rm erg}$, dissipates in a sequence of flares \citep{Svinkin2021}. These flares are sufficiently powerful to accelerate nuclei to energies $\gtrsim150\,{\rm EeV}$, especially if, as is likely, they are heavy. However, the luminosity associated with the decay of magnetar magnetospheric field is only $\sim10^{-38}\,{\rm erg\,cm}^{-}\,{\rm s}^{-1}$. This is nearly a factor ten smaller than the luminosity density needed to account for the source of the highest energy cosmic rays. It is therefore required that the magnetic field below the surface of a magnetar be much larger than that inferred above the surface  --- $\sim10^{15}\,{\rm G}$ --- if they are to account for the acceleration of the highest energy cosmic rays. If, despite this requirement, they are the main source, then they are unlikely to produce temporal doublets, though they could provide statistical positional identifications with galaxies in the Local Group.

An improved mass resolution, on an event by event basis, may be achievable at existing facilities through applying machine learning techniques to shower data \citep[e.g.,][]{2019APh...111...12G} and by developing better particle physics  models that successfully confront the ``muon excess'' \citep{2022Ap&SS.367...27A}. If this happens and, especially, if the event rate increases at a future, larger facility, then there are many ways that we might be able to discriminate between alternative source models. 

For example, if a single, medium mass, like nitrogen, is detected at $\sim300\,{\rm EeV}\cong40\,{\rm EV}$, then it must have originated in the Local Group from a single source as the uncertainty in the GMF model will be small. This would rule out powerful AGN jets or clusters of galaxies. The uncertainty in the GMF model and the positions should be small at these large energies and therefore identifications with 10-100 galaxies within the horizon should be secure.

When considering three or more EECRs from the same transient source, a richer analysis of the directions, rigidities and arrival times can be a powerful tool in validating a transient source candidate. For example, the failure to see any of the temporal correlations outlined in this paper would support the view that the sources are essentially continuous.

It is hoped that the prospect of detecting doublets and higher multiplet cosmic ray events at the highest energy, correlated in time and rigidity, as well as position, as outlined in this paper, helps motivate the development of larger cosmic ray detector arrays and better approaches to mass measurement.  

\section*{Acknowledgements}
N.G.’s research is supported by the Simons Foundation, the Chancellor Fellowship at UCSC and the Vera Rubin Presidential Chair. A.F.\ acknowledges the hospitality within the group of Hiroyuki Sagawa at the ICRR, where he completed initial parts of this work as a JSPS International Research Fellow (JSPS KAKENHI Grant Number 19F19750).

\bibliography{refs}

\begin{thebibliography}{}
\expandafter\ifx\csname natexlab\endcsname\relax\def\natexlab#1{#1}\fi
\providecommand{\url}[1]{\href{#1}{#1}}
\providecommand{\dodoi}[1]{doi:~\href{http://doi.org/#1}{\nolinkurl{#1}}}
\providecommand{\doeprint}[1]{\href{http://ascl.net/#1}{\nolinkurl{http://ascl.net/#1}}}
\providecommand{\doarXiv}[1]{\href{https://arxiv.org/abs/#1}{\nolinkurl{https://arxiv.org/abs/#1}}}

\bibitem[{Aab {et~al.}(2015)}]{PierreAuger:2014yba}
Aab, A., {et~al.} 2015, Astrophys. J., 804, 15,
  \dodoi{10.1088/0004-637X/804/1/15}

\bibitem[{Aab {et~al.}(2017)}]{PierreAuger:2017tlx}
---. 2017, Phys. Rev. D, 96, 122003, \dodoi{10.1103/PhysRevD.96.122003}

\bibitem[{Abbasi {et~al.}(2008)}]{Abbasi:2007sv}
Abbasi, R.~U., {et~al.} 2008, Phys. Rev. Lett., 100, 101101,
  \dodoi{10.1103/PhysRevLett.100.101101}

\bibitem[{{Abbasi} {et~al.}(2014){Abbasi}, {Abe}, {Abu-Zayyad}, {Allen},
  {Anderson}, {Azuma}, {Barcikowski}, {Belz}, {Bergman}, {Blake}, {Cady},
  {Chae}, {Cheon}, {Chiba}, {Chikawa}, {Cho}, {Fujii}, {Fukushima}, {Goto},
  {Hanlon}, {Hayashi}, {Hayashida}, {Hibino}, {Honda}, {Ikeda}, {Inoue},
  {Ishii}, {Ishimori}, {Ito}, {Ivanov}, {Jui}, {Kadota}, {Kakimoto},
  {Kalashev}, {Kasahara}, {Kawai}, {Kawakami}, {Kawana}, {Kawata}, {Kido},
  {Kim}, {Kim}, {Kim}, {Kitamura}, {Kitamura}, {Kuzmin}, {Kwon}, {Lan}, {Lim},
  {Lundquist}, {Machida}, {Martens}, {Matsuda}, {Matsuyama}, {Matthews},
  {Minamino}, {Mukai}, {Myers}, {Nagasawa}, {Nagataki}, {Nakamura}, {Nonaka},
  {Nozato}, {Ogio}, {Ogura}, {Ohnishi}, {Ohoka}, {Oki}, {Okuda}, {Ono},
  {Oshima}, {Ozawa}, {Park}, {Pshirkov}, {Rodriguez}, {Rubtsov}, {Ryu},
  {Sagawa}, {Sakurai}, {Sampson}, {Scott}, {Shah}, {Shibata}, {Shibata},
  {Shimodaira}, {Shin}, {Smith}, {Sokolsky}, {Springer}, {Stokes}, {Stratton},
  {Stroman}, {Suzawa}, {Takamura}, {Takeda}, {Takeishi}, {Taketa}, {Takita},
  {Tameda}, {Tanaka}, {Tanaka}, {Tanaka}, {Thomas}, {Thomson}, {Tinyakov},
  {Tkachev}, {Tokuno}, {Tomida}, {Troitsky}, {Tsunesada}, {Tsutsumi},
  {Uchihori}, {Udo}, {Urban}, {Vasiloff}, {Wong}, {Yamane}, {Yamaoka},
  {Yamazaki}, {Yang}, {Yashiro}, {Yoneda}, {Yoshida}, {Yoshii}, {Zollinger}, \&
  {Zundel}}]{abbasi2014}
{Abbasi}, R.~U., {Abe}, M., {Abu-Zayyad}, T., {et~al.} 2014, \apjl, 790, L21,
  \dodoi{10.1088/2041-8205/790/2/L21}

\bibitem[{Abbasi {et~al.}(2014)}]{TelescopeArray:2014tsd}
Abbasi, R.~U., {et~al.} 2014, Astrophys. J. Lett., 790, L21,
  \dodoi{10.1088/2041-8205/790/2/L21}

\bibitem[{Abreu {et~al.}(2022)}]{PierreAuger:2022axr}
Abreu, P., {et~al.} 2022, Astrophys. J., 935, 170,
  \dodoi{10.3847/1538-4357/ac7d4e}

\bibitem[{{Albrecht} {et~al.}(2022){Albrecht}, {Cazon}, {Dembinski},
  {Fedynitch}, {Kampert}, {Pierog}, {Rhode}, {Soldin}, {Spaan}, {Ulrich}, \&
  {Unger}}]{2022Ap&SS.367...27A}
{Albrecht}, J., {Cazon}, L., {Dembinski}, H., {et~al.} 2022, \apss, 367, 27,
  \dodoi{10.1007/s10509-022-04054-5}

\bibitem[{{Alcock} \& {Hatchett}(1978)}]{Alcock1978}
{Alcock}, C., \& {Hatchett}, S. 1978, \apj, 222, 456, \dodoi{10.1086/156159}

\bibitem[{{Allard}(2012)}]{Allard2012}
{Allard}, D. 2012, Astroparticle Physics, 39, 33,
  \dodoi{10.1016/j.astropartphys.2011.10.011}

\bibitem[{{Allard} {et~al.}(2008){Allard}, {Busca}, {Decerprit}, {Olinto}, \&
  {Parizot}}]{allard2008}
{Allard}, D., {Busca}, N.~G., {Decerprit}, G., {Olinto}, A.~V., \& {Parizot},
  E. 2008, \jcap, 2008, 033, \dodoi{10.1088/1475-7516/2008/10/033}

\bibitem[{Alves~Batista {et~al.}(2022)}]{AlvesBatista:2022vem}
Alves~Batista, R., {et~al.} 2022.
\newblock \doarXiv{2208.00107}

\bibitem[{{Arons}(2003)}]{Arons2003}
{Arons}, J. 2003, \apj, 589, 871, \dodoi{10.1086/374776}

\bibitem[{{Bell} \& {Matthews}(2022)}]{2022MNRAS.511..448B}
{Bell}, A.~R., \& {Matthews}, J.~H. 2022, \mnras, 511, 448,
  \dodoi{10.1093/mnras/stac031}

\bibitem[{{Blandford}(2000)}]{2000PhST...85..191B}
{Blandford}, R.~D. 2000, Physica Scripta Volume T, 85, 191,
  \dodoi{10.1238/Physica.Topical.085a00191}

\bibitem[{{Cuoco} {et~al.}(2009){Cuoco}, {Hannestad}, {Haugb{\o}lle},
  {Kachelrie{\ss}}, \& {Serpico}}]{2009ApJ...702..825C}
{Cuoco}, A., {Hannestad}, S., {Haugb{\o}lle}, T., {Kachelrie{\ss}}, M., \&
  {Serpico}, P.~D. 2009, \apj, 702, 825, \dodoi{10.1088/0004-637X/702/2/825}

\bibitem[{{Ding} {et~al.}(2021){Ding}, {Globus}, \& {Farrar}}]{DGF20}
{Ding}, C., {Globus}, N., \& {Farrar}, G.~R. 2021, \apjl, 913, L13,
  \dodoi{10.3847/2041-8213/abf11e}

\bibitem[{{Fang} {et~al.}(2013){Fang}, {Kotera}, \& {Olinto}}]{Fang2013}
{Fang}, K., {Kotera}, K., \& {Olinto}, A.~V. 2013, \jcap, 2013, 010,
  \dodoi{10.1088/1475-7516/2013/03/010}

\bibitem[{{Farrar} {et~al.}(2015){Farrar}, {Awal}, {Khurana}, \&
  {Sutherland}}]{Farrar2015optics}
{Farrar}, G.~R., {Awal}, N., {Khurana}, D., \& {Sutherland}, M. 2015, in
  International Cosmic Ray Conference, Vol.~34, 34th International Cosmic Ray
  Conference (ICRC2015), 560

\bibitem[{{Farrar} \& {Gruzinov}(2009)}]{Farrar2009}
{Farrar}, G.~R., \& {Gruzinov}, A. 2009, \apj, 693, 329,
  \dodoi{10.1088/0004-637X/693/1/329}

\bibitem[{Globus {et~al.}(2015)Globus, Allard, Mochkovitch, \&
  Parizot}]{stv893}
Globus, N., Allard, D., Mochkovitch, R., \& Parizot, E. 2015, Monthly Notices
  of the Royal Astronomical Society, 451, 751, \dodoi{10.1093/mnras/stv893}

\bibitem[{{Globus} {et~al.}(2008){Globus}, {Allard}, \&
  {Parizot}}]{2008A&A...479...97G}
{Globus}, N., {Allard}, D., \& {Parizot}, E. 2008, \aap, 479, 97,
  \dodoi{10.1051/0004-6361:20078653}

\bibitem[{{Globus} \& {Eichler}(2016)}]{globus2016}
{Globus}, N., \& {Eichler}, D. 2016, \apjl, 833, L17,
  \dodoi{10.3847/2041-8213/833/2/L17}

\bibitem[{{Globus} \& {Piran}(2017)}]{2017ApJ...850L..25G}
{Globus}, N., \& {Piran}, T. 2017, \apjl, 850, L25,
  \dodoi{10.3847/2041-8213/aa991b}

\bibitem[{G\'orski {et~al.}(2005)G\'orski, Hivon, Banday, Wandelt, Hansen,
  Reinecke, \& Bartelman}]{Gorski:2004by}
G\'orski, K.~M., Hivon, E., Banday, A.~J., {et~al.} 2005, Astrophys. J., 622,
  759, \dodoi{10.1086/427976}

\bibitem[{Greisen(1966)}]{Greisen:1966jv}
Greisen, K. 1966, Phys. Rev. Lett., 16, 748, \dodoi{10.1103/PhysRevLett.16.748}

\bibitem[{{Guill{\'e}n} {et~al.}(2019){Guill{\'e}n}, {Bueno}, {Carceller},
  {Mart{\'\i}nez-Vel{\'a}zquez}, {Rubio}, {Todero Peixoto}, \&
  {Sanchez-Lucas}}]{2019APh...111...12G}
{Guill{\'e}n}, A., {Bueno}, A., {Carceller}, J.~M., {et~al.} 2019,
  Astroparticle Physics, 111, 12, \dodoi{10.1016/j.astropartphys.2019.03.001}

\bibitem[{Hanlon(2019)}]{Hanlon:2018dvd}
Hanlon, W. 2019, EPJ Web Conf., 208, 02001,
  \dodoi{10.1051/epjconf/201920802001}

\bibitem[{Harari {et~al.}(2000)Harari, Mollerach, \& Roulet}]{harari2000}
Harari, D., Mollerach, S., \& Roulet, E. 2000, Journal of High Energy Physics,
  2000, 035, \dodoi{10.1088/1126-6708/2000/02/035}

\bibitem[{{He} {et~al.}(2016){He}, {Kusenko}, {Nagataki}, {Zhang}, {Yang}, \&
  {Fan}}]{He2016}
{He}, H.-N., {Kusenko}, A., {Nagataki}, S., {et~al.} 2016, \prd, 93, 043011,
  \dodoi{10.1103/PhysRevD.93.043011}

\bibitem[{Heinze {et~al.}(2019)Heinze, Fedynitch, Boncioli, \&
  Winter}]{Heinze:2019jou}
Heinze, J., Fedynitch, A., Boncioli, D., \& Winter, W. 2019, Astrophys. J.,
  873, 88, \dodoi{10.3847/1538-4357/ab05ce}

\bibitem[{{Jansson} \& {Farrar}(2012)}]{JF12}
{Jansson}, R., \& {Farrar}, G.~R. 2012, \apj, 757, 14,
  \dodoi{10.1088/0004-637X/757/1/14}

\bibitem[{Karachentsev {et~al.}(2013)Karachentsev, Makarov, \&
  Kaisina}]{Karachentsev:2013cva}
Karachentsev, I.~D., Makarov, D.~I., \& Kaisina, E.~I. 2013, Astron. J., 145,
  101, \dodoi{10.1088/0004-6256/145/4/101}

\bibitem[{{Lemoine} {et~al.}(1997){Lemoine}, {Sigl}, {Olinto}, \&
  {Schramm}}]{Lemoine1997}
{Lemoine}, M., {Sigl}, G., {Olinto}, A.~V., \& {Schramm}, D.~N. 1997, \apjl,
  486, L115, \dodoi{10.1086/310847}

\bibitem[{{Lunardini} {et~al.}(2019){Lunardini}, {Vance}, {Emig}, \&
  {Windhorst}}]{2019JCAP...10..073L}
{Lunardini}, C., {Vance}, G.~S., {Emig}, K.~L., \& {Windhorst}, R.~A. 2019,
  \jcap, 2019, 073, \dodoi{10.1088/1475-7516/2019/10/073}

\bibitem[{{Lundquist} \& {Sokolsky}(2019)}]{2019EPJWC.21001006L}
{Lundquist}, J.~P., \& {Sokolsky}, P.~V. 2019, in European Physical Journal Web
  of Conferences, Vol. 210, European Physical Journal Web of Conferences,
  01006, \dodoi{10.1051/epjconf/201921001006}

\bibitem[{{Milgrom} \& {Usov}(1995)}]{1995ApJ...449L..37M}
{Milgrom}, M., \& {Usov}, V. 1995, \apjl, 449, L37, \dodoi{10.1086/309633}

\bibitem[{Nakar(2015)}]{Nakar_2015}
Nakar, E. 2015, The Astrophysical Journal, 807, 172,
  \dodoi{10.1088/0004-637X/807/2/172}

\bibitem[{Pfeffer {et~al.}(2017)Pfeffer, Kovetz, \& Kamionkowski}]{pfeffer2017}
Pfeffer, D.~N., Kovetz, E.~D., \& Kamionkowski, M. 2017, Monthly Notices of the
  Royal Astronomical Society, 466, 2922

\bibitem[{{Pierre Auger Collaboration} {et~al.}(2017){Pierre Auger
  Collaboration}, {Aab}, {Abreu}, {Aglietta}, {Samarai}, {Albuquerque},
  {Allekotte}, {Almela}, {Alvarez Castillo}, {Alvarez-Mu{\~n}iz}, {Anastasi},
  {Anchordoqui}, {Andrada}, {Andringa}, {Aramo}, {Arqueros}, {Arsene},
  {Asorey}, {Assis}, {Aublin}, {Avila}, {Badescu}, {Balaceanu}, {Barbato},
  {Barreira Luz}, {Beatty}, {Becker}, {Bellido}, {Berat}, {Bertaina}, {Bertou},
  {Biermann}, {Billoir}, {Biteau}, {Blaess}, {Blanco}, {Blazek}, {Bleve},
  {Boh{\'a}{\v{c}}ov{\'a}}, {Boncioli}, {Bonifazi}, {Borodai}, {Botti},
  {Brack}, {Brancus}, {Bretz}, {Bridgeman}, {Briechle}, {Buchholz}, {Bueno},
  {Buitink}, {Buscemi}, {Caballero-Mora}, {Caccianiga}, {Cancio}, {Canfora},
  {Caramete}, {Caruso}, {Castellina}, {Cataldi}, {Cazon}, {Chavez},
  {Chinellato}, {Chudoba}, {Clay}, {Cobos}, {Colalillo}, {Coleman}, {Collica},
  {Coluccia}, {Concei{\c{c}}{\~a}o}, {Consolati}, {Contreras}, {Cooper},
  {Coutu}, {Covault}, {Cronin}, {D'Amico}, {Daniel}, {Dasso}, {Daumiller},
  {Dawson}, {de Almeida}, {de Jong}, {De Mauro}, {de Mello Neto}, {De Mitri},
  {de Oliveira}, {de Souza}, {Debatin}, {Deligny}, {Di Giulio}, {Di Matteo},
  {D{\'\i}az Castro}, {Diogo}, {Dobrigkeit}, {D'Olivo}, {Dorosti}, {dos Anjos},
  {Dova}, {Dundovic}, {Ebr}, {Engel}, {Erdmann}, {Erfani}, {Escobar},
  {Espadanal}, {Etchegoyen}, {Falcke}, {Farrar}, {Fauth}, {Fazzini}, {Fenu},
  {Fick}, {Figueira}, {Filip{\v{c}}i{\v{c}}}, {Fratu}, {Freire}, {Fujii},
  {Fuster}, {Gaior}, {Garc{\'\i}a}, {Garcia-Pinto}, {Gat{\'e}}, {Gemmeke},
  {Gherghel-Lascu}, {Ghia}, {Giaccari}, {Giammarchi}, {Giller}, {G{\l}as},
  {Glaser}, {Golup}, {G{\'o}mez Berisso}, {G{\'o}mez Vitale}, {Gonz{\'a}lez},
  {Gorgi}, {Gorham}, {Grillo}, {Grubb}, {Guarino}, {Guedes}, {Hampel},
  {Hansen}, {Harari}, {Harrison}, {Harton}, {Haungs}, {Hebbeker}, {Heck},
  {Heimann}, {Herve}, {Hill}, {Hojvat}, {Holt}, {Homola}, {H{\"o}randel},
  {Horvath}, {Hrabovsk{\'y}}, {Huege}, {Hulsman}, {Insolia}, {Isar}, {Jandt},
  {Jansen}, {Johnsen}, {Josebachuili}, {Jurysek}, {K{\"a}{\"a}p{\"a}},
  {Kambeitz}, {Kampert}, {Katkov}, {Keilhauer}, {Kemmerich}, {Kemp}, {Kemp},
  {Kieckhafer}, {Klages}, {Kleifges}, {Kleinfeller}, {Krause}, {Krohm},
  {Kuempel}, {Kukec Mezek}, {Kunka}, {Kuotb Awad}, {LaHurd}, {Lauscher},
  {Legumina}, {Leigui de Oliveira}, {Letessier-Selvon}, {Lhenry-Yvon}, {Link},
  {Lo Presti}, {Lopes}, {L{\'o}pez}, {L{\'o}pez Casado}, {Luce}, {Lucero},
  {Malacari}, {Mallamaci}, {Mandat}, {Mantsch}, {Mariazzi}, {Mari{\c{s}}},
  {Marsella}, {Martello}, {Martinez}, {Mart{\'\i}nez Bravo}, {Mas{\'\i}as
  Meza}, {Mathes}, {Mathys}, {Matthews}, {Matthews}, {Matthiae}, {Mayotte},
  {Mazur}, {Medina}, {Medina-Tanco}, {Melo}, {Menshikov}, {Merenda}, {Michal},
  {Micheletti}, {Middendorf}, {Miramonti}, {Mitrica}, {Mockler}, {Mollerach},
  {Montanet}, {Morello}, {Mostaf{\'a}}, {M{\"u}ller}, {M{\"u}ller}, {Muller},
  {M{\"u}ller}, {Mussa}, {Naranjo}, {Nellen}, {Nguyen}, {Niculescu-Oglinzanu},
  {Niechciol}, {Niemietz}, {Niggemann}, {Nitz}, {Nosek}, {Novotny},
  {No{\v{z}}ka}, {N{\'u}{\~n}ez}, {Ochilo}, {Oikonomou}, {Olinto}, {Palatka},
  {Pallotta}, {Papenbreer}, {Parente}, {Parra}, {Paul}, {Pech}, {Pedreira},
  {Pkala}, {Pelayo}, {Pe{\~n}a-Rodriguez}, {Pereira}, {Perl{\'\i}n}, {Perrone},
  {Peters}, {Petrera}, {Phuntsok}, {Piegaia}, {Pierog}, {Pieroni}, {Pimenta},
  {Pirronello}, {Platino}, {Plum}, {Porowski}, {Prado}, {Privitera}, {Prouza},
  {Quel}, {Querchfeld}, {Quinn}, {Ramos-Pollan}, {Rautenberg}, {Ravignani},
  {Revenu}, {Ridky}, {Riehn}, {Risse}, {Ristori}, {Rizi}, {Rodrigues de
  Carvalho}, {Rodriguez Fernandez}, {Rodriguez Rojo}, {Rogozin}, {Roncoroni},
  {Roth}, {Roulet}, {Rovero}, {Ruehl}, {Saffi}, {Saftoiu}, {Salamida},
  {Salazar}, {Saleh}, {Salesa Greus}, {Salina}, {S{\'a}nchez}, {Sanchez-Lucas},
  {Santos}, {Santos}, {Sarazin}, {Sarmento}, {Sarmiento}, {Sato}, {Schauer},
  {Scherini}, {Schieler}, {Schimp}, {Schmidt}, {Scholten}, {Schov{\'a}nek},
  {Schr{\"o}der}, {Schulz}, {Schumacher}, {Sciutto}, {Segreto}, {Settimo},
  {Shadkam}, {Shellard}, {Sigl}, {Silli}, {Sima}, {{\'S}mia{\l}kowski},
  {{\v{S}}m{\'\i}da}, {Snow}, {Sommers}, {Sonntag}, {Sorokin}, {Squartini},
  {Stanca}, {Stani{\v{c}}}, {Stasielak}, {Stassi}, {Strafella}, {Suarez},
  {Suarez Dur{\'a}n}, {Sudholz}, {Suomij{\"a}rvi}, {Supanitsky},
  {{\v{S}}up{\'\i}k}, {Swain}, {Szadkowski}, {Taboada}, {Taborda}, {Tapia},
  {Theodoro}, {Timmermans}, {Todero Peixoto}, {Tomankova}, {Tom{\'e}},
  {Torralba Elipe}, {Travnicek}, {Trini}, {Ulrich}, {Unger}, {Urban},
  {Vald{\'e}s Galicia}, {Vali{\~n}o}, {Valore}, {van Aar}, {van Bodegom}, {van
  den Berg}, {van Vliet}, {Varela}, {Vargas C{\'a}rdenas}, {Varner},
  {V{\'a}zquez}, {Veberi{\v{c}}}, {Ventura}, {Vergara Quispe}, {Verzi},
  {Vicha}, {Villase{\~n}or}, {Vorobiov}, {Wahlberg}, {Wainberg}, {Walz},
  {Watson}, {Weber}, {Weindl}, {Wiencke}, {Wilczy{\'n}ski}, {Wirtz},
  {Wittkowski}, {Wundheiler}, {Yang}, {Yushkov}, {Zas}, {Zavrtanik},
  {Zavrtanik}, {Zepeda}, {Zimmermann}, {Ziolkowski}, {Zong}, \&
  {Zuccarello}}]{2017Sci...357.1266P}
{Pierre Auger Collaboration}, {Aab}, A., {Abreu}, P., {et~al.} 2017, Science,
  357, 1266, \dodoi{10.1126/science.aan4338}

\bibitem[{{Planck Collaboration} {et~al.}(2016){Planck Collaboration}, {Adam},
  {Ade}, {Alves}, {Ashdown}, {Aumont}, {Baccigalupi}, {Banday}, {Barreiro},
  {Bartolo}, {Battaner}, {Benabed}, {Benoit-L{\'e}vy}, {Bernard}, {Bersanelli},
  {Bielewicz}, {Bonavera}, {Bond}, {Borrill}, {Bouchet}, {Boulanger}, {Bucher},
  {Burigana}, {Butler}, {Calabrese}, {Cardoso}, {Catalano}, {Chiang},
  {Christensen}, {Colombo}, {Combet}, {Couchot}, {Crill}, {Curto}, {Cuttaia},
  {Danese}, {Davis}, {de Bernardis}, {de Rosa}, {de Zotti}, {Delabrouille},
  {Dickinson}, {Diego}, {Dolag}, {Dor{\'e}}, {Ducout}, {Dupac}, {Elsner},
  {En{\ss}lin}, {Eriksen}, {Ferri{\`e}re}, {Finelli}, {Forni}, {Frailis},
  {Fraisse}, {Franceschi}, {Galeotta}, {Ganga}, {Ghosh}, {Giard}, {Gjerl{\o}w},
  {Gonz{\'a}lez-Nuevo}, {G{\'o}rski}, {Gregorio}, {Gruppuso}, {Gudmundsson},
  {Hansen}, {Harrison}, {Hern{\'a}ndez-Monteagudo}, {Herranz}, {Hildebrandt},
  {Hobson}, {Hornstrup}, {Hurier}, {Jaffe}, {Jaffe}, {Jones}, {Juvela},
  {Keih{\"a}nen}, {Keskitalo}, {Kisner}, {Knoche}, {Kunz}, {Kurki-Suonio},
  {Lamarre}, {Lasenby}, {Lattanzi}, {Lawrence}, {Leahy}, {Leonardi}, {Levrier},
  {Liguori}, {Lilje}, {Linden-V{\o}rnle}, {L{\'o}pez-Caniego}, {Lubin},
  {Mac{\'\i}as-P{\'e}rez}, {Maggio}, {Maino}, {Mandolesi}, {Mangilli}, {Maris},
  {Martin}, {Mart{\'\i}nez-Gonz{\'a}lez}, {Masi}, {Matarrese}, {Melchiorri},
  {Mennella}, {Migliaccio}, {Miville-Desch{\^e}nes}, {Moneti}, {Montier},
  {Morgante}, {Munshi}, {Murphy}, {Naselsky}, {Nati}, {Natoli},
  {N{\o}rgaard-Nielsen}, {Oppermann}, {Orlando}, {Pagano}, {Pajot}, {Paladini},
  {Paoletti}, {Pasian}, {Perotto}, {Pettorino}, {Piacentini}, {Piat},
  {Pierpaoli}, {Plaszczynski}, {Pointecouteau}, {Polenta}, {Ponthieu}, {Pratt},
  {Prunet}, {Puget}, {Rachen}, {Reinecke}, {Remazeilles}, {Renault}, {Renzi},
  {Ristorcelli}, {Rocha}, {Rossetti}, {Roudier}, {Rubi{\~n}o-Mart{\'\i}n},
  {Rusholme}, {Sandri}, {Santos}, {Savelainen}, {Scott}, {Spencer},
  {Stolyarov}, {Stompor}, {Strong}, {Sudiwala}, {Sunyaev}, {Suur-Uski},
  {Sygnet}, {Tauber}, {Terenzi}, {Toffolatti}, {Tomasi}, {Tristram}, {Tucci},
  {Valenziano}, {Valiviita}, {Van Tent}, {Vielva}, {Villa}, {Wade}, {Wandelt},
  {Wehus}, {Yvon}, {Zacchei}, \& {Zonca}}]{Planck2016}
{Planck Collaboration}, {Adam}, R., {Ade}, P.~A.~R., {et~al.} 2016, \aap, 596,
  A103, \dodoi{10.1051/0004-6361/201528033}

\bibitem[{{Rouill{\'e} d'Orfeuil} {et~al.}(2014){Rouill{\'e} d'Orfeuil},
  {Allard}, {Lachaud}, {Parizot}, {Blaksley}, \&
  {Nagataki}}]{2014A&A...567A..81R}
{Rouill{\'e} d'Orfeuil}, B., {Allard}, D., {Lachaud}, C., {et~al.} 2014, \aap,
  567, A81, \dodoi{10.1051/0004-6361/201423462}

\bibitem[{{Sagawa}(2022)}]{sagawa2022}
{Sagawa}, H. 2022, arXiv e-prints, arXiv:2209.03591.
\newblock \doarXiv{2209.03591}

\bibitem[{Sommers(2001)}]{Sommers:2000us}
Sommers, P. 2001, Astropart. Phys., 14, 271,
  \dodoi{10.1016/S0927-6505(00)00130-4}

\bibitem[{{Svinkin} {et~al.}(2021){Svinkin}, {Frederiks}, {Hurley}, {Aptekar},
  {Golenetskii}, {Lysenko}, {Ridnaia}, {Tsvetkova}, {Ulanov}, {Cline},
  {Mitrofanov}, {Golovin}, {Kozyrev}, {Litvak}, {Sanin}, {Goldstein}, {Briggs},
  {Wilson-Hodge}, {von Kienlin}, {Zhang}, {Rau}, {Savchenko}, {Bozzo},
  {Ferrigno}, {Ubertini}, {Bazzano}, {Rodi}, {Barthelmy}, {Cummings}, {Krimm},
  {Palmer}, {Boynton}, {Fellows}, {Harshman}, {Enos}, \& {Starr}}]{Svinkin2021}
{Svinkin}, D., {Frederiks}, D., {Hurley}, K., {et~al.} 2021, \nat, 589, 211,
  \dodoi{10.1038/s41586-020-03076-9}

\bibitem[{{Takami} \& {Sato}(2008)}]{2008ApJ...678..606T}
{Takami}, H., \& {Sato}, K. 2008, \apj, 678, 606, \dodoi{10.1086/533522}

\bibitem[{{Telescope Array Collaboration} {et~al.}(2020){Telescope Array
  Collaboration}, {Abbasi}, {Abe}, {Abu-Zayyad}, {Allen}, {Azuma},
  {Barcikowski}, {Belz}, {Bergman}, {Blake}, {Cady}, {Cheon}, {Chiba},
  {Chikawa}, {di Matteo}, {Fujii}, {Fujisue}, {Fujita}, {Fujiwara},
  {Fukushima}, {Furlich}, {Hanlon}, {Hayashi}, {Hayashida}, {Hibino},
  {Higuchi}, {Honda}, {Ikeda}, {Inadomi}, {Inoue}, {Ishii}, {Ishimori}, {Ito},
  {Ivanov}, {Iwakura}, {Jeong}, {Jeong}, {Jui}, {Kadota}, {Kakimoto},
  {Kalashev}, {Kasahara}, {Kasami}, {Kawai}, {Kawakami}, {Kawana}, {Kawata},
  {Kido}, {Kim}, {Kim}, {Kim}, {Kim}, {Kim}, {Kishigami}, {Kuzmin},
  {Kuznetsov}, {Kwon}, {Lee}, {Lubsandorzhiev}, {Lundquist}, {Machida},
  {Matsumiya}, {Matsuyama}, {Matthews}, {Mayta}, {Minamino}, {Mukai}, {Myers},
  {Nagataki}, {Nakai}, {Nakamura}, {Nakamura}, {Nakamura}, {Nonaka}, {Oda},
  {Ogio}, {Ohnishi}, {Ohoka}, {Oku}, {Okuda}, {Omura}, {Ono}, {Onogi},
  {Oshima}, {Ozawa}, {Park}, {Pshirkov}, {Remington}, {Rodriguez}, {Rubtsov},
  {Ryu}, {Sagawa}, {Sahara}, {Saito}, {Sakaki}, {Sako}, {Sakurai}, {Sano},
  {Seki}, {Sekino}, {Shah}, {Shibata}, {Shibata}, {Shimodaira}, {Shin}, {Shin},
  {Smith}, {Sokolsky}, {Sone}, {Stokes}, {Stroman}, {Suzawa}, {Takagi},
  {Takahashi}, {Takamura}, {Takeishi}, {Taketa}, {Takita}, {Tameda}, {Tanaka},
  {Tanaka}, {Tanaka}, {Tanoue}, {Thomas}, {Thomson}, {Tinyakov}, {Tkachev},
  {Tokuno}, {Tomida}, {Troitsky}, {Tsunesada}, {Uchihori}, {Udo}, {Uehama},
  {Urban}, {Wong}, {Yada}, {Yamamoto}, {Yamazaki}, {Yang}, {Yashiro}, {Yosei},
  {Zhezher}, \& {Zundel}}]{2020arXiv200507312T}
{Telescope Array Collaboration}, {Abbasi}, R.~U., {Abe}, M., {et~al.} 2020,
  arXiv e-prints, arXiv:2005.07312.
\newblock \doarXiv{2005.07312}

\bibitem[{{Terral} \& {Ferri{\`e}re}(2017)}]{TF17}
{Terral}, P., \& {Ferri{\`e}re}, K. 2017, \aap, 600, A29,
  \dodoi{10.1051/0004-6361/201629572}

\bibitem[{{Thielheim} \& {Langhoff}(1968)}]{Thielheim1968}
{Thielheim}, K.~O., \& {Langhoff}, W. 1968, Journal of Physics A Mathematical
  General, 1, 694, \dodoi{10.1088/0305-4470/1/6/308}

\bibitem[{{van Velzen} {et~al.}(2012){van Velzen}, {Falcke}, {Schellart},
  {Nierstenh{\"o}fer}, \& {Kampert}}]{vanVelzen2012}
{van Velzen}, S., {Falcke}, H., {Schellart}, P., {Nierstenh{\"o}fer}, N., \&
  {Kampert}, K.-H. 2012, \aap, 544, A18, \dodoi{10.1051/0004-6361/201219389}

\bibitem[{{van Velzen} \& {Farrar}(2014)}]{2014ApJ...792...53V}
{van Velzen}, S., \& {Farrar}, G.~R. 2014, \apj, 792, 53,
  \dodoi{10.1088/0004-637X/792/1/53}

\bibitem[{{Wanderman} \& {Piran}(2010)}]{2010MNRAS.406.1944W}
{Wanderman}, D., \& {Piran}, T. 2010, \mnras, 406, 1944,
  \dodoi{10.1111/j.1365-2966.2010.16787.x}

\bibitem[{{Williamson}(1974)}]{Williamson1974}
{Williamson}, I.~P. 1974, \mnras, 166, 499, \dodoi{10.1093/mnras/166.3.499}

\bibitem[{Zatsepin \& Kuzmin(1966)}]{Zatsepin:1966jv}
Zatsepin, G.~T., \& Kuzmin, V.~A. 1966, JETP Lett., 4, 78

\end{thebibliography}

\appendix
\section{Source counts and treasure maps for an EGMF strength of 0.1 \nG}

 \begin{figure}[h!]
\centering
\includegraphics[width=0.45\textwidth]{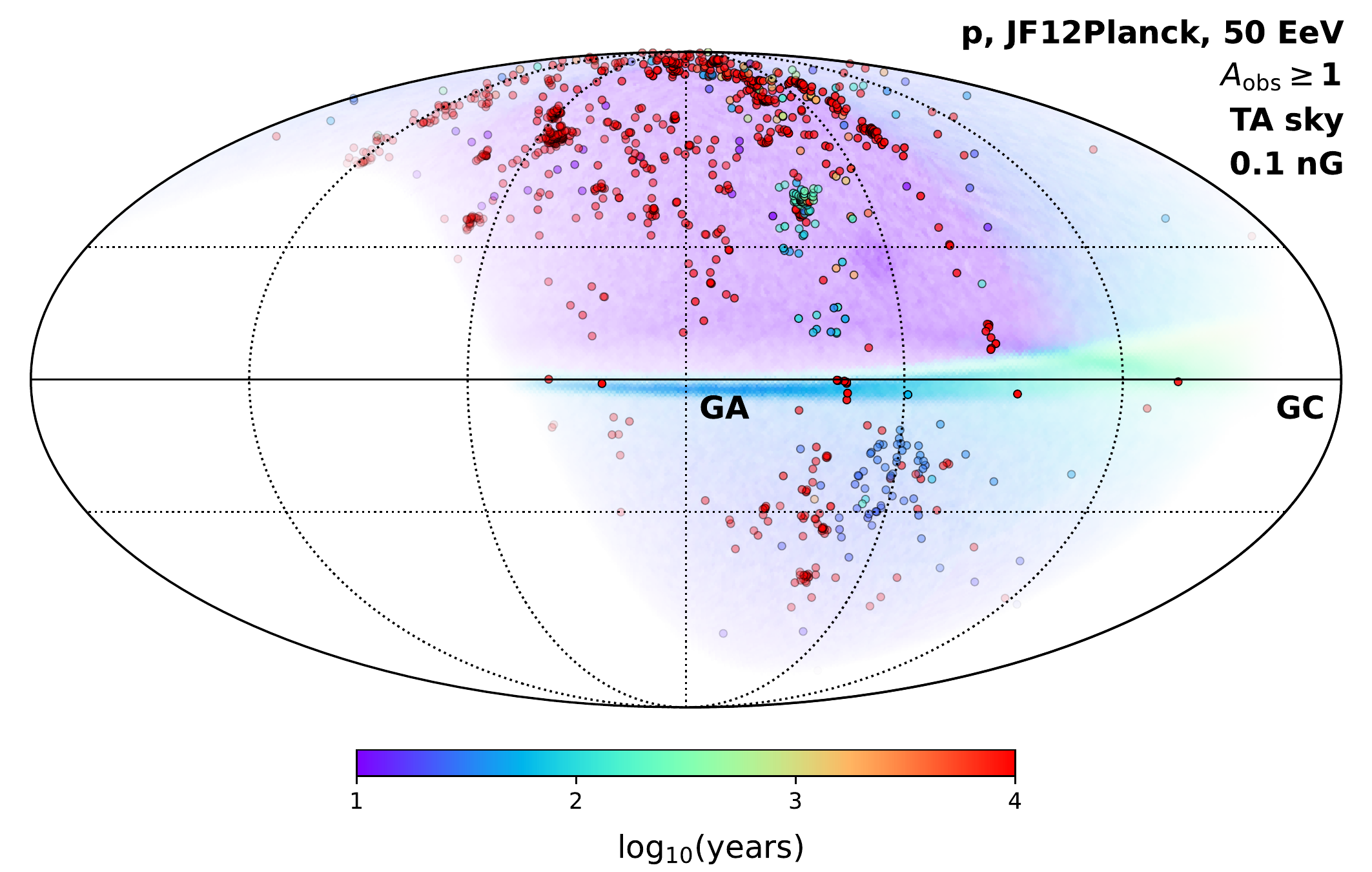}
\includegraphics[width=0.45\textwidth]{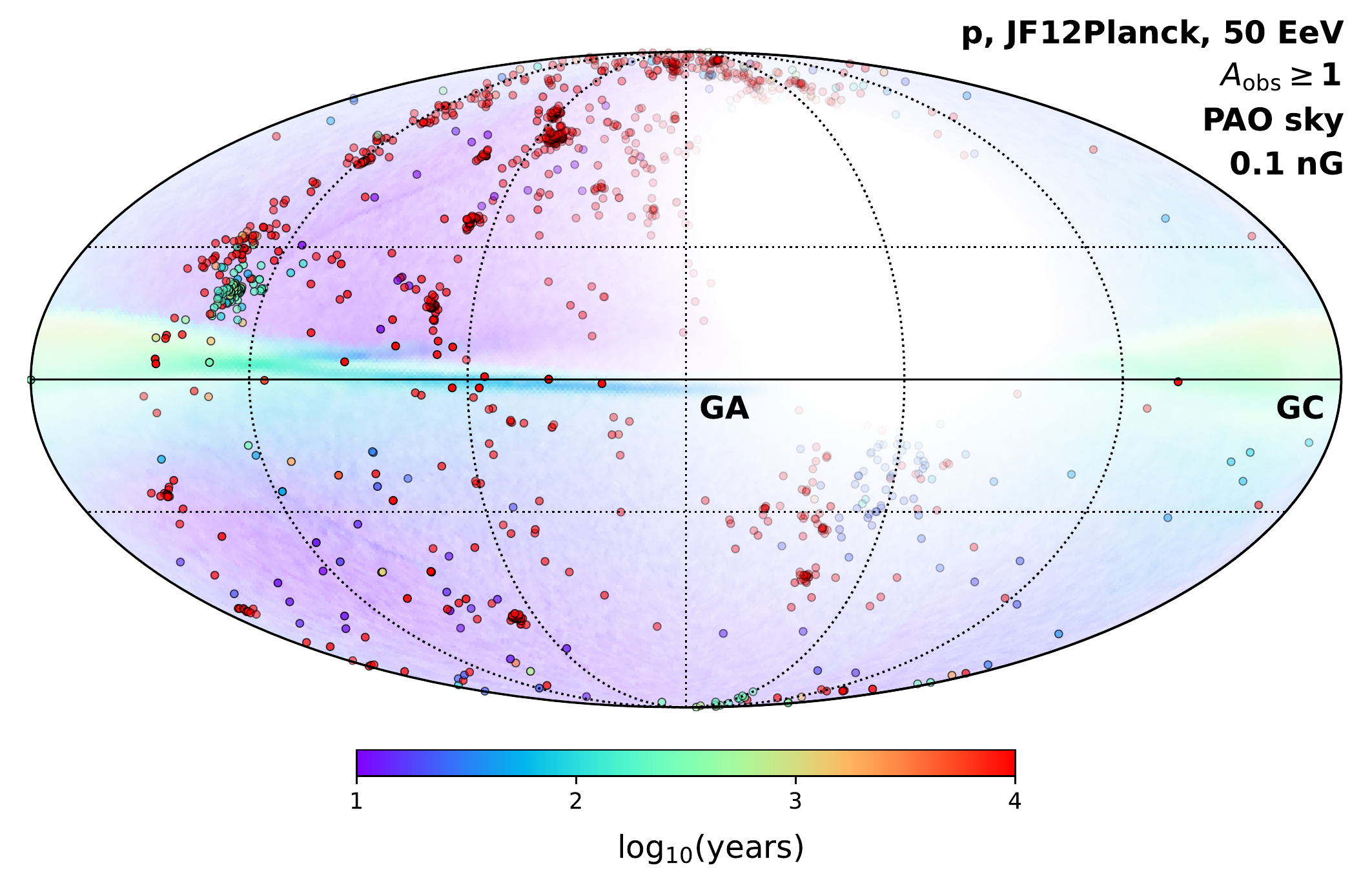}
\includegraphics[width=0.45\textwidth]{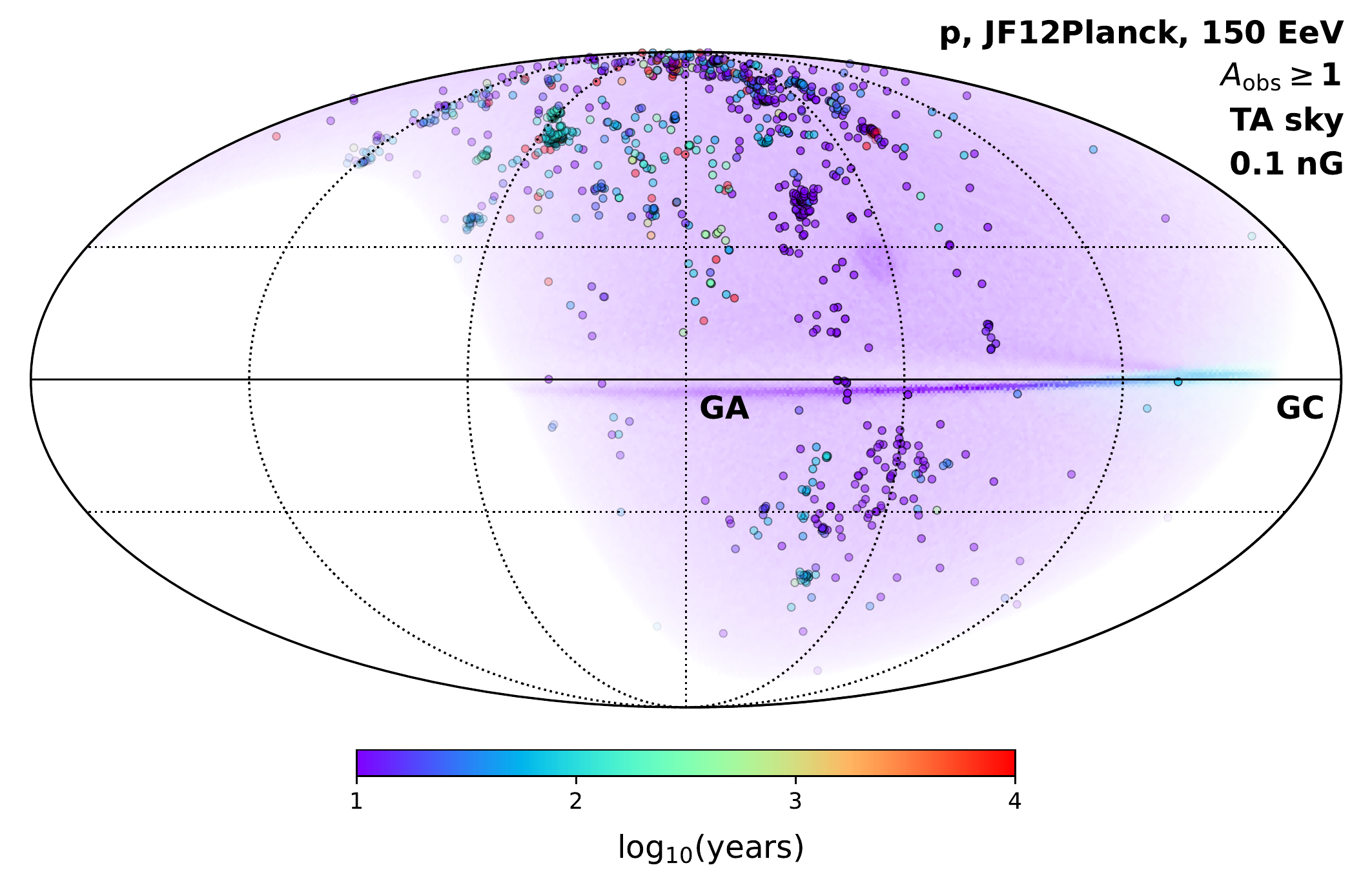}
\includegraphics[width=0.45\textwidth]{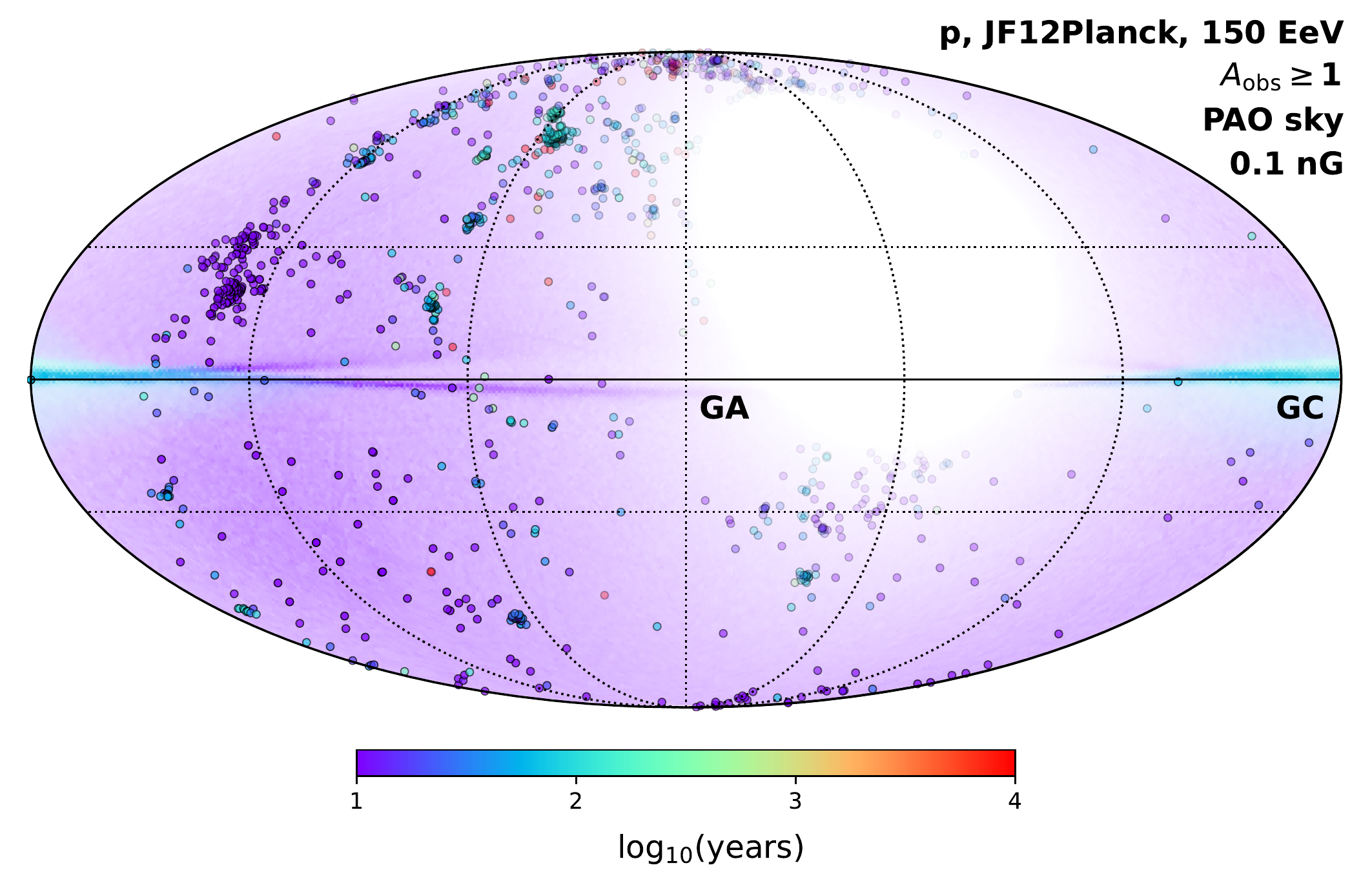}
\includegraphics[width=0.45\textwidth]{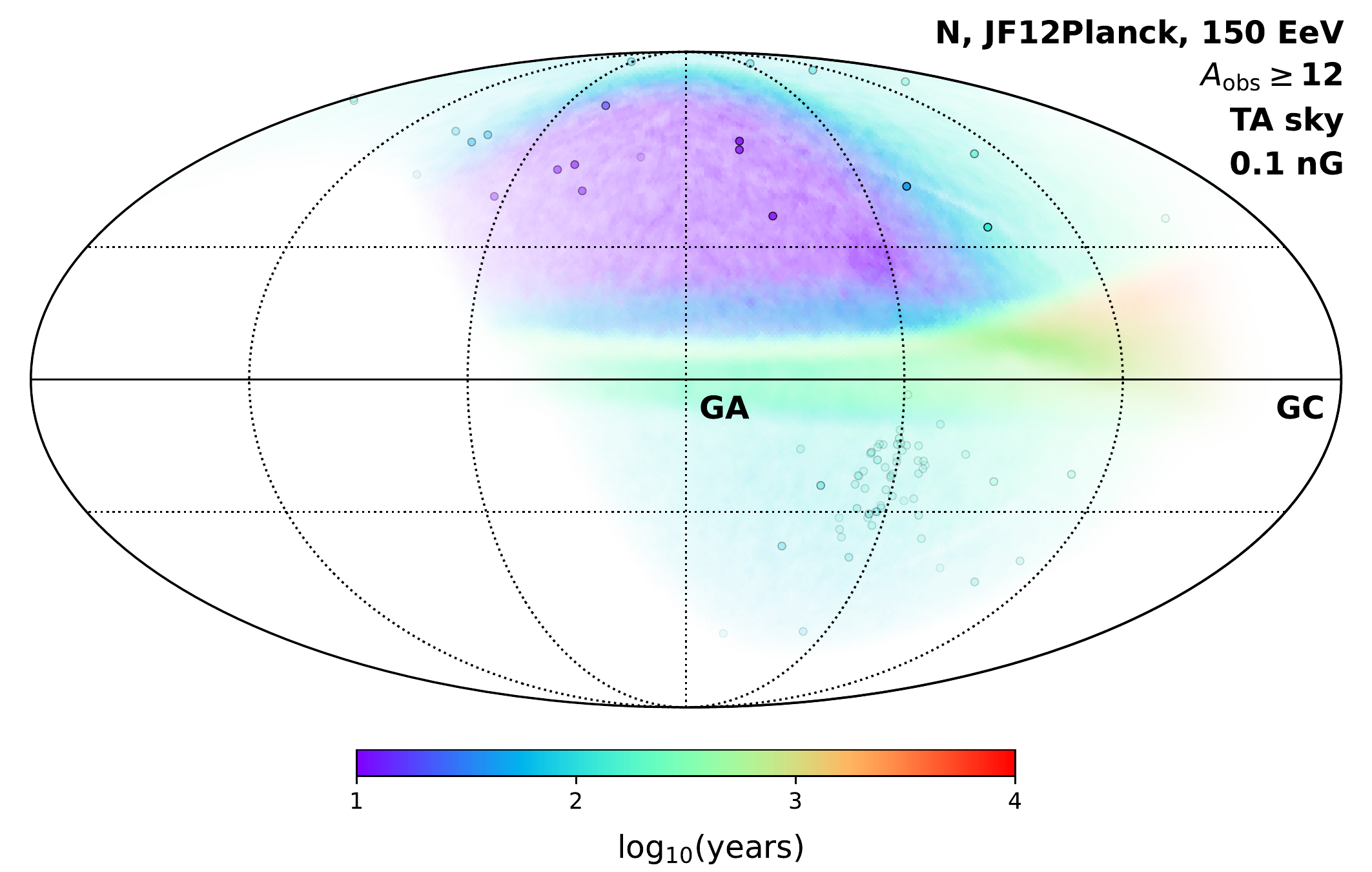}
\includegraphics[width=0.45\textwidth]{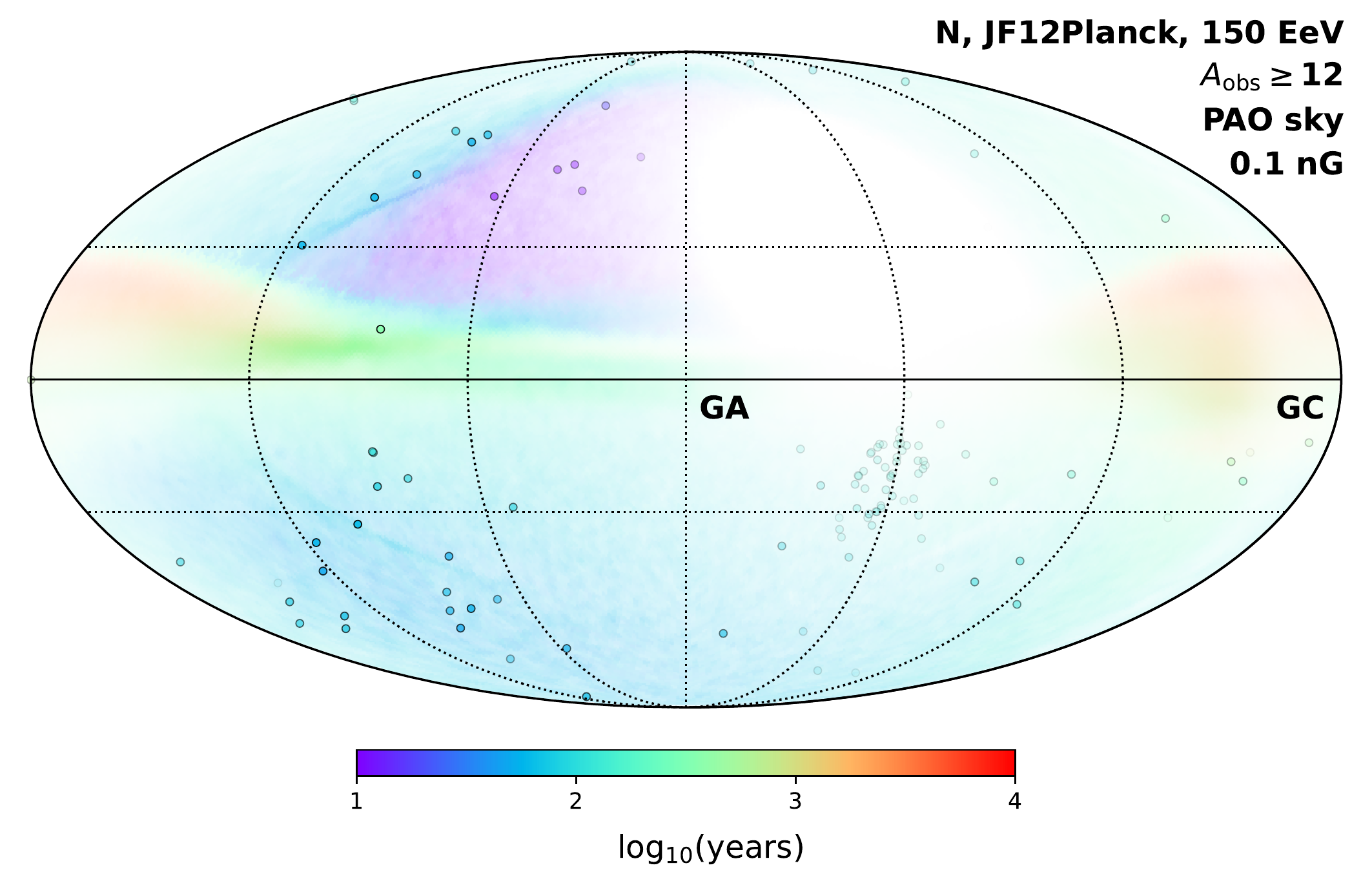}
\includegraphics[width=0.45\textwidth]{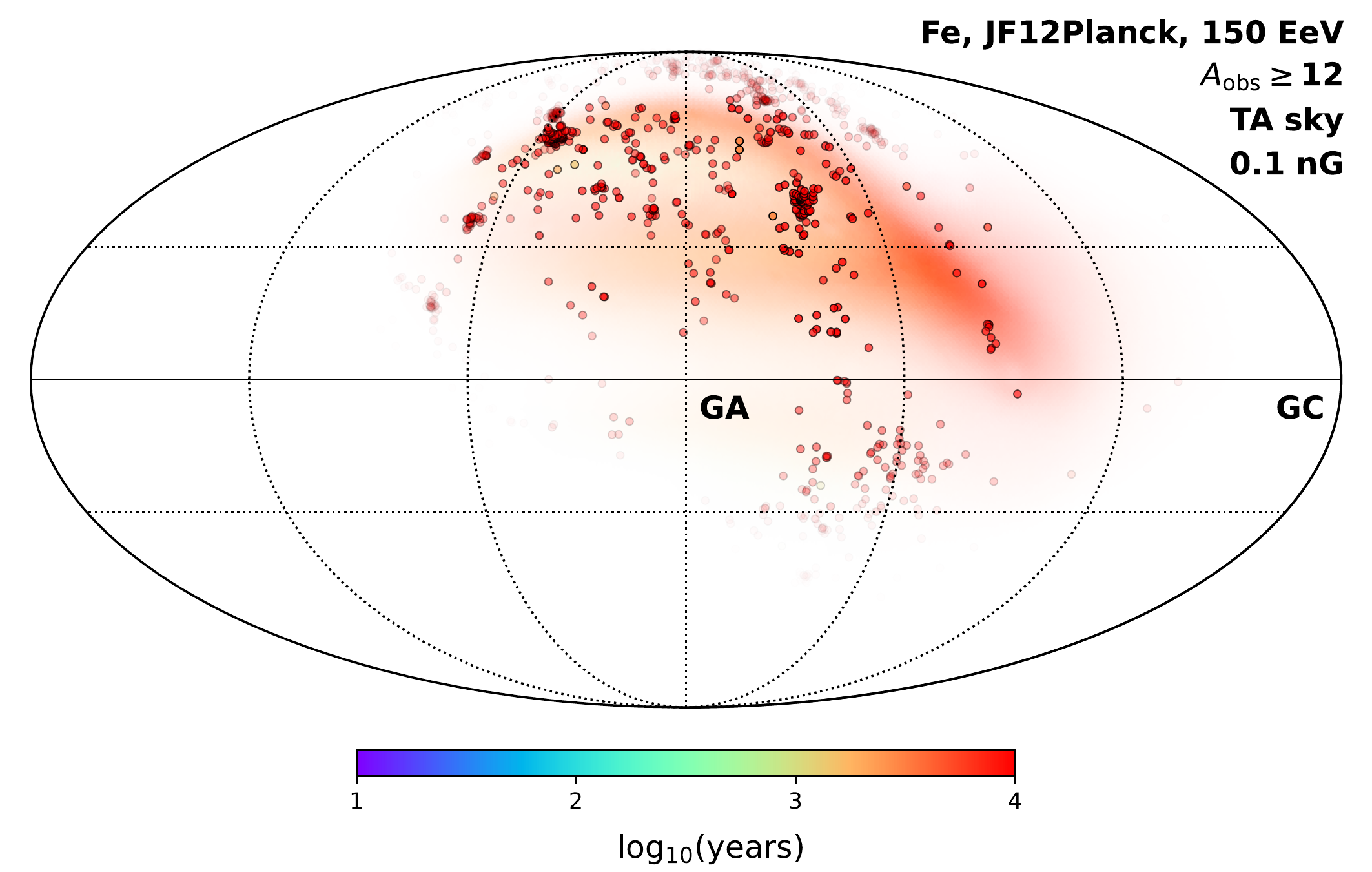}
\includegraphics[width=0.45\textwidth]{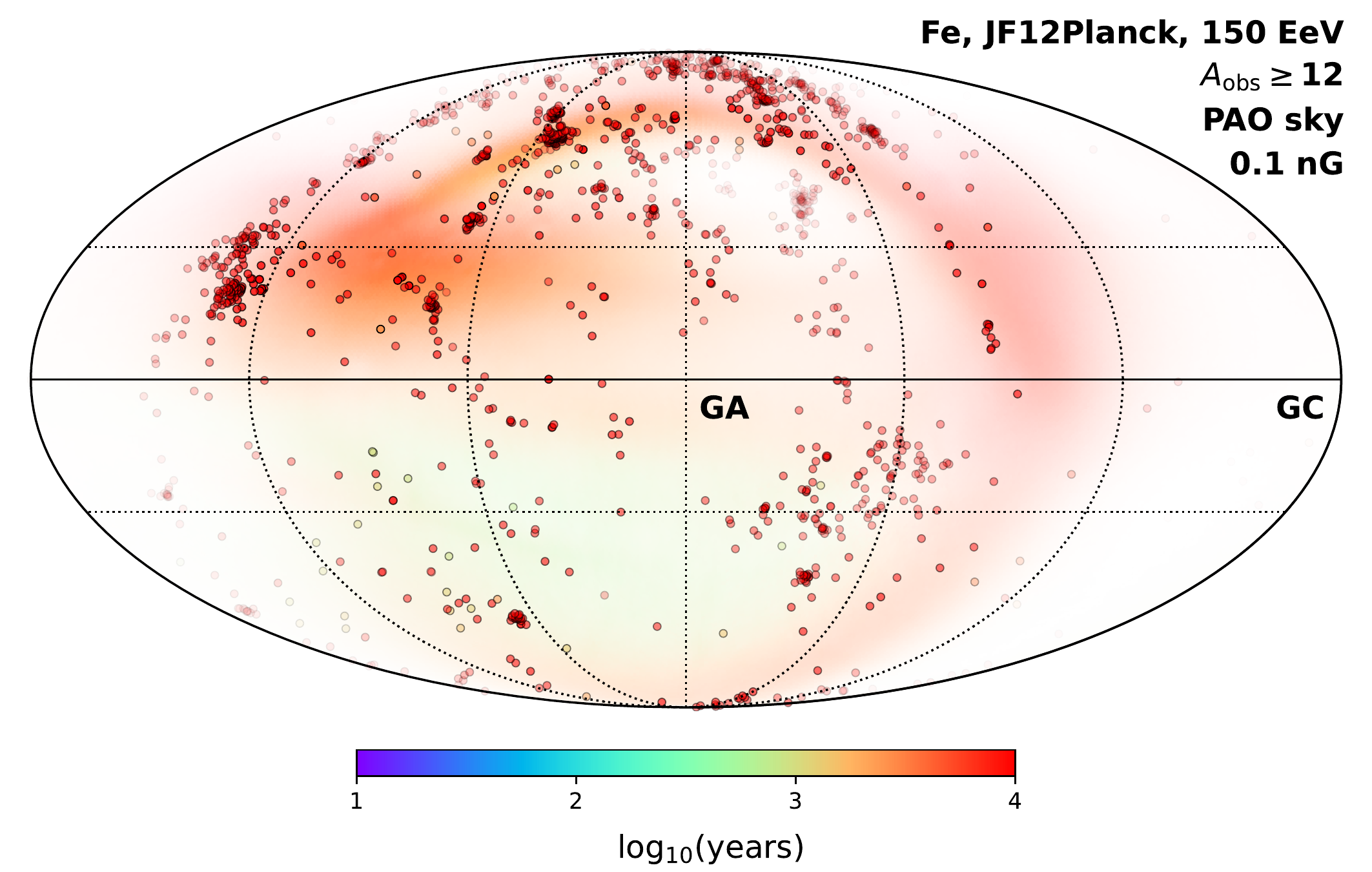}
\caption{Treasure maps for TA (left) and PAO (right), for EGMF of 0.1 nG and the JF12Planck GMF model for different cases of ($E_{\rm obs}$, $A_{\rm obs}$) as indicated. The color of the gradients is assigned to $\tau_{d, \text{GMF}}$. The gradient's opacity is controlled by the truncated magnification maps $\min(M, 1)$, which include the detector exposure function. The source colors are assigned to the total $\tau_{d}$. The source marker opacity is set to $\min(a_\text{GZK}, M)$, i.e.~within the detector's exposure the markers fade mostly due to $a_\text{GZK}$ and due to $M$ outside of it.}
\label{fig:EGMF01}
\end{figure}

\begin{figure}
\centering
\includegraphics[width=0.45\textwidth]{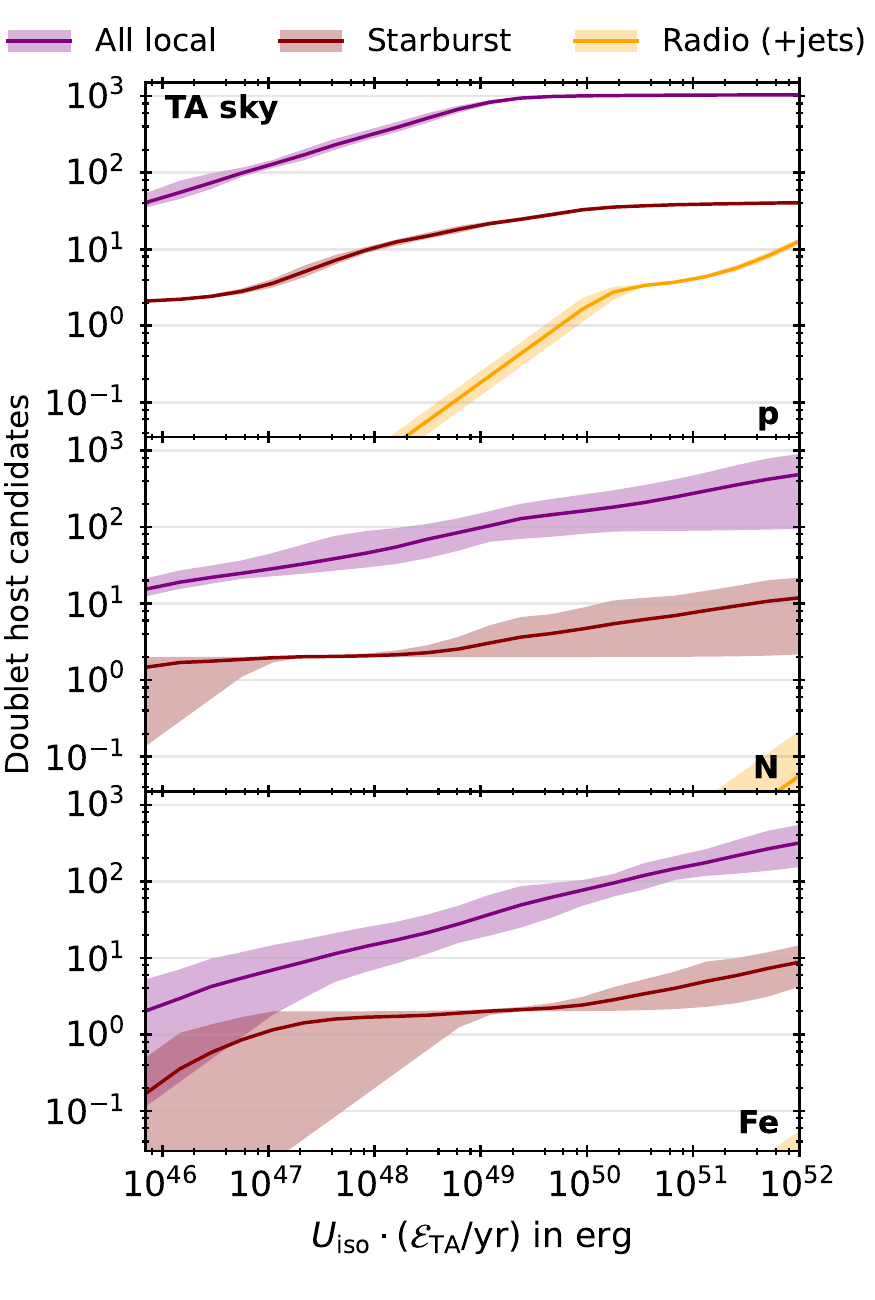}
\includegraphics[width=0.45\textwidth]{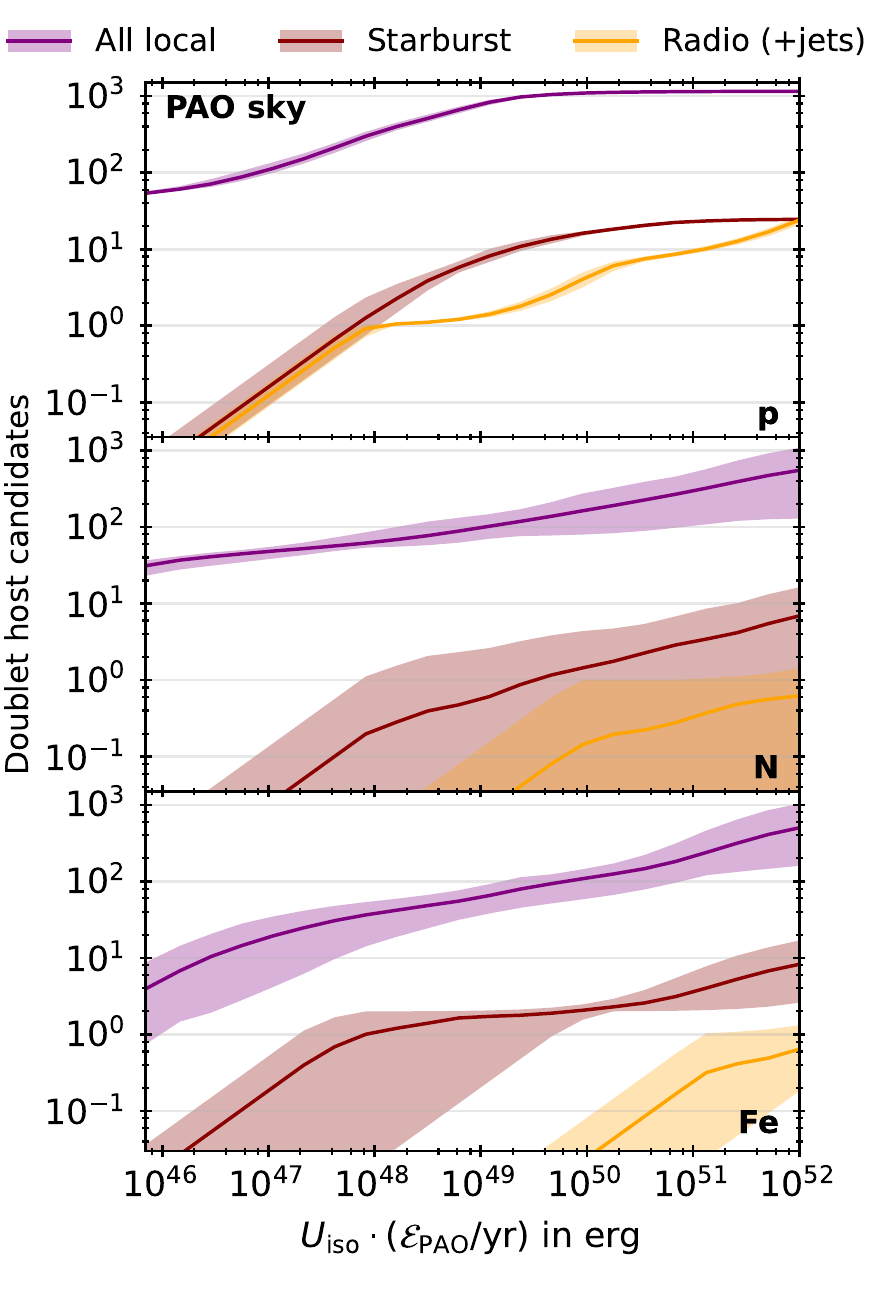}
\caption{Same as Fig.~\ref{fig:doubletsallsky}. The EGMF strength is 0.1~nG.}
\label{fig:doublets01}
\end{figure}

\begin{figure*}
\centering
\includegraphics[width=0.35\textwidth,trim=6.5cm 6.5cm 6.5cm 0.8cm,clip=true]{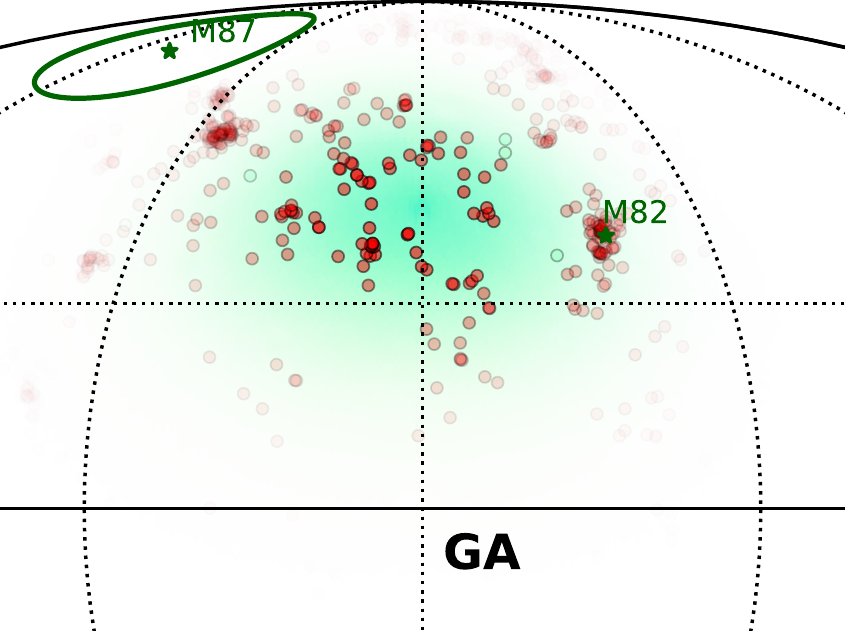}
\includegraphics[width=0.35\textwidth,trim=6.5cm 6.5cm 6.5cm 0.8cm,clip=true]{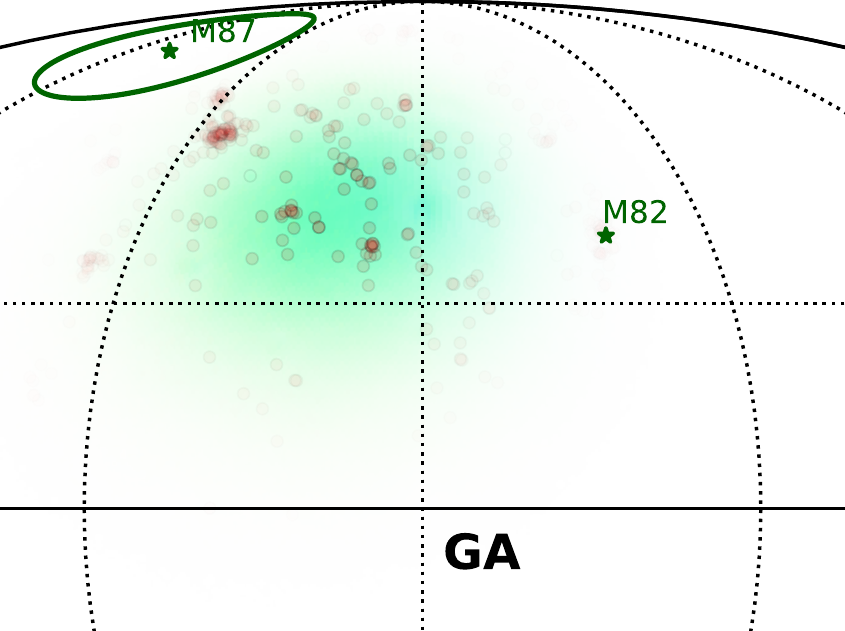}
\includegraphics[width=0.35\textwidth,trim=6.5cm 6.5cm 6.5cm 0.8cm,clip=true]{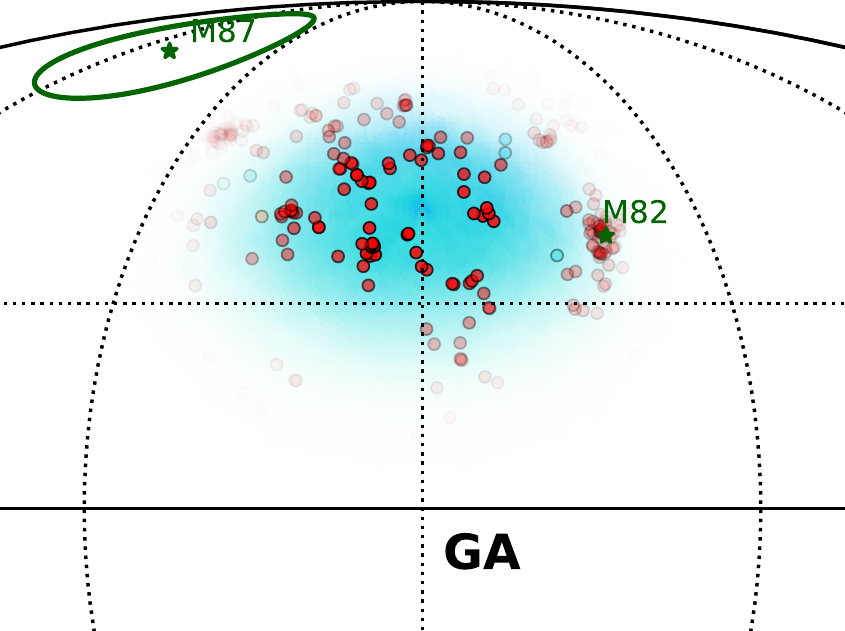}
\includegraphics[width=0.35\textwidth,trim=6.5cm 6.5cm 6.5cm 0.8cm,clip=true]{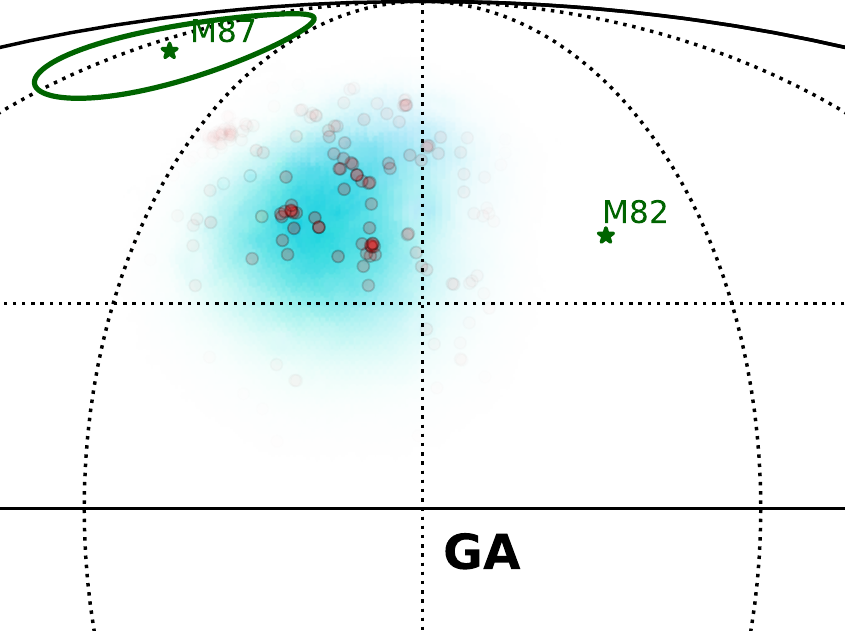}
\includegraphics[width=0.35\textwidth,trim=6.5cm 6.5cm 6.5cm 0.8cm,clip=true]{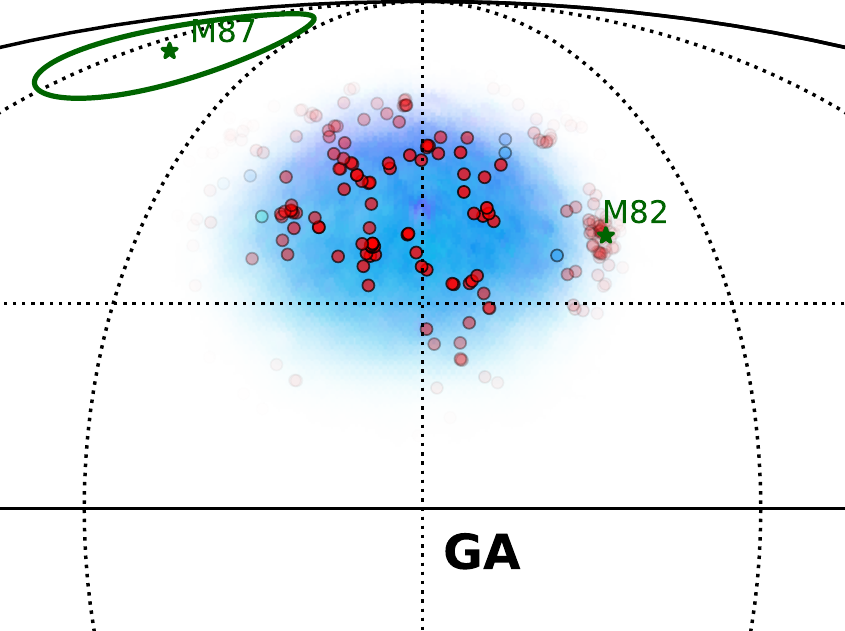}
\includegraphics[width=0.35\textwidth,trim=6.5cm 6.5cm 6.5cm 0.8cm,clip=true]{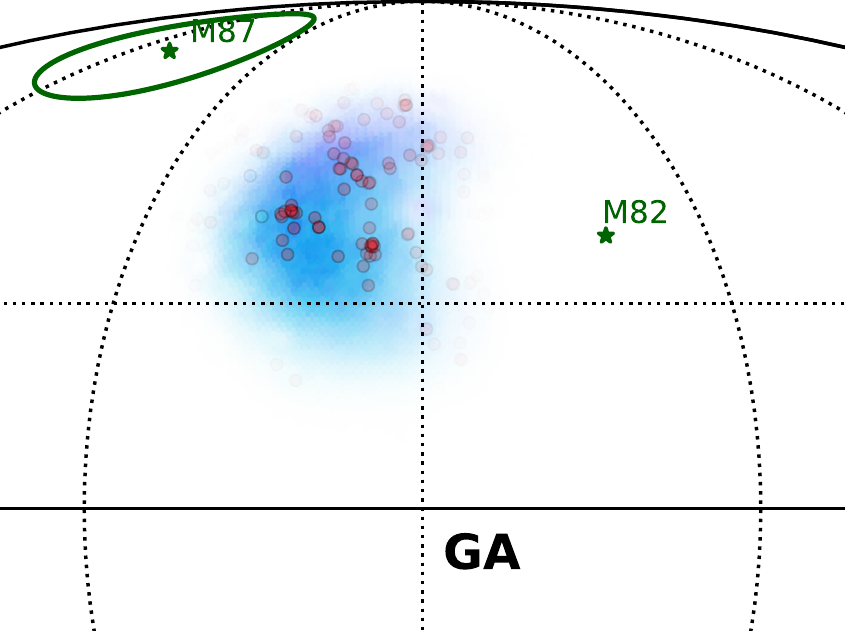}
\includegraphics[width=0.35\textwidth,trim=6.5cm 6.5cm 6.5cm 0.8cm,clip=true]{TA_gmf_delay_JF12planck_p_Ao1_0.1nG_10.0EeV.pdf}
\includegraphics[width=0.35\textwidth,trim=6.5cm 6.5cm 6.5cm 0.8cm,clip=true]{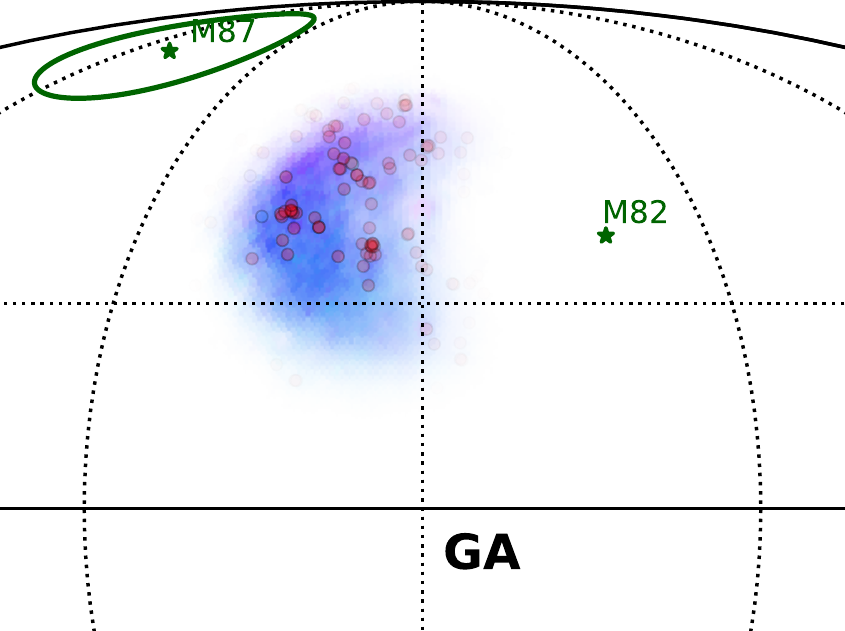}
\includegraphics[width=0.35\textwidth,trim=6.5cm 6.5cm 6.5cm 0.8cm,clip=true]{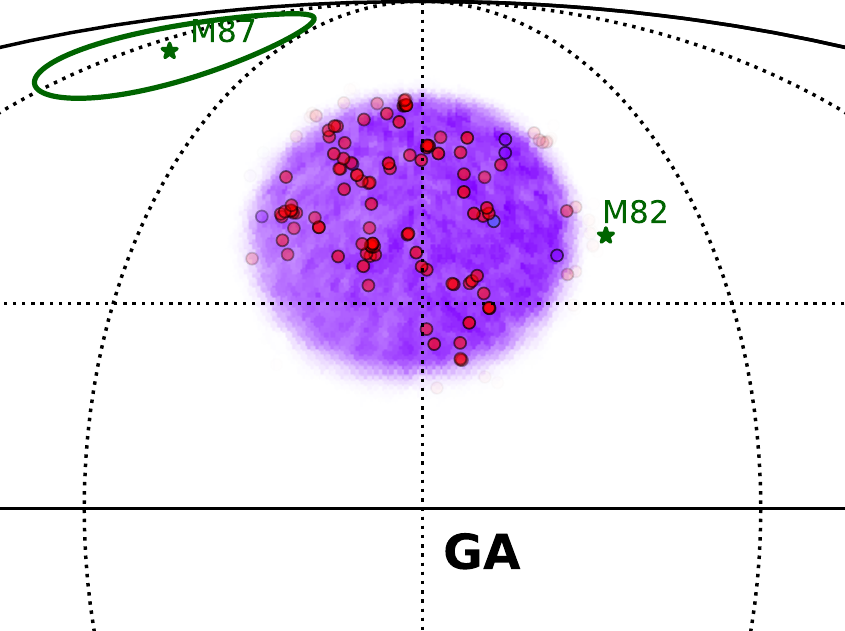}
\includegraphics[width=0.35\textwidth,trim=6.5cm 6.5cm 6.5cm 0.8cm,clip=true]{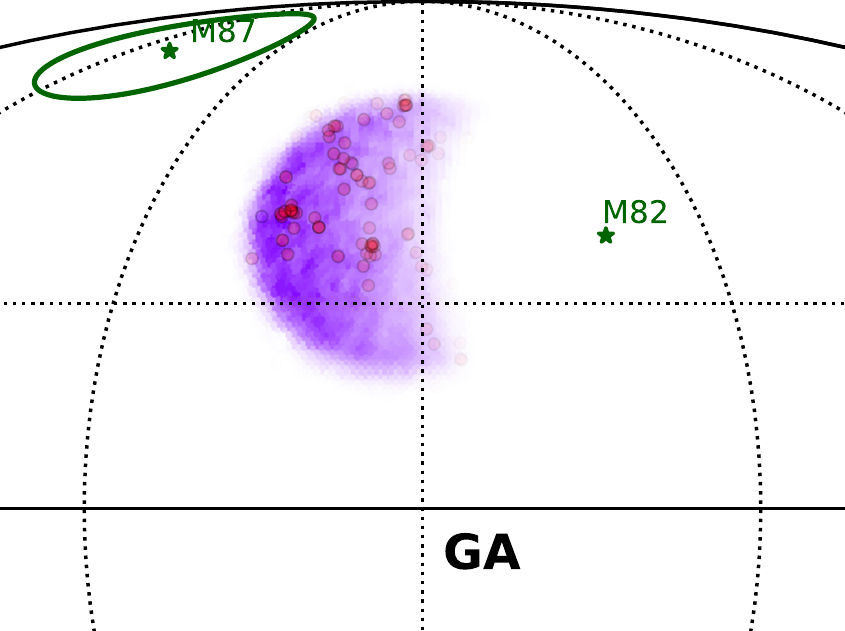}
\caption{Same as Figure \ref{fig:hotspot} but zoom into the back-projected TA hotspot region. The color legend are the same as in Fig.~\ref{fig:hotspot}. The rigidities are, from top to bottom: 2.5, 5, 7.5, 10, 25 EV, respectively. Left panel: TA sky. Right panel: PAO sky.}
\label{fig:TA_01}
\end{figure*}

\end{document}